\documentclass[aip,amsmath,amssymb,reprint,cha]{revtex4-1}

\usepackage[margin=.75in]{geometry}

\usepackage{graphicx}
\usepackage{caption}
\usepackage{comment}
\usepackage{subcaption}
\usepackage{ragged2e}
\DeclareCaptionJustification{justified}{\justifying}
\usepackage{dcolumn}

\usepackage[utf8]{inputenc}
\usepackage[T1]{fontenc}
\usepackage{mathptmx}
\usepackage{etoolbox}

\usepackage{bm}
\usepackage{nicematrix}

\usepackage{color}

\definecolor{indigo}{RGB}{0,0,120}
\usepackage[colorlinks=true, linkcolor=indigo, citecolor=blue, urlcolor=indigo,backref=page]{hyperref}

\newcommand{\tl}[1]{\tilde{#1}}

\newcommand{\pdr}{\partial}
\newcommand{\DD}[2]{\frac {d #1}{d #2}}

\def\diag{\text{diag}}

\newcommand{\beq}{\begin{equation}}
\newcommand{\eeq}{\end{equation}}
\newcommand{\beqs}{\begin{eqnarray}}
\newcommand{\eeqs}{\end{eqnarray}}

\newcommand{\half}{\frac{1}{2}}
\newcommand{\ov}[1]{\frac{1}{#1}}
\newcommand{\fr}[2]{\frac{#1}{#2}}

\def\tr{{\rm tr}\:}

\def\fl{\noindent}

\def\al{\alpha}
\def\g{\gamma}

\def\del{\delta}
\def\D{\Delta}
\def\eps{\epsilon}

\def\ka{\kappa}
\def\la{\lambda}

\def\tht{\theta}	
		
\def\om{\omega}

\def\vphi{\varphi}

\def\IS{{\rm IS}}
\def\PD{{\rm PD}}
\def\Es{{\rm Es}}
\def\Ec{{\rm Ec}}
\def\Ms{{\rm Ms}}
\def\Mc{{\rm Mc}}
\newcommand{\Lc}{{\rm \: Lc}}
\newcommand{\Ls}{{\rm \: Ls}}
\newcommand{\sn}{{\rm \: sn}}

\usepackage{titlesec}
\titleformat{\section}{\normalsize\bfseries}{\thesection}{1em}{}
\titleformat{\subsection}{\small\bfseries}{\thesubsection}{1em}{}
\titleformat{\subsubsection}{\small\bfseries}{\thesubsubsection}{1em}{}

\newcount\colveccount  
\newcommand*\colvec[1]{\global\colveccount#1  \begin{pmatrix} \colvecnext} \def\colvecnext#1{#1 \global\advance\colveccount-1
        \ifnum\colveccount>0 \\ \expandafter\colvecnext
        \else \end{pmatrix} \fi}

\newenvironment{smmat}
  {\left(\begin{smallmatrix}}
  {\end{smallmatrix}\right)}

\usepackage{calligra} 
\DeclareMathAlphabet{\mathcalligra}{T1}{calligra}{m}{n}
\DeclareFontShape{T1}{calligra}{m}{n}{<->s*[2.2]callig15}{}

\begin{document}

\title[Bifurcation cascade, self-similarity and duality in the 3-rotor problem \hfill {\tt \hfill\href{https://arxiv.org/abs/2303.01057}{arXiv:2303.01057[nlin.CD]}}]
{Bifurcation cascade, self-similarity and duality in the 3-rotor problem}

\author{Govind S. Krishnaswami}
\email{govind@cmi.ac.in}
\author{Ankit Yadav}
\email{ankit@cmi.ac.in}
\affiliation{Physics Department, Chennai Mathematical Institute, SIPCOT IT Park, Siruseri 603103, India}

\date{1 June 2023}

\begin{abstract}

Published in \href{https://doi.org/10.1063/5.0160496}{Chaos {\bf 33},  083101 (2023); DOI 10.1063/5.0160496}. \\

\fl The three-rotor system concerns equally massive point particles moving on a circle subject to attractive cosine potentials of strength $g$. The quantum theory models chains of coupled Josephson junctions. Classically, it displays order-chaos-order behavior with increasing energy $E$ along with a seemingly globally chaotic phase for $5.33g \lesssim E \lesssim 5.6g$. It is also known to admit pendulum and isosceles breather families of periodic orbits at all energies. While pendula display a doubly infinite sequence of stability transitions accumulating at their libration to rotation threshold at $E = 4g$, breathers undergo only one stability transition. Here, we show that these stability transitions are associated with forward and reverse fork-like isochronous and period-doubling bifurcations. The new family of periodic orbits born at each of these bifurcations is found using an efficient search algorithm starting from a transverse perturbation to the parent orbit. The graphs of stability indices of various classes of orbits born at pendulum bifurcations meet at $E = 4g$ forming `fans'. The transitions in the librational and rotational phases are related by an asymptotic duality between bifurcation energies and shapes of newly born periodic orbits. The latter are captured by solutions to a Lam\'e equation. We also find and numerically validate values of scaling constants for self-similarity in (a) stability indices of librational and rotational pendula and (b) shapes of newly born orbits as $E \to 4g$. Finally, we argue that none of the infinitely many families of periodic orbits we have found is stable for $5.33g \lesssim E \lesssim 5.6g$, providing further evidence for global chaos in this energy band.

\end{abstract}

\maketitle

\footnotesize \tableofcontents

\vspace{1cm}

\normalsize

\begin{quotation}


The idea that periodic orbits can serve as a tool in the study of classical systems has been recognized since the time of Poincar\'e. On the other hand, bifurcations encode qualitative changes in a dynamical system and often involve universal behavior. Interestingly, bifurcations of a family of periodic orbits can be used to find new periodic orbits. The infinite sequence of period-doubling bifurcations associated with the onset of chaos in the logistic map is a well-researched example. Bifurcation cascades of periodic orbits have also been discovered in Hamiltonian systems such as H\'enon-Heiles. In the latter, the cascade displays scale-invariance and fan-like structures although the connection to chaos is still unclear. Thus, it is important to understand the common/distinctive phenomena surrounding such cascades in other interesting examples. Here we investigate a doubly infinite sequence of bifurcations at stability transitions in the `pendulum' family of periodic orbits of the 3-rotor system. In the latter, neighboring rotors interact via the cosine of the relative angle. It arises as a classical limit of a cyclic chain of coupled Josephson junctions used in superconducting qubits. We develop a search algorithm that exploits orbital symmetries to find periodic orbits that germinate at fork-like isochronous and period-doubling bifurcations of pendula. The bifurcations accumulate geometrically at the energy threshold between librational and rotational pendula. We estimate scaling exponents that characterize the self-similarity in stability indices of pendula and shapes of periodic orbits born at the bifurcations. Remarkably, we also discover a duality that relates bifurcation energies and shapes of new periodic orbits in the librational and rotational phases. Moreover, stability indices of classes of newly born orbits form forward and backward fans that meet at the self-dual energy, which intriguingly is also the energy around which widespread chaos sets in. One hopes that a common framework may be developed to describe these bifurcation cascades, scaling symmetries, fan-like structures and their possible implications for chaos in a variety of few-degree of freedom Hamiltonian systems.

\end{quotation}

\section{Introduction}
\label{intro}

It is well known that periodic orbits play a prominent role in nonlinear and chaotic systems. In fact, Poincar\'e suggested that periodic orbits can help to understand the dynamics and that they could be plentiful or even dense (especially for bounded motions) in the space of trajectories\cite{gutzwiller-book}. In the quantum theory, periodic orbits enter via semi-classical trace formulas \cite{gutzwiller}. Thus, it is of interest to find and classify periodic orbits in a dynamical system. On the other hand, families of periodic orbits at their stability transitions are known to undergo bifurcations producing new families of periodic orbits\cite{isoch-per-dou-bifur}. Thus, bifurcations of a known family of periodic orbits may be used as a tool to discover new periodic orbits. What is more, bifurcations of families of periodic orbits are interesting even in the quantum theory as they introduce subtleties in semiclassical trace formulas \cite{brck-omega, ellip-bill}.

The system we study in this paper concerns the conservative dynamics of three coupled rotors: point particles of equal mass moving on a circle subject to attractive cosine inter-particle potentials. This 3-rotor problem was introduced in Ref.~\cite{gskhs-3rotor} and arises as a classical limit of a cyclic chain of 3 coupled Josephson junctions\cite{shnirman, mooij,sondhi-girvin,class-JJ}. Since the rotor angles represent superconducting phases of distinct metallic segments, they can coincide. Thus, it is reasonable for the rotors to pass through each other without any collisional singularities. The 3-rotor system has been shown\cite{gskhs-3rotor,gskhs-3rotor-ergodicity} to display rich dynamics including families of periodic orbits (pendula and isosceles breathers), order-chaos-order behavior with increasing energy and a band of energies where the dynamics appears to be globally chaotic and displays ergodicity and mixing. What is more, the onset of widespread chaos seemed to coincide with an accumulation of stability transitions in the pendulum family of periodic orbits. Thus, the 3-rotor problem offers an arena to study these and related phenomena without having to deal with collisions or escape to infinity.

In this paper, we propose and use an accurate and efficient search algorithm to find the new periodic orbits born at stability transitions of pendula and isosceles breathers. By examining the properties of these new families of orbits, we characterize the isochronous and period-doubling  bifurcations at the stability transitions as forward and reverse fork-like and also discover several remarkable phenomena associated with the bifurcation cascade of pendula. These include scale-invariance and scaling constants, `fans', a libration-rotation duality and a period-doubling analog of the fork-like bifurcation (FLB) slope theorem\cite{brck-fork}. Although doubly infinite, as it involves both libration and rotation, the pendulum cascade is reminiscent of that in the H\'enon-Heiles system \cite{church-HH-survey} with pendula playing the role of orbit A. However, the numerical challenges here are greater since the cascade begins closer to the accumulation point. Among other things, it is important to have examples of such bifurcation cascades in Hamiltonian systems to determine which phenomena are common/system-specific and also to discover any universal features or quantities that one may compute. As a by-product of our investigation of pendulum and isosceles breather bifurcations, we argue that none of the parent or daughter families of periodic orbits is stable in the energy band identified in Ref.\cite{gskhs-3rotor}, giving further evidence for global chaos in this regime. This is particularly interesting since there are hardly any examples of physically realizable continuous time Hamiltonian systems without specular reflections that display global chaos.

\section{Formulation of the problem}
\label{s:formulation-3-rotor}

\begin{figure}[!h]
\centering
	\begin{subfigure}{0.23\textwidth}
	\includegraphics[width = \textwidth]{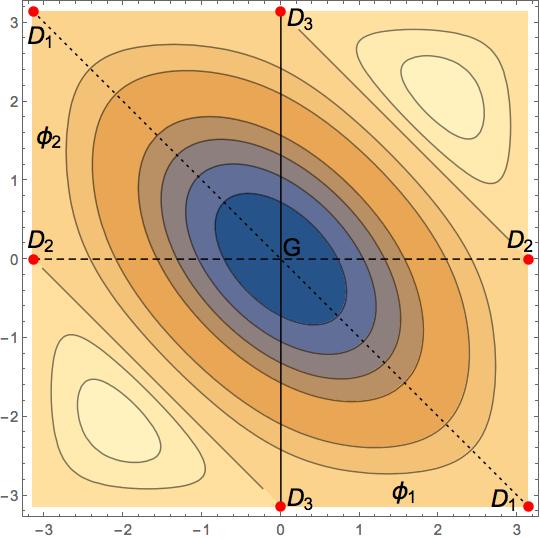}
	\caption{}
	\end{subfigure}
\hfil
	\begin{subfigure}{0.23\textwidth}
	\includegraphics[width = \textwidth]{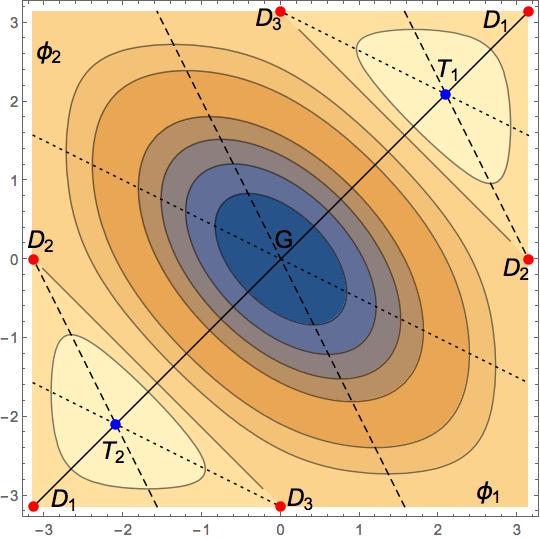}
	\caption{}
	\end{subfigure}
	\caption{\small Equipotentials and minima, saddles and maxima ($G, D$ and $T$) of $V$ on the fundamental $\vphi_1-\vphi_2$ square. (a) The three types of pendula each occupies a portion of the $E = 4g$ homoclinic orbit (denoted by dotted, dashed or solid lines) through one of the three saddle points. (b) Similarly, isosceles breathers lie along an $E = 4.5g$ heteroclinic orbit between maxima.}
	\label{f:equipotn-pend-breather}
\end{figure}

The classical 3-rotor problem introduced in Ref.~\cite{gskhs-3rotor} concerns the conservative dynamics of three coupled rotors, point particles of equal mass $m$ moving without collisions on a circle of radius $r$ subject to attractive cosine inter-particle potentials of coupling strength $g$. If the $2\pi$-periodic rotor angles are denoted $\tht_{1,2,3}$, the potential energy $V(\tht_1, \tht_2, \tht_3)$ is
	\beq
	V = g[3 - \cos(\tht_1 - \tht_2) - \cos(\tht_2 - \tht_3) - \cos(\tht_3 - \tht_1)].
	\eeq
The associated Lagrangian
	\beq
	L = \half m r^2 (\dot \tht_1^2 + \dot \tht_2^2 + \dot \tht_3^2) - V(\tht_1, \tht_2, \tht_3),
	\eeq
is invariant under transformations of the group $\rm S_3 \times \mathbb Z_2$ of all permutations and reflections ($\tht_j \to - \tht_j$) of the rotor angles. It is convenient to introduce center of mass and relative angles
	\beq
	\vphi_0 = \fr{\tht_1 + \tht_2 + \tht_3}{3}, \quad \vphi_1 = \tht_1 - \tht_2 \quad \text{and} \quad \vphi_2 = \tht_2 - \tht_3.
	\eeq
In terms of these, the Lagrangian becomes
	\beqs
	L &=& \fr{3}{2} m r^2 \dot \vphi_0^2 + \fr{m r^2}{3}(\dot \vphi_1^2 + \dot \vphi_2^2 + \dot \vphi_1 \dot \vphi_2) - V(\vphi_1, \vphi_2) \quad \text{where} \cr
	V &=& g[3 - \cos \vphi_1 - \cos \vphi_2 - \cos (\vphi_1 + \vphi_2)].
	\label{e:lagr-poten-phi1-phi2}
	\eeqs
The center of mass (CM) angle $\vphi_0$ is cyclic and the dynamics of the relative angles $\vphi_1$ and $\vphi_2$ decouples from that of $\vphi_0$:
	\beq
	3 m r^2 \ddot \vphi_0 = 0, \quad 
	m r^2 (2 \ddot \vphi_1 + \ddot \vphi_2) = -3g[\sin \vphi_1 + \sin(\vphi_1 + \vphi_2)]
	\eeq
and the equation obtained from $1 \leftrightarrow 2$. The conserved total energy is a sum of CM and relative contributions:
	\beqs
	E_{\rm tot} &=& E_{\rm CM}(\vphi_0) + E_{\rm rel}(\vphi_1, \vphi_2) \cr &=& \fr{3}{2} m r^2 \dot \vphi_0^2 
	+ \fr{m r^2}{3}(\dot \vphi_1^2 + \dot \vphi_2^2 + \dot \vphi_1 \dot \vphi_2) + V(\vphi_1, \vphi_2).
	\label{e:egy-CM-rel}
	\eeqs
Remarkably, the relative energy $E_{\rm rel}$ (hereafter referred to as $E$) is closely related to that of a three Josephson junction qubit. This is clarified in Appendix \ref{s:JJ-qubit}. Henceforth, we focus on the two-degree-of-freedom relative motion on the $\vphi_1$-$\vphi_2$ torus ($0 \leq \vphi_{1,2} \leq 2\pi$). To begin with, the potential energy $ V(\vphi_1, \vphi_2)$ (\ref{e:lagr-poten-phi1-phi2}) has three kinds of extrema: a local minimum $G$ at $(0, 0)$, 2 local maxima $T_{1,2}$ at $(\pm 2 \pi/3, \pm 2 \pi/3)$  and 3 saddle points $D_{1,2,3}$ at $(\pi, \pi)$,  $(\pi, 0)$ and $(0, \pi)$, corresponding to the energies $0$, $4.5g$ and $4g$ respectively (see Fig~\ref{f:equipotn-pend-breather}). In Ref.~\cite{gskhs-3rotor, him-sen-thesis} it was shown that the relative dynamics is integrable at asymptotically low as well as high energies. There is a transition to widespread chaos around $E = 4g$ with a band of seemingly global chaos for energies in the range $5.33 g \lesssim E \lesssim 5.6 g$. Moreover, the relative dynamics was shown to admit three simple types of periodic orbits: pendula, isosceles breathers and choreographies. Pendula and breathers are relevant to the present work; they exist at all $E$ and transition from libration to rotation with increasing $E$.

\paragraph*{Pendula.} In pendula, two of the three rotors always coincide. Depending on which pair are bound together (i.e., $\tht_1 \equiv \tht_2$ etc.), there are three types of pendulum solutions which are permuted among each other by the symmetries of the potential. As shown in Fig.~\ref{f:equipotn-pend-breather}, they lie along portions of the three straight lines on the $\vphi_1-\vphi_2$ square: $\vphi_1 \equiv 0$, $\vphi_2 \equiv 0$ and $\vphi_1 + \vphi_2 \equiv 0$. 
For definiteness, we focus on the $\vphi_1 \equiv 0$ type of pendula. In this case, the nontrivial equation of motion (EOM) reduces to that of a pendulum: $m r^2 \ddot\vphi_2 = - 3 g \sin \vphi_2$, justifying the name. For $E < 4g$, pendula are librational and have two turning points $\pm \vphi_2^*$ which approach the saddle point $D_3$ as $E \to 4g$. For $E > 4g$, the pendula are rotational with trajectories winding around the $\vphi_2$ cycle of the torus. The corresponding librational and rotational pendulum solutions are
$\bar \vphi_1 \equiv 0$ and
	\beq
	\bar{\vphi}_2 = \begin{cases} 2 \arcsin ( k \sn( \om_0 t,k)) \quad &\text{for} \quad 0 \leq E < 4g,
	\\ 2 \arcsin ( \tanh( \om_0 t)) \quad &\text{when} \quad  E = 4g,
	\\ 2 \arcsin (  \sn( \om_0 t / \ka, \ka)) \quad &\text{for} \quad E > 4g,
	\end{cases}
	\label{phi1=0-pend}
	\eeq
where $\om_0 = \sqrt{3g / m r^2}$ and $\sn$ is the Jacobi elliptic sine function with elliptic modulus $k = 1/\ka = \sqrt{E/4g}$. The periods
	\beq
	\tau_\ell = 4 K(k)/ \om_0 \quad \text{and} \quad
	\tau_r = 2 \ka K(\ka) / \om_0
	\label{e:time-per-lib-rot-pend}
	\eeq
both diverge logarithmically [$\tau \approx \om_0^{-1} \log(1-E/4g)$] as $E \to 4g$. $K$ is the complete elliptic integral of the first kind.

\paragraph*{Isosceles Breathers.} In breathers, one rotor is always at the CM which is midway between the other two rotors so that they always lie at the vertices of an isosceles triangle. The `peripheral' rotors oscillate symmetrically about the CM justifying the name breather (see Fig.~3(b) of Ref.~\cite{gskhs-3rotor}). There are three types of breathers depending on which of the three rotors is equidistant from the other two. As shown in Fig.~\ref{f:equipotn-pend-breather}, they lie along portions of the three closed curves made of straight lines on the $\vphi_1-\vphi_2$ square that join the two maxima ($T_{1,2}$) of $V$ while passing through $G$ and one among $D_{1,2,3}$. They can be represented by the lines $\vphi_1 = \vphi_2$, $\vphi_1 + 2 \vphi_2 = 0$ and $2 \vphi_1 + \vphi_2 = 0$ passing through $D_1$, $D_3$ and $D_2$. In each type we may distinguish two librational and one rotational family depending on initial conditions (ICs): LG for $E \leq 4.5g$, LD for $4g \leq E \leq 4.5g$ and R for $E \geq 4.5g$ (see Fig.~5 of Ref.~\cite{gskhs-3rotor}).

A distinction between pendula and breathers is that the separatrix pendulum (at the libration to rotation threshold energy $4g$) is a homoclinic orbit at one of the saddle points $D$ of $V$, while the separatrix breather (with $E = 4.5g$) is a heteroclinic orbit joining the maxima $T_1$ and $T_2$. Remarkably, pendula display a geometric accumulation of stability transitions as $E \to 4g^\pm$, which is also the energy at which widespread chaos sets in. By contrast, breathers display only one stability transition, which occurs at $E \approx 8.97g$. In Ref.\cite{gskhs-3rotor}, questions were raised about the nature of possible bifurcations, scaling and self-similar behavior at these stability transitions. In this paper we address several of these and related questions.


\paragraph*{Comparison with cascades in H\'enon-Heiles and an anharmonic oscillator.} Before summarizing our results, we note that the above-mentioned behavior of pendula is reminiscent of the cascade of stability transitions known to occur in a 2d anharmonic oscillator studied by Yoshida \cite{anhrm-oscl} and in the H\'enon-Heiles system\cite{church-HH-survey, brck-omega}. The former system is governed by
	\beq
	H_{\rm anharm} = \fr{p_1^2 + p_2^2}{2} + \fr{q_1^4 + q_2^4}{4} + \al q_1^2 q_2^2
	\eeq
and admits a family of periodic orbits (with $q_1 \equiv 0$) for all values of the coupling constant $\al$. The family undergoes an infinite sequence of stability transitions as $\al$ goes from $0$ to $\infty$. Similarly, in the H\'enon-Heiles system with energy
	\beq
	E_{\rm HH} = \half (\dot x^2 + \dot y^2) + \half (x^2 + y^2) + \eps \left( x^2 y - \ov{3} y^3 \right),
	\eeq
there is a family of straight-line periodic trajectories (the so-called orbit A with, say, $x \equiv 0$) that shows a geometric accumulation of stability transitions as the energy $E$ goes from $0$ to the  saddle point energy $E^* = 1/(6 \eps^2)$. In fact, in Ref.\cite{church-HH-per-orb}, the authors propose conditions   for a one-parameter family of periodic orbits labeled by energy to undergo an infinite sequence of stability transitions. Using the numerical methods developed in Ref.~\cite{brng-pert}, the authors of Ref.~\cite{brng-HH, brck-omega} found the new families of periodic orbits born at these transitions and thereby determined the nature of the corresponding bifurcations. The stability of the newly born orbits was investigated and scaling constants were estimated.

Inspired in part by these developments, in this paper, we look for new periodic orbits of the three-rotor system, which germinate at the stability transitions of pendula and breathers and study their properties. In fact, we may view pendula as 3-rotor analogs of the librational A orbits of H\'enon-Heiles. Each family of orbits lies along a straight line (on the corresponding configuration space) that ends at a saddle point of the potential. However, there are differences. To begin with, the 3-rotor equations have trigonometric nonlinearities unlike the quadratic ones in H\'enon-Heiles. Moreover, the stability transitions in the 3-rotor system begin much closer to the saddle energy compared with H\'enon-Heiles. Consequently, the associated magnification constant $\del$ is larger for the 3-rotor system, necessitating greater numerical precision to establish scale-invariance. Furthermore, while the motion in the 3-rotor system is always bounded on its configuration space, in H\'enon-Heiles, there is a transition from bound to unbound motion when the energy $E$ exceeds $E^*$. As a consequence, the A orbits cease to be periodic when $E > E^*$. By contrast, pendula are periodic both in their librational $(E < 4g)$ and rotational $(E > 4g)$ phases. Thus, unlike Yoshida's oscillator or H\'enon-Heiles, the 3-rotor system displays a doubly infinite sequence of bifurcations that accumulate from both sides at $E = 4g$, leading to new phenomena. In addition, while the H\'enon-Heiles and the anharmonic oscillator are systems with two degrees of freedom, the 3-rotor problem goes from three to two degrees of freedom upon restricting to relative motion of rotors.

\section{Summary of results} 
\label{s:summary-results}

In Ref.~\cite{gskhs-3rotor} the stability of pendula (with $\vphi_1 \equiv 0$) was examined by numerical diagonalization of the monodromy matrix $M$, whose trace determines linear stability, with transitions occurring when $\tr M$ crosses $0$ or $4$. We revisit this problem in \S \ref{s:pert-pend-Lame-M} with an eye toward more precise numerical determination of transition energies, which are required to find new families of periodic orbits. It turns out that the pendulum perturbation equations in the $\vphi_1$-$\vphi_2$ variables are coupled Lam\'e equations. Here we propose new angular variables $(\al_1, \al_2)$ to decouple them into transverse and longitudinal Lam\'e equations which block diagonalizes the monodromy matrix $M = \diag(M_1 = M_\perp, M_2 = M_\parallel)$. Since\cite{gutzwiller} $\tr M_\parallel = 2$, it suffices to calculate $\tr M_\perp$ to detect stability transitions. In addition to numerically evaluating transition energies to high precision, we use this decoupling in Appendix \ref{s:new-lame-fns}, to propose an expression for $\tr M$ in terms of suitably defined Lam\'e functions.


It is expected from Ref.~\cite{isoch-per-dou-bifur} that stability transitions at $\tr M = 4$ are associated with isochronous ($\IS$) bifurcations while those at $\tr M = 0$ correspond to period-doubling ($\PD$) bifurcations. In \S \ref{s:Pendulum stability transition energies and time periods}, we use the asymptotic periodicity of $\tr M(\tau)$ to classify these bifurcations of librational ($\ell$) and rotational ($r$) pendula based on the nature of the stability transitions. The latter is encoded in the slope of $\tr M(\tau)$. This leads to seven classes of bifurcations that we label $\IS^{\ell,r}_{2n-1}$, $\IS^{\ell,r}_{2n}$, $\PD^r_{2n-1}$, $\PD^r_{2n}$ and $\PD^\ell_n$ (for $n = 1,2, \ldots$) and indicate in Fig.~\ref{f:TrMvT-pendl}. We numerically determine the energies and time periods of pendula at the first few transitions in each class. Moreover, in each class, the differences $|4g - E_n|$ (where $E_n$ are transition energies) form an asymptotically geometric sequence (Fig.~\ref{f:fgb-lib-rot-self-simi-trM-vs-E}). This leads us to a scaling constant $\del$ for each class, which on account of the asymptotic periodicity of $\tr M(\tau)$ is common to all classes of bifurcations in each phase. Thus, the bifurcation cascades of pendula are characterized by two scaling constants: $\del^\ell$ and $\del^r$. The latter are estimated numerically and compared with our predictions $\del^\ell = e^{\sqrt{3} \pi}$ and $\del^r = (\del^\ell)^2$. Finally, we find the spectrum of $M$ at the transitions: it is $(1,1,1,1)$ at $\IS$ and $(-1,-1,1,1)$ at $\PD$ bifurcations. The transverse eigenvectors (at the point $G$ on the orbit) at all bifurcations of a given class are identical while the two longitudinal eigenvectors are common to all pendulum orbits. The transverse eigenvectors are then used to explore the vicinity of these bifurcations.

In \S \ref{s:search-method-pend}, we present our search algorithm to find newly born families of periodic trajectories at each of the isochronous and period-doubling bifurcations. The algorithm exploits the expected time periods of newly born trajectories and the coordinate system in which the perturbation equations to pendula decouple into longitudinal and transverse Lam\'e equations. This allows us to look for new families of periodic orbits by perturbing in the direction transverse to the pendulum family. This gives us an approximately periodic orbit, which allows us to make educated guesses about the shape and symmetries of the newly born orbit. These are exploited to improve the  approximately periodic orbit through an efficient search procedure that is confined to a single direction. We find that this type of search algorithm is much faster than a multi-dimensional search. 

In \S \ref{s:features-newly-born-pend}, we study various properties of the new periodic orbits that are born at the above $\IS$ and $\PD$ bifurcations of pendula. The salient ones are enumerated here. (i) These bifurcations are shown to be forward fork-like. (ii) As Fig.~\ref{f:pend-lib-new-traj} and Fig.~\ref{f:pend-rot-new-traj} indicate, orbits born at successive bifurcations of the same type ($\IS^\ell_{4n-k}, \PD^\ell_n, \IS^r_{2n}, \IS^r_{2n-1}, \PD^r_{2n-1}, \PD^r_{2n}$ for $n=1,2, \ldots$ and $0 \leq k \leq 3$) have similar gross shapes in the $\al_1$-$\al_2$ plane but display additional oscillations as $E \to 4g$. (iii) All orbits from a given class of bifurcations display a similar dependence of the monodromy trace on the time period $\tau$ (see Fig.~\ref{f:trM-v-T-lib-rot-pend-IS}). (iv) At $\IS$ bifurcations, we find that the slopes of $\tr M(E)$ for pendula and the newly born orbits are related by a fixed multiple given by the fork-like bifurcation slope theorem\cite{brck-fork}. Interestingly, we find an analog of this slope theorem that applies to $\PD$ bifurcations. (v) The graphs of $\tr M$ vs $E$ for a given class (e.g. $\IS^\ell_{2n-1}$ for $n = 1,2, \ldots$) are found to meet at $E = 4g$ forming a fan-like structure (see Fig.~\ref{f:fan-IS-lib-rot}). (vi) With the exception of the $\PD^\ell_n$ class, we represent $\al_1(t)$ for all newly born orbits as periodic Lam\'e functions \cite{Ince, erdelyi}. (vii) In another direction, we define two additional scaling constants $\al^\ell, \beta^\ell$ associated with self-similarity in the shapes of newly born orbits in the librational regime. Their values are estimated numerically and compared with analytical predictions. These definitions do not directly extend to the rotational regime: they are modified to enable us to estimate $\al^r$ and $\beta^r$. (viii) Finally, we discover an asymptotic duality as $E \to 4g^\pm$ between the sequence of $\IS^\ell$ bifurcations on the one hand and the sequence of $\IS^r$ and $\PD^r$ bifurcations (with the exception of $\PD^r_{1,2}$) on the other. The energies and elliptic moduli at corresponding bifurcations are asymptotically related in a simple manner. Furthermore, we find that the periodic Lam\'e functions associated with the newly born orbits at dual bifurcations are also related.

In \S \ref{s:breather}, we turn our attention to the breather family of periodic orbits which has a single stability transition at $E \approx 8.97g$. As with pendula, the perturbation equations decouple upon transforming to new angular variables but by contrast, the bifurcation at the transition is shown to be reverse fork-like and period-doubling (see Fig~\ref{f:breather-pd-monod-vs-E}).

In \S \ref{s:global-chaos-band}, we argue using our results on $\tr M(E)$ for newly born families of periodic orbits (at pendulum and breather bifurcations) that none of them is stable for $5.33g \lesssim E \lesssim 5.6g$. This provides further evidence for global chaos in this band of energies supplementing the Poincar\'e plot data given in Fig.~12 of Ref.\cite{gskhs-3rotor}. We conclude in \S \ref{Discussion} with some open questions arising from this work.

\section{Perturbations to pendula: Lam\'e equation and monodromy}
\label{s:pert-pend-Lame-M}

In Ref.~\cite{gskhs-3rotor} the stability index of pendula was calculated numerically directly from the associated $4 \times 4$ monodromy matrices. Here we simplify the calculations by decoupling the perturbation equations into transverse and longitudinal parts. This separation will then be exploited in our search for new periodic orbits born at stability transitions of pendula and to obtain expressions for them in terms of periodic Lam\'e functions \cite{Ince,erdelyi}. The perturbation equations to the pendulum family $\vphi_1 = 0$ (see Eqn.~(36) of Ref. \cite{gskhs-3rotor}) can be written as
	\begin{widetext}
	\beq
	\fr{m r^2}{g} \fr{d^2}{d t^2} \begin{pmatrix} \del \vphi_1 \\ \del \vphi_2 \end{pmatrix} 
	= \begin{cases} 
	- \begin{pmatrix} 3 - 2\, k^2 \sn^2(\om_0 t, k)  & 0 \\  - 2\, k^2 \sn^2(\om_0 t, k)  & 3 - 6\, k^2 \sn^2(\om_0 t, k) \end{pmatrix} \begin{pmatrix} \del \vphi_1 \\ \del \vphi_2 \end{pmatrix} \quad \text{for} \quad 0 \leq E \leq 4g, \\ 
	- \begin{pmatrix} 3 - 2 \sn^2(\om_0 t / \ka, \ka)  & 0 \\  - 2 \sn^2(\om_0 t / \ka, \ka)  & 3 - 6 \sn^2(\om_0 t / \ka, \ka) \end{pmatrix} \begin{pmatrix} \del \vphi_1 \\ \del \vphi_2 \end{pmatrix} \quad \text{for} \quad E \geq 4g. \end{cases}
	\eeq
	\end{widetext}
These are a pair of coupled Lam\'e equations, since in Jacobi elliptic form, the Lam\'e equation is \cite{Ince}
	\beq
	y''(z) + [h - n (n + 1)\, k^2 \sn^2 (z,k)] y(z) = 0.
	\label{e:std-lame-eqn}
	\eeq
They may be decoupled by changing variables to $\al_1 = \vphi_1/2$ and $\al_2 = \vphi_1/2 + \vphi_2$. The fundamental square $[0, 2 \pi] \times [0, 2 \pi]$ in the $(\vphi_1, \vphi_2)$ variables is mapped to a parallelogram with vertices at $(0, 0)$, $(\pi, \pi)$, $(\pi, 3 \pi)$ and $(0, 2 \pi)$ in the $(\al_1, \al_2)$ variables. In terms of these variables, the Lagrangian becomes
	\beqs
	L_{\rm rel} & = & \ov{3} m r^2 ( 3 \dot \al_1^2 + \dot \al_2^2) - V(\al_1, \al_2) \quad \text{where} \cr
	V(\al_1, \al_2) & = & g [ 3 - \cos 2 \al_1 - \cos (\al_2 - \al_1)- \cos(\al_1 + \al_2)]. \qquad
	\label{e:lagr-V-3rot-al1-al2}
	\eeqs
For future reference, the relative energy of the system in these variables is
	\beq
	E = \ov{3} m r^2 ( 3 \dot \al_1^2 + \dot \al_2^2) + V(\al_1, \al_2).
	\label{e:egy-V-3rot-al1-al2}
	\eeq
The equations of motion are
	\beqs
	m r^2 \ddot \al_1 &=& - \half g (2 \sin 2 \al_1 - \sin( \al_2 - \al_1 ) + \sin( \al_1 + \al_2 )), \cr
	m r^2 \ddot \al_2 &=& - (3/2) g ( \sin( \al_2 - \al_1 ) + \sin( \al_1 + \al_2 )).
	\label{sec-ordr-eom}
	\eeqs
The conjugate momenta are $\pi_1 = 2 m r^2 \dot \al_1$ and $\pi_2 = \fr{2}{3} m r^2 \dot \al_2$.
The pendulum orbit $\vphi_1 \equiv 0$ and $\vphi_2 \equiv \bar \vphi_2$ (\ref{phi1=0-pend}), in these new variables is
	\beq
	\al_1(t) \equiv 0, \quad \al_2(t) = \bar \al_2(t) \equiv \bar \vphi_2(t).
	\label{al1=0-pend}
	\eeq
Thus, we will regard $\al_2$ as the `longitudinal' coordinate (along the `sliding' direction) of pendula and $\al_1$ as the `transverse' coordinate. Let us define the dimensionless variables $\tl t = \sqrt{g/m r^2} t = \om_0 t/\sqrt{3}$,
	\beq
	\tl \pi_1 = 2 \DD{\al_1}{\tl t} = \fr{\pi_1}{\sqrt{g m r^2}} 
	\quad \text{and} 	\quad 
	\tl \pi_2 = \frac{2}{3} \DD{\al_2}{\tl t} = \fr{\pi_2}{\sqrt{g m r^2}}.
	\label{e:dim-less-t-mom}
	\eeq
Then, $\del \al_{1,2}$ satisfy the decoupled Lam\'e equations
	\begin{widetext}
	\beq
	\fr{d^2}{d\tl t^2} \begin{pmatrix} \del \al_1 \\ \del \al_2 \end{pmatrix} 
	= \begin{cases} - \begin{pmatrix} 3 - 2\, k^2 \sn^2(\sqrt{3} \tl t, k)  & 0 \\  0  & 3 - 6\, k^2 \sn^2(\sqrt{3} \tl t, k) \end{pmatrix} 
	\begin{pmatrix} \del \al_1 \\ \del \al_2 \end{pmatrix} \quad \text{for} \quad 0 \leq E \leq 4g, 
	\\ - \begin{pmatrix} 3 - 2 \sn^2(\sqrt{3} \tl t / \ka, \ka)  & 0 \\  0  & 3 - 6 \sn^2(\sqrt{3} \tl t / \ka, \ka) \end{pmatrix} \begin{pmatrix} \del \al_1 \\ \del \al_2 \end{pmatrix} \quad \text{for} \quad E \geq 4g. \end{cases}
	\label{e:decoupled-lame-al1-al2}
	\eeq
	\end{widetext}


Having decoupled the equations, let us write them in first order form by introducing the angular momentum perturbations $\del \tl \pi_{1,2}$:
 	\beqs
	\DD{}{\tl t} \begin{smmat} \del \al_1 \\ \del \tl \pi_1\\ \del \al_2 \\ \del \tl \pi_2  \end{smmat} 
	&=& A \begin{smmat} \del \al_1 \\ \del \tl \pi_1\\ \del \al_2 \\ \del \tl \pi_2  \end{smmat} \quad \text{where}  \cr  
	A &=& -  \begin{smmat} 0 &  -1/2 & 0 & 0 \\ 2(2 + \cos \bar \al_2) & 0 &  0 & 0 \\ 0 & 0 & 0 & -3/2 \\ 0 & 0 & 2 \cos \bar \al_2 & 0 \end{smmat}.
	\label{e:pend-pert-al-12-coeff-mat}
	\eeqs
Since the coefficient matrix is block diagonal, the pendulum monodromy matrix (see \S IV.A.1 of Ref.~\cite{gskhs-3rotor}) $M = \diag(M_1, M_2) = \diag(M_\perp, M_\parallel)$ is also block diagonal in this basis and $\tr M = \tr M_\perp + \tr M_\parallel$.

As noted in \S \ref{s:summary-results}, the eigenvalues of $M_\parallel$ are $(1, 1)$ at all energies while those of $M_\perp$ are nontrivial. Thus, the study of the stability of pendula is reduced to a single (transverse) Lam\'e equation for $\del \al_1$. In standard form (\ref{e:std-lame-eqn}), for libration ($E \leq 4g$) and $z = \sqrt 3 \tl t$, it is
	\beq
	\del \al_1''(z) = - (1 - (2/3) k^2 \sn^2(z , k)) \del \al_1,
	\label{e:lib-pend-lame-eqn}
	\eeq
while for rotation ($E \geq 4g$) and $z = \sqrt 3 \tl t / \ka$, it is
	\beq
	\del \al_1''(z)  = - (\ka^2 - (2/3) \ka^2 \sn^2(z , \ka)) \del \al_1.
	\label{e:rot-pend-lame-eqn}
	\eeq
This corresponds to the parameter values $n(n+1) = 2/3$ and $h = 1$ for libration and $h = \ka^2$ for rotation.
The coefficients in (\ref{e:lib-pend-lame-eqn}) and (\ref{e:rot-pend-lame-eqn}) are periodic in $z$ with period $2 K(k)$ and $2 K(\ka)$ in the librational and rotational phases. However, there is a distinction between the two phases. The librational pendulum has twice the period of the Lam\'e equation coefficients, while the rotational pendula and the corresponding Lam\'e equation have the same period (the latter property is shared by the A orbits of H\'enon-Heiles \cite{brck-omega}). Finally, for the purpose of determining the stability of pendula ($\tr M_\perp = \tr M - 2$), it is convenient to formulate the transverse Lam\'e equations (\ref{e:lib-pend-lame-eqn}) and (\ref{e:rot-pend-lame-eqn}) as first order systems with $\bar \al_2$ given in (\ref{al1=0-pend}):
	\beq
	\DD{}{\tl t} \begin{pmatrix} \del \al_1 \\ \del \tl \pi_1\end{pmatrix}
	= - \begin{pmatrix} 0 & -1/2 \\ 2(2 + \cos \bar \al_2) & 0  \end{pmatrix} \begin{pmatrix} \del \al_1 \\ \del \tl \pi_1 
	\end{pmatrix}.
	\label{e:lame-eq-transv-al1-pi1}
	\eeq
In Appendix \ref{s:new-lame-fns}, we derive an analytic expression for $\tr M_\perp$ in terms of a special class of Lam\'e functions. However, for practical purposes we find it convenient to evaluate $\tr M_\perp$ numerically. The results are reported in the in Section \S \ref{s:Pendulum stability transition energies and time periods}.

\section{Pendulum stability transition energies and scaling constants}
\label{s:Pendulum stability transition energies and time periods}

In Ref.~\cite{gskhs-3rotor} it was numerically observed that $\tr M$ is asymptotically periodic in $\log|E-4g|$ as $E \to 4g$. Here we exploit the decoupling of the perturbation equations to calculate $\tr M_\perp$ to much greater precision (six significant figures) than in Ref.~\cite{gskhs-3rotor} and identify the transition energies. Moreover, viewing $\tr M$ as a function of time period $\tau$ of pendula helps us provide an explanation of the above asymptotic periodicity and estimate it. It also helps in classifying the bifurcations at stability transitions as isochronous and period-doubling. These improvements in precision and classification have additional payoffs: they help us discover and formulate a duality between librational and rotational bifurcations in \S \ref{s:duality-lib-rot}.

\begin{table*}
\begin{center}
\begin{tabular}{|c|c|c|c|c|c|c|c|c|}
\hline
	\multicolumn{3}{|c|}{$\PD^\ell$ (stable $\to$ stable)} &
	\multicolumn{3}{|c|}{$\IS^\ell_{\rm odd}$ (stable $\to$ unstable)} &
	\multicolumn{3}{|c|}{$\IS^\ell_{\rm even}$ (unstable $\to$ stable)}  \\
\hline
$n$ & $\log(4$ - $E)$ & $\tau$ & $n$ & $\log(4$ - $E)$ & $\tau$ & $n$ & $\log(4$ - $E)$ & $\tau$ \\
\hline
1 & -2.4268 & 7.6341 & 1 & -4.6658 & 10.1945 & 2 & -5.7819 & 11.4804 \\
2 & -7.9522 & 13.9849 & 3 & -10.1183 & 16.4859 & 4 & -11.2287 & 17.7681 \\
3 & -13.3943 & 20.2687 &  5 & -15.5598 & 22.7692 & 6 & -16.6702 & 24.0514 \\
4 & -18.8356 & 26.5517 & 7 & -21.0012 & 29.0524 & 8 & -22.1115 & 30.3344 \\	
5 & -24.2772 & 32.8352 & 9 & -26.4426 & 33.3356 & 10 & -27.5529 & 36.6175 \\
\hline
\end{tabular}
\end{center}
\caption{\small Energies and time periods (in units where $m = g = r = 1$) of first few period-doubling and isochronous bifurcations of librational pendula.}
\label{t:lib-pend-trans-E-T}
\end{table*}

\begin{table}
\begin{center}
\begin{tabular}{|c|c|c|c|c|c|}
\hline
	\multicolumn{3}{|c|}{$\IS^r_{\rm odd}$ (unstable $\to$ stable)} &
	\multicolumn{3}{|c|}{$\IS^r_{\rm even}$ (stable $\to$ unstable)}  \\
\hline
$n$ & $\log(E - 4)$ & $\tau$ & $n$ & $\log(E - 4)$ & $\tau$ \\
\hline
1 & -4.6680 & 5.1055 & 2 & -5.7927 & 5.7447 \\
3 & -15.5598 & 11.3846 & 4 & -16.6701 & 12.0256 \\
5 & -26.4426 & 17.6678 & 6 & -27.5529 & 18.3090 \\
\hline
	\multicolumn{3}{|c|}{$\PD^r_{\rm odd}$ (unstable $\to$ stable)} &
	\multicolumn{3}{|c|}{$\PD^r_{\rm even}$ (stable $\to$ unstable)}  \\
\hline
1 & 0.4682 & 2.0475 & 2 & -0.7395 & 2.7809 \\
3 & -10.1185 & 8.243 & 4 & -11.2288 & 8.8841 \\
5 & -21.0012 & 14.5262 & 6 & -22.1116 & 15.1673 \\ 
\hline
\end{tabular}
\end{center}
\caption{\small Energies and time periods (for $m = g = r = 1$) of first few period-doubling and isochronous bifurcations in the rotational regime of pendula.}
\label{t:rot-pend-trans-E-T}
\end{table}

\begin{figure}
\centering
	\begin{subfigure}{0.45\textwidth}
	\includegraphics[width = \textwidth]{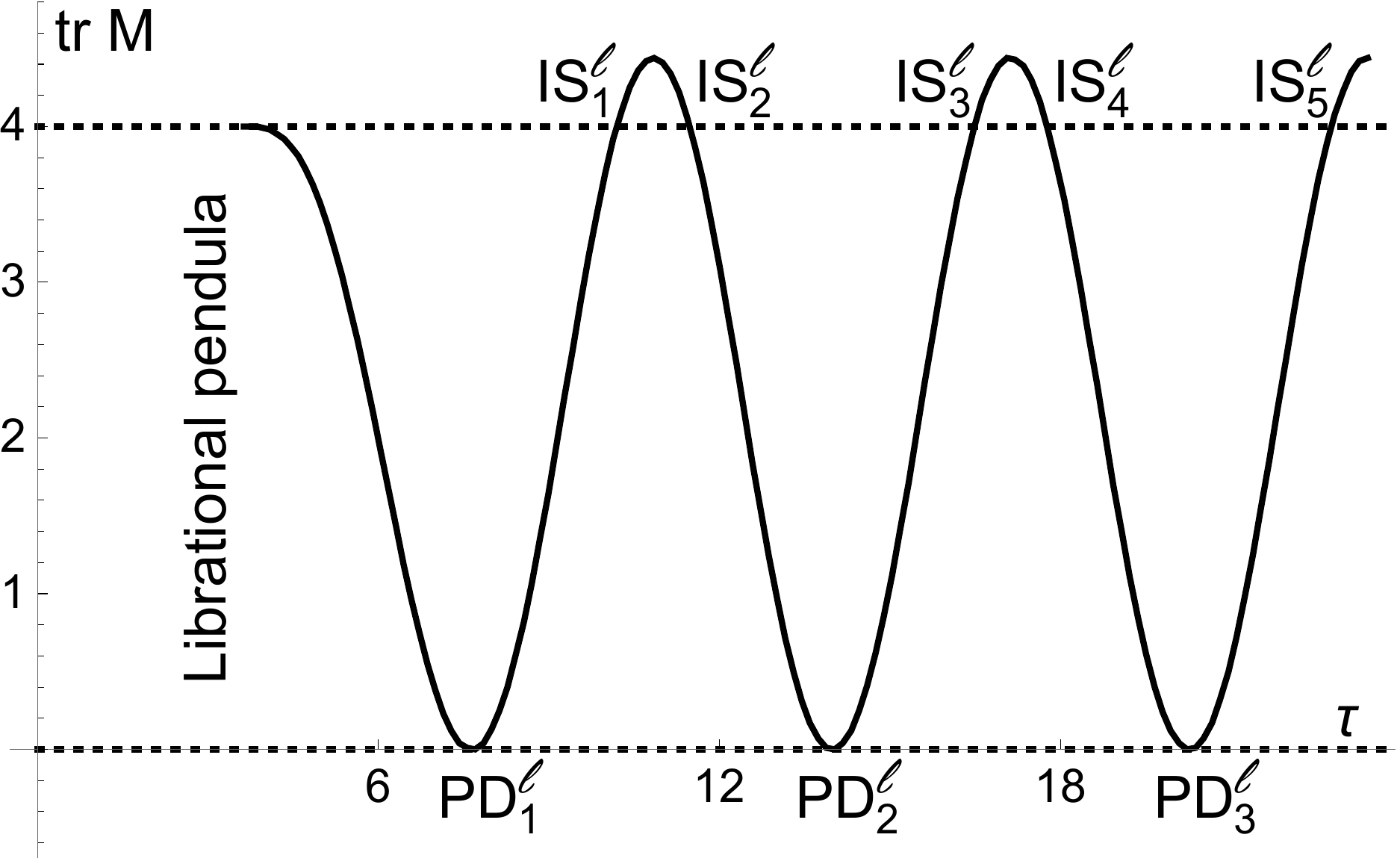}
	\caption{}
	\end{subfigure}
\hfil
	\begin{subfigure}{0.45\textwidth}
	\includegraphics[width = \textwidth]{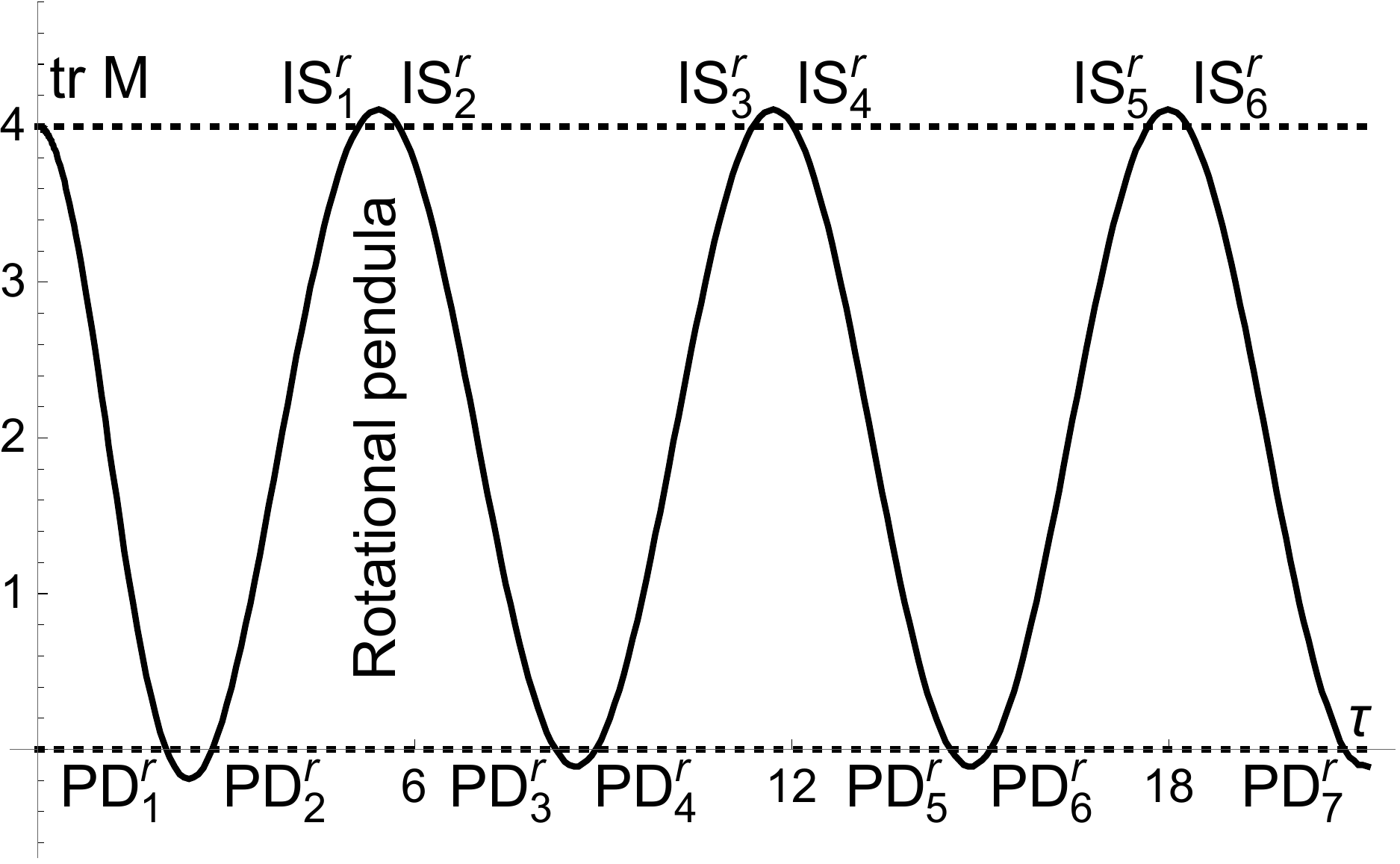}
	\caption{}
	\end{subfigure}
	\caption{\small Numerically obtained $\tr M$ vs time period $\tau$ for (a) librational and (b) rotational pendula showing asymptotically periodic behavior as $\tau \to \infty$ ($E \to 4g^\mp$). Pendula are stable when $0 < \tr M < 4$. Coincidentally, the band of global chaos \cite{gskhs-3rotor} lies in the energy interval between $\PD^r_1$ and $\PD^r_2$, where pendula are unstable.}
	\label{f:TrMvT-pendl}
\end{figure}

As Fig.~\ref{f:TrMvT-pendl} indicates, pendula are stable $(0 < \tr M = 2 + \tr M_\perp  < 4)$ at low $(0 \leq E \lesssim 3.9g)$ as well as at high energies $(E \gtrsim 5.6g)$. In between, they undergo a doubly infinite sequence of stability transitions that accumulate from both the librational $(\ell)$ and rotational $(r)$ phases at $E = 4g$ where the time period $\tau$ (\ref{e:time-per-lib-rot-pend}) of pendula diverges. In the stable windows, the eigenvalues $\mu_{1,2}$ of $M_\perp$ are of the form $e^{\pm i\tht}$ while in the unstable windows they take the form $(\mu, 1/\mu)$ for real $\tht$ and $\mu$. Consequently, the transverse Lyapunov exponents\cite{gskhs-3rotor} of pendula ($\la_{1,2} = (1/\tau) \log |\mu_{1,2}| $) vanish in the stable windows while they are nonzero in the unstable windows. The stable and unstable windows are also analogous to bands and bandgaps in the Bloch energy spectrum of an electron in an ionic lattice\cite{ashcroft-mermin}. In fact, (\ref{e:lib-pend-lame-eqn}) and (\ref{e:rot-pend-lame-eqn}) are analogs of the Schr\"odinger equation for an electron in a 1d periodic potential. The distinction is that while we solve the perturbation equation as an initial value problem and determine the eigenvalues of $M$ using $\del \al_1(\tau)$, the Schr\"odinger equation is solved as an eigenvalue problem using the Bloch wave ansatz which is assumed to be an eigenfunction of the monodromy matrix.

In Tables \ref{t:lib-pend-trans-E-T} and \ref{t:rot-pend-trans-E-T} we give the numerically obtained energies $E$ and time periods $\tau$ of pendula at their first few stability transitions determined by solving (\ref{e:lame-eq-transv-al1-pi1}). These transitions correspond to two types of bifurcations \cite{isoch-per-dou-bifur}: (a) isochronous ($\IS^{\ell, r}_n$) when $\tr M = 4$ since a new family of orbits is born with initial time period equal to that of the parent pendulum and (b) period-doubling ($\PD^{\ell, r}_n$) when $\tr M = 0$ since the new orbits have twice the period of the parent. The integer $n = 1,2,3, \ldots$ increases/decreases with energy in the librational/rotational phases. Fig.~\ref{f:TrMvT-pendl} also indicates that $\tr M(\tau)$ becomes periodic as $\tau \to \infty$ in both phases. This will be established in (\ref{e:trM1-asymp-cos-om-tau}) of \S \ref{s:asymp-beha-time-per-egy-pend-bifur}. We use this asymptotic periodicity of $\tr M(\tau)$ to classify these bifurcations into seven infinite sequences: $\IS^\ell_{2n}$, $\IS^\ell_{2n-1}$, $\PD^\ell_n$, $\IS^r_{2n}$, $\IS^r_{2n-1}$, $\PD^r_{2n}$ and $\PD^r_{2n-1}$. The classification is based on the nature of the transition, as encoded in the slope of $\tr M(\tau)$. Period-doubling bifurcations in the librational and rotational phases are slightly different. In the former, $\tr M(\tau)$ has a double zero so that pendula are stable both before and after each such bifurcation. It is as if two PD bifurcations have coalesced. On the other hand, at PD bifurcations in the rotational phase, pendula switch stabilities. By contrast, at IS bifurcations in both the rotational and librational regimes, $\tr M - 4$ has a simple zero and pendula undergo stability transitions whose nature is determined by the slope of $\tr M$. 

Moreover, as $E \to 4g^\pm $, we find that the sequences of time periods of pendula at stability transitions in each of the 7 classes asymptotically approach arithmetic progressions with the same common difference. For instance, the time period differences corresponding to successive stable to unstable isochronous transitions of librational pendula approach a constant: $\tau(\IS^\ell_{3}) - \tau(\IS^\ell_{1}) = 6.2914$, $\tau(\IS^\ell_5) - \tau(\IS^\ell_3) = 6.2833$, $\tau(\IS^\ell_7) - \tau(\IS^\ell_5) = 6.2832$, \ldots,  converging to the asymptotic value $2 \pi$ (in units where $m = g = r =1$) as shown in \S \ref{s:asymp-beha-time-per-egy-pend-bifur}.


\subsection{Scaling constant associated to geometric cascade}
\label{s:feigenbaum-const-geom-cascade}

\begin{figure*}
\centering
	\begin{subfigure}{0.32\textwidth}
	\includegraphics[width = \textwidth]{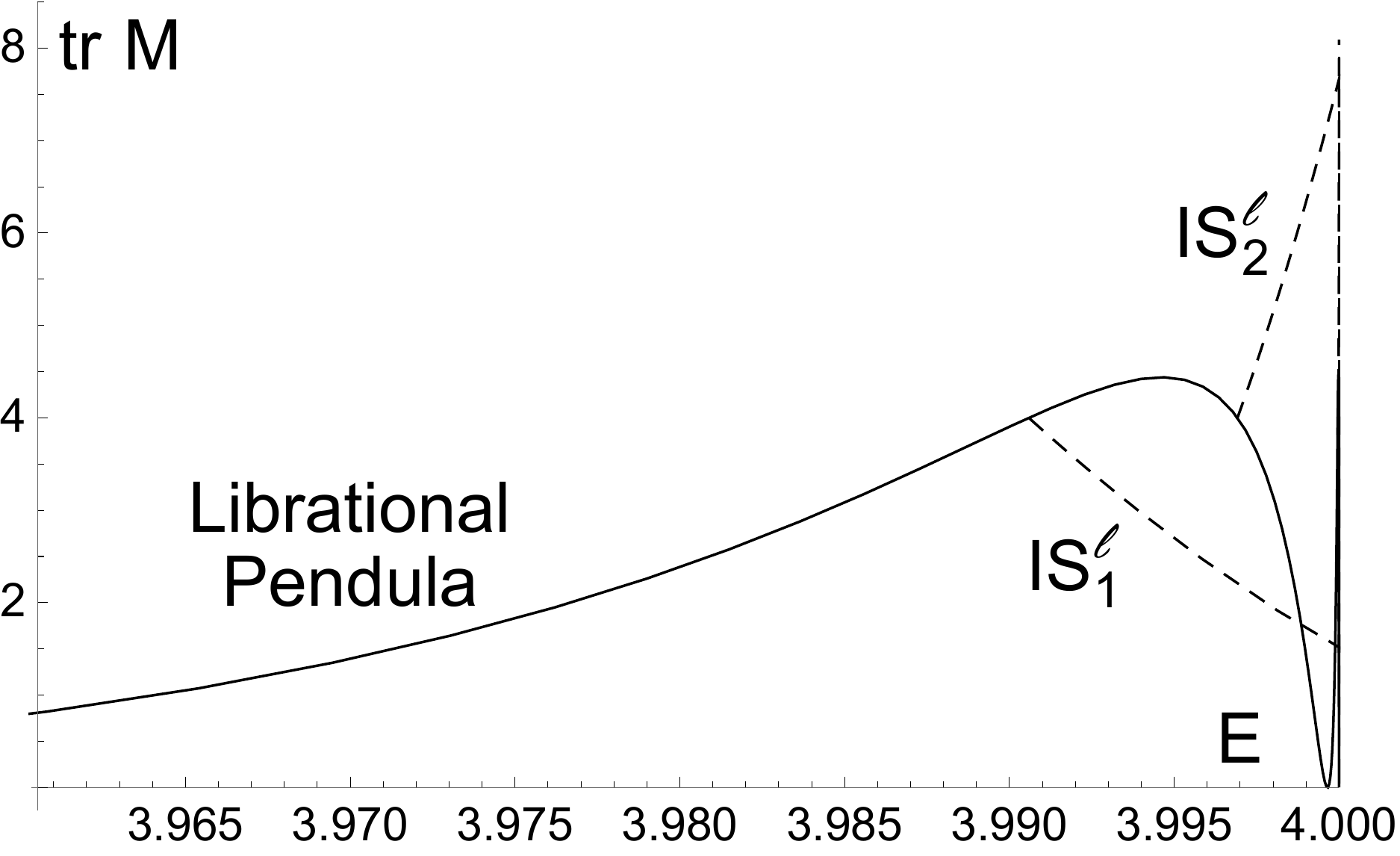}
	\caption{}
	\label{f:fgb-lib-self-simi-trM-vs-E-1}
	\end{subfigure}
\hfil
	\begin{subfigure}{0.32\textwidth}
	\includegraphics[width = \textwidth]{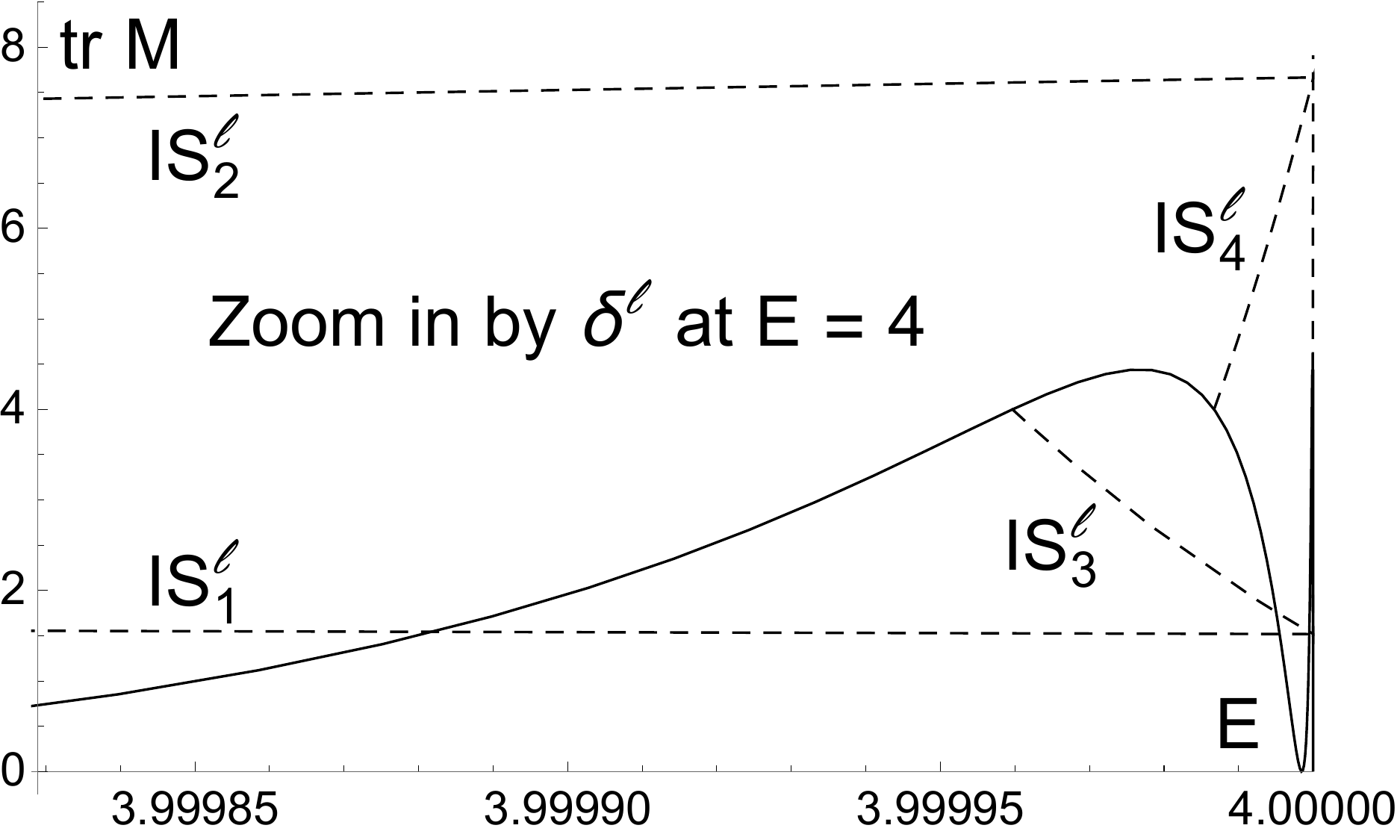}
	\caption{}
	\label{f:fgb-lib-self-simi-trM-vs-E-2}
	\end{subfigure}
\hfil
	\begin{subfigure}{0.32\textwidth}
	\includegraphics[width = \textwidth]{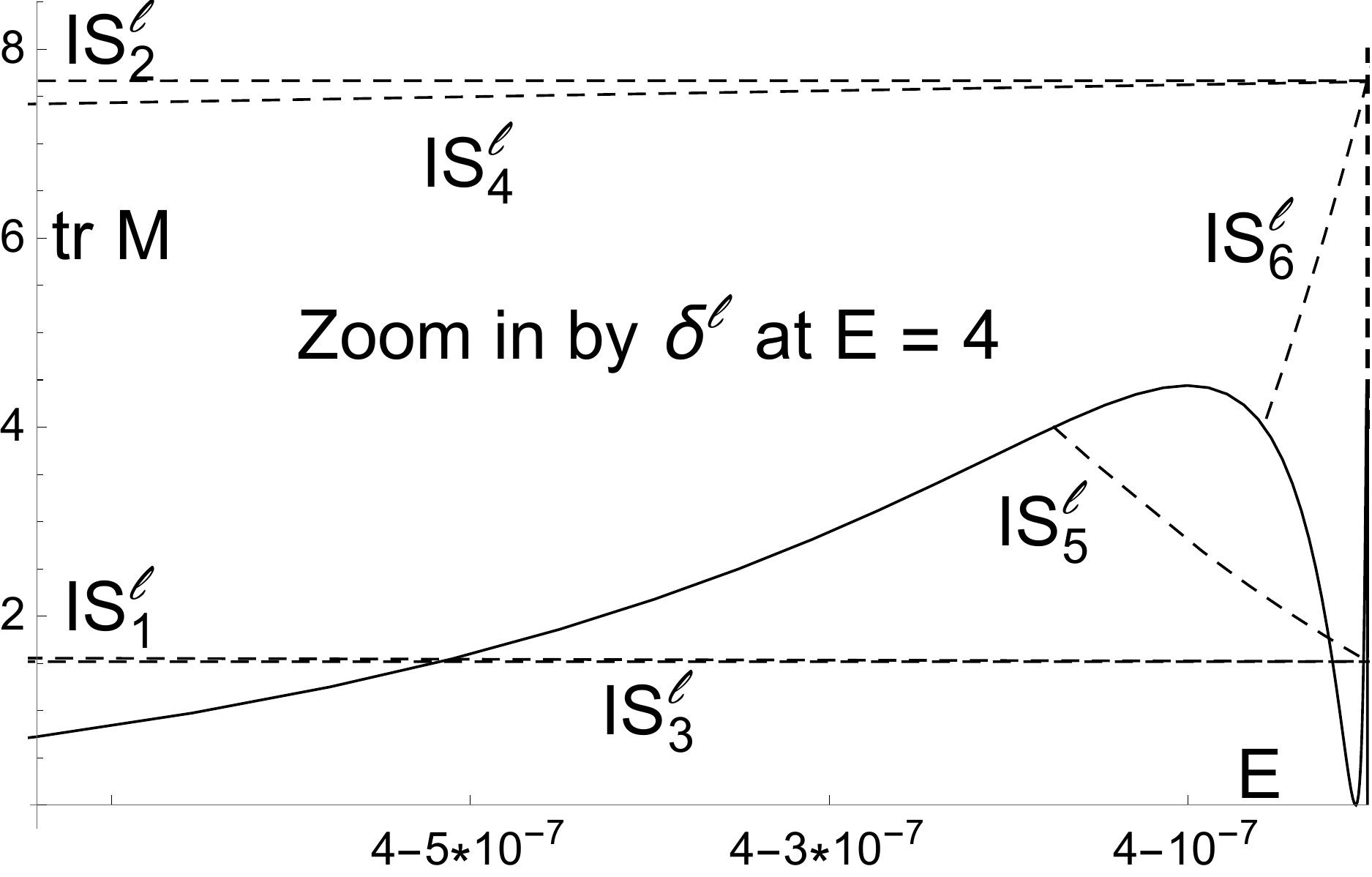}
	\caption{}
	\label{f:fgb-lib-self-simi-trM-vs-E-3}
	\end{subfigure}

	\begin{subfigure}{0.32\textwidth}
	\includegraphics[width = \textwidth]{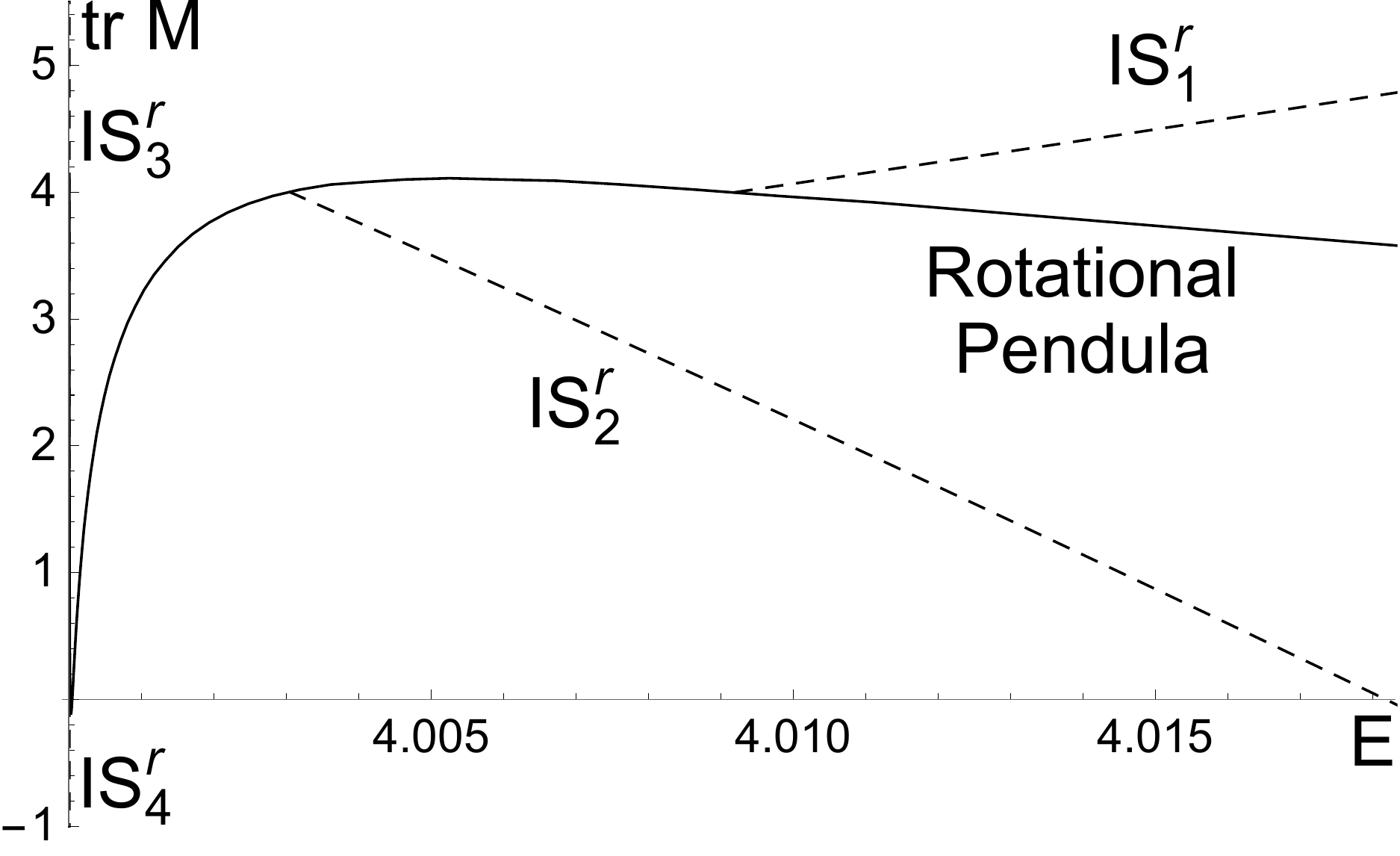}
	\caption{}
	\label{f:fgb-rot-self-simi-trM-vs-E-1}
	\end{subfigure}
\hfil
	\begin{subfigure}{0.32\textwidth}
	\includegraphics[width = \textwidth]{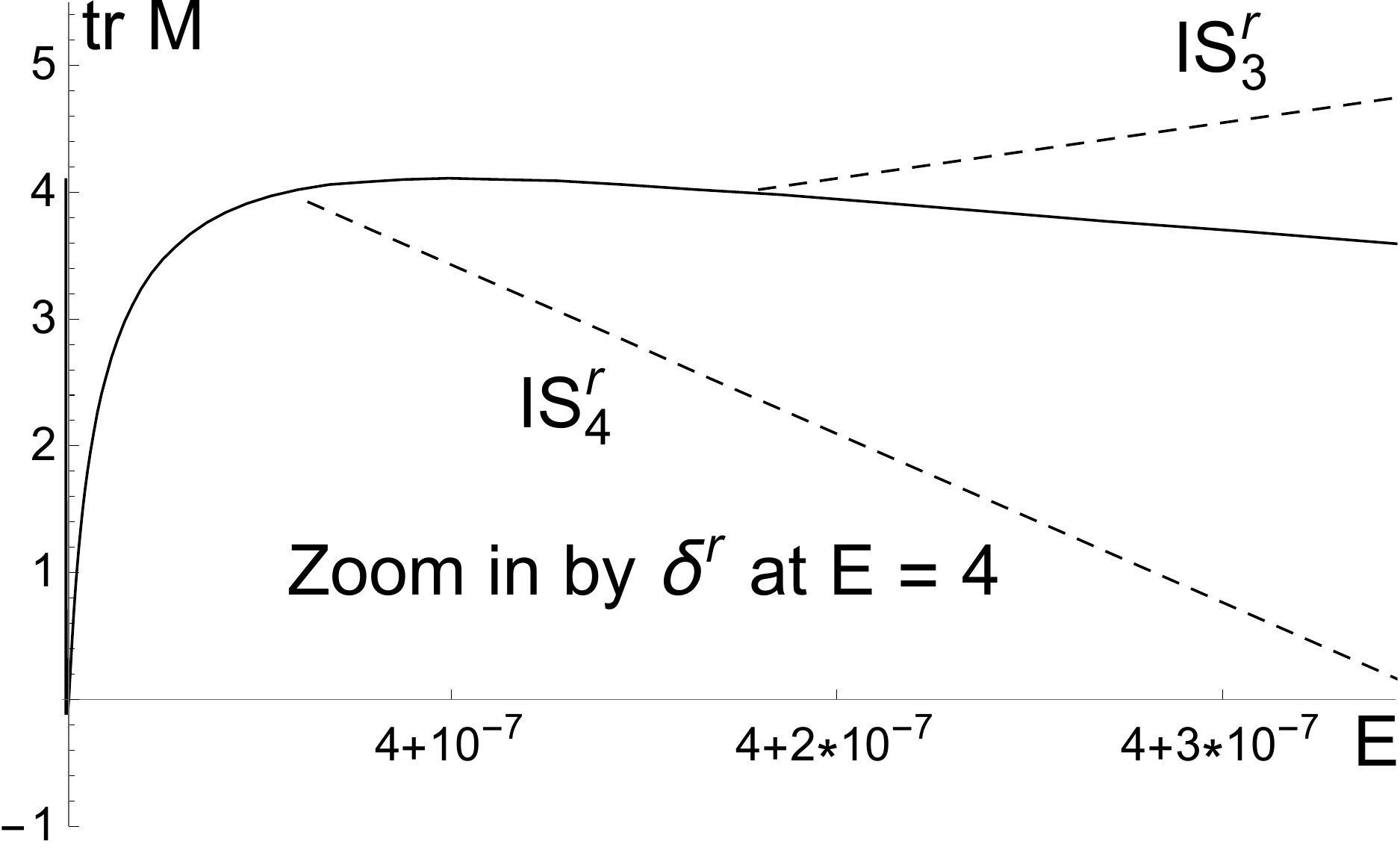}
	\caption{}
	\label{f:fgb-rot-self-simi-trM-vs-E-2}
	\end{subfigure}
	\caption{\small Self-similarity under magnification by the scale factor $\del^\ell$ (in (b) and (c)) and $\del^r$ (in (e)) in the $\tr M$ vs $E$ graph for librational and rotational pendula (solid) and the families of periodic orbits (dashed) born at their isochronous bifurcations. The minimum energy in (b) is $E_b = 4 - E_a/\del^\ell$, where $E_a$ is the minimum energy in (a). Similarly, $E_c = 4 - E_b/\del^\ell$. We observe two fans arising from the confluence of $\IS^\ell_{2,4,6, \ldots}$ and $\IS^\ell_{1,3,5, \ldots}$ at $E \approx 4$. In (c) the expected curve for $\IS^\ell_8$ is indicated by a vertical line at $E = 4$. In (e) the maximum energy is $E_e = 4+ E_d/\del^r$ where $E_d$ is the maximum energy in (d). Throughout, we work in units where $m = g = r =1$.}
	\label{f:fgb-lib-rot-self-simi-trM-vs-E}
\end{figure*}

Given the relation (\ref{e:time-per-lib-rot-pend}) between pendulum time periods and energies, we should expect the asymptotic arithmetic sequence of time periods to imply a geometric approach to $4g$ of the transition energies in each bifurcation sequence. For instance, consider the $\IS^\ell_{\rm odd}$ family of bifurcations of pendula which occur at energies $E_n = E(\IS^\ell_{2n-1})$ for $n = 1,2, \ldots$. Then, asymptotically $4g - E_n$ forms a geometric sequence. Thus, it is natural to define the sequence 
	\beq
	\del_n(\IS^\ell_{\rm odd}) = \frac{4g - E_n}{4g - E_{n+1}} 
	\label{fgb-def}
	\eeq
which in the limit leads us to our first scaling constant $\del(\IS^\ell_{\rm odd}) \equiv \lim_{n \to \infty} \del_n(\IS^\ell_{\rm odd})$. Alternatively, we may arrive at the same scaling constant by considering the limit of the sequence of spacing ratios
	\beq
	\del(\IS^\ell_{\rm odd}) = \lim_{n \to \infty} \frac{E_{n+1} - E_n}{E_{n+2} - E_{n+1}}.
	\eeq
Similar constants can be defined for the other families of bifurcations: $\IS^\ell_{\rm even}$, $\PD^\ell_n$, $\PD^r_{\rm odd}$, $\PD^r_{\rm even}$, $\IS^r_{\rm even}$ and $\IS^r_{\rm odd}$. Numerically, we find that all librational families have a common scaling constant $\del^\ell$ as do the rotational families, with 
	\beq
	\log \del^r = 2 \log \del^\ell \approx 10.883.
	\label{e:fgb-const-lib-rot-relation}
	\eeq
For instance, the sequence $\log \del^\ell_n(\PD)$ for $n = 1,2, \cdots, 6$ is given by 
	\beq
	5.5254, \quad 5.4421, \quad 5.4413, \quad 5.4416, \quad 5.4413, \quad 5.4414
	\label{e:feigenbaum-seq-for-PDl}
	\eeq
In fact, the `tail' of the graph (see Figs.~\ref{f:fgb-lib-self-simi-trM-vs-E-1}, \ref{f:fgb-lib-self-simi-trM-vs-E-2}, \ref{f:fgb-lib-self-simi-trM-vs-E-3}) of $\tr M$ vs $E$ in the window $E_{n+1} < E < 4g$ upon magnification by the factor $\del^\ell$ resembles that in the previous window $E_{n} < E < 4g$. This self-similarity in the $\tr M$ vs $E$ graph also applies to the rotational regime $E > 4g$ as shown in Figs.~\ref{f:fgb-rot-self-simi-trM-vs-E-1} and \ref{f:fgb-rot-self-simi-trM-vs-E-2}.


\subsection{Asymptotic behavior of time periods and energies of bifurcations}
\label{s:asymp-beha-time-per-egy-pend-bifur}

\paragraph*{Arithmetic progression of bifurcation time periods.} Inspired by our numerical observations, we will now argue from the perturbation equations (\ref{e:lib-pend-lame-eqn}) and (\ref{e:rot-pend-lame-eqn}) that the sequence of pendulum time periods for each bifurcation family (e.g., $\IS^\ell_{2n}$ or $\PD^r_{2n-1}$) forms an asymptotic arithmetic progression with common difference $\D \tau = 2\pi r \sqrt{m/g}$. As $k \to 1^\pm$ $(E \to 4g^\pm)$, the pendulum trajectory approaches a separatrix spending most of its time at the ``bottlenecks'' near the saddle point of the potential ($\al_2 = \pm \pi, \al_1 = 0$). Putting $\bar \al_2 = \pm \pi$ as $k \to 1$, the perturbation equation for $\del \al_1$ (\ref{e:decoupled-lame-al1-al2}) becomes the harmonic oscillator equation:
	\beq
	\fr{d^2 \del \al_1}{d t^2} = -\fr{g}{m r^2}( 2 + \cos \bar \al_2) \del \al_1 
	\approx - {\om_\perp^2} \del \al_1,
	\label{shm-apprx}
	\eeq
where $\om_\perp^2 = g/m r^2$. We may interpret $\om_\perp^2 = (1/2m r^2) \pdr^2 V/\pdr \al_1^2$ as the curvature of the 3-rotor potential (\ref{e:lagr-V-3rot-al1-al2}) in the $\al_1$ direction at the saddle point $D_3$ (see Fig.~\ref{f:equipotn-pend-breather}). The monodromy matrix $M_\perp(\tau)$ asymptotically approaches the time evolution operator of a harmonic oscillator with angular frequency $\om_\perp$, evaluated at the time period $\tau$ of the pendulum orbit being perturbed. Thus,
	\beq
	\tr M_\perp(\tau) \to \tr \colvec{2}{\cos(\om_\perp \tau) & -\sin(\om_\perp \tau)/2}{2\sin(\om_\perp \tau) & \cos(\om_\perp \tau)} = 2 \cos(\om_\perp \tau).
	\label{e:trM1-asymp-cos-om-tau}
	\eeq
Consequently, $\tr M_\perp$ is asymptotically a periodic function of $\tau$ with period $2\pi/\om_\perp$. In particular, the asymptotic value of the time period difference $\D \tau$ between two successive bifurcations in each class (e.g., $\IS^\ell_{2n-1}$) is 
	\beq
	\D \tau = 2 \pi/\om_\perp = 2\pi r \sqrt{m/g}.
	\label{tau-pend}
	\eeq
In units where $m=g=r=1$, this agrees with the numerical $\D \tau \approx 6.283$ from Tables \ref{t:lib-pend-trans-E-T} and \ref{t:rot-pend-trans-E-T}. 


\paragraph*{Scaling constant for geometric cascade of bifurcations.} We can estimate the scaling constant $\del$ using the above formula for $\D \tau$ (\ref{tau-pend}). As $k \to 1^-$, the time periods (\ref{e:time-per-lib-rot-pend}) of librational pendula (whose energies are $E(k) = 4 g k^2$) can be approximated by 
	\beqs
	\tau_\ell &=& \fr{4}{\om_0} K(k) \approx - \fr{2}{\om_0} \log\left[\fr{1-k^2}{16}\right] = - \fr{2}{\om_0} \log\left[\fr{1 - E/4g}{16}\right]. \;\;\;\, \quad
	\label{asy-TvE-lib}
	\eeqs
Equivalently, the asymptotic energies of pendula are 
	\beq
	E \approx 4g(1- 16 e^{-\om_0 \tau_\ell / 2}) \quad \text{as} \quad \tau_\ell \to \infty.
	\label{egy-asymp}
	\eeq
This agrees fairly well with our numerical results. For instance it predicts that $4-E(\PD^\ell_5) = 2.8611 \times 10^{-11}$ using $\tau_\ell = 32.8352$ from Table \ref{t:lib-pend-trans-E-T} while the corresponding numerical value from the same table is $2.8612 \times 10^{-11}$. It follows that our first scaling constant, say for the $\PD^\ell_n$ family is
	\beq
	\del^\ell = \lim_{n \to \infty} \fr{4g - E_n}{4g - E_{n+1}} = e^{\om_0 \D \tau/2} = e^{\sqrt 3 \pi} \approx e^{5.4414} \approx 230.8,
	\eeq
in agreement with our numerical estimate in (\ref{e:feigenbaum-seq-for-PDl}). Note that $\del^\ell$ depends on the differences $\D \tau$ in the asymptotic time periods in a sequence. Since all three librational bifurcation sequences have the same $\D \tau = 2\pi r \sqrt{m/g}$ (which is the asymptotic period of $\tr M(\tau)$), they share the same value of $\del^\ell$, as we observe numerically.

Analogously, one may obtain the scaling constant in the rotational regime. As $\ka = 1/k \to 1^-$, the asymptotic time period becomes, {\small
	\beq
	\tau_r = \fr{2 \ka}{\om_0} K(\ka) \approx -\fr{\ka}{\om_0} \log\left[\fr{1 - \ka^2}{16}\right] = -\fr{2 }{\om_0} \sqrt{\fr{g}{E}} \log\left[\fr{1 - 4g/E}{16} \right].
	\label{asy-TvE-rot}
	\eeq}
The energies of pendula in the rotational regime as $\ka \to 1^-$ are
	\beq
	E \approx 4g(1 + 16 e^{-\om_0 \tau_r }) \quad \text{as} \quad \tau_r \to \infty.
	\label{egy-asymp-rot}
	\eeq
Therefore, the scaling constant in the rotational regime (e.g., for $\IS^r_{2n-1}$) is
	\beq
	\del^r = \lim_{n \to \infty} \fr{ E_n - 4g}{E_{n+1} - 4g} = e^{\om_0 \D \tau} = e^{2\sqrt 3 \pi} \approx e^{10.8828}.
	\eeq
As expected from our numerical results in (\ref{e:fgb-const-lib-rot-relation}), $\del^r = (\del^\ell)^2$. This is because asymptotically (as $E \to 4g$), a librational pendulum orbit has twice the period of a rotational one: $\tau_\ell/\tau_r \to 2$ from (\ref{asy-TvE-lib}) and (\ref{asy-TvE-rot}).


\subsection{Monodromy eigenvectors at stability transitions of pendula}
\label{evecs-monod}

Here we discuss the eigenvectors of the monodromy matrix of pendula at stability transitions. They will help us discover new periodic orbits born at these transitions. The pendulum monodromy matrix $M = \diag(M_\perp, M_\parallel)$ is block diagonal in the $(\del \al_1, \del \tl \pi_1, \del \al_2, \del \tl \pi_2)$ basis. Here and elsewhere, we evaluate $M$ with respect to the basepoint $G$ on pendula. The eigenvalues of $M_\parallel$ are always $(1,1)$ with a single linearly independent eigenvector $(1,0)^t$, which leads to the sliding eigenvector $(0,0,1,0)$ of $M$. The nontrivial eigenvalues are those of $M_\perp$ whose eigenvectors contribute to the transverse eigenvectors of $M$.

\paragraph*{Librational pendula.} At stability transitions where $\tr M = 4$ (bifurcations points $\IS^\ell_1, \IS^\ell_2, \cdots$), we find that all eigenvalues of $M$ are 1 and $M$ has only one linearly independent transverse eigenvector. It is either $(1,0,0,0)^t$ (for stable to unstable transitions) or  $(0,1,0,0)^t$ (for unstable to stable transitions). On the other hand, at transitions where $\tr M = 0$ ($\PD^\ell_1, \PD^\ell_2, \cdots$), $M$ has two linearly independent transverse eigenvectors corresponding to the eigenvalues $-1$. They span the $\del \al_1-\del \tl \pi_1$ plane and can be taken as $(1,0,0,0)^t$ and $(0,1,0,0)^t$. However, we find that both eigenvectors lead to the same family of newly born periodic trajectories to be discussed in \S \ref{s:search-method-pend} and \S \ref{s:features-newly-born-pend}.

\paragraph*{Rotational pendula.} At all stability transitions of rotational pendula (both $\IS$ and $\PD$ where $\tr M = 4, 0$), there is just one linearly independent `transverse' eigenvector. It may be taken as $(1,0,0,0)^t$ for unstable $\to$ stable transitions and $(0,1,0,0)^t$ for stable $\to$ unstable transitions with increasing energy.

\section{Generating new periodic trajectories at bifurcations of pendula}
\label{s:search-method-pend}

The idea underlying our search algorithm for new families of periodic trajectories was outlined in \S \ref{s:summary-results}. Here we present the details of the search method.

{\fl \bf 1.} First we pick a pendulum bifurcation energy $E$ and a basepoint on the corresponding pendulum orbit. We then compute the eigenvectors of the monodromy matrix (\ref{e:pend-pert-al-12-coeff-mat}) and identify the transverse eigenvector(s). These eigenvectors depend on the choice of the basepoint, which is taken to be the one at which $V = 0$ along the trajectory.

{\fl \bf 2.} We solve the first order nonlinear EOM for $\al_1, \al_2, \pi_1$ and $\pi_2$ obtained from (\ref{sec-ordr-eom}) \& (\ref{e:dim-less-t-mom})
with ICs given by a small perturbation to the pendulum IC at the basepoint $[\al_1(0) = 0, \pi_1(0) = 0, \al_2(0) = 0, \pi_2(0) = \sqrt{4 m r^2 E/ 3}]$ in the direction of the transverse eigenvector, of amplitude $\del \al_1$ or $\del \pi_1$ depending on the bifurcation point, as discussed in \S \ref{evecs-monod}.

{\fl \bf 3.} We evolve the new trajectory till a time $t_*$ when $\al_2$ completes either one (for \IS) or two (for \PD) cycles. The resulting trajectory is typically approximately periodic with $t_* \approx \tau$ or $2 \tau$ where $\tau$ is the period of the unperturbed pendulum orbit. In order to make this trajectory periodic, we adjust the value of $\pi_2(0)$ to minimize the `departure from periodicity'
	\beqs
	d(\pi_2(0)) &=& [ (\al_1(t_*) - \al_1(0))^2 + (\pi_1(t_*) - \pi_1(0))^2
	\cr & & + (\pi_2(t_*) - \pi_2(0))^2 ]^{1/2}.
	\eeqs
The obtained periodic trajectory is found to be independent of the choice of the basepoint. 

{\fl \bf 4.} By varying the amplitude of the perturbation to the pendulum IC, we may generate a family of newly born periodic trajectories. To go further down this family, it is advantageous to use the ICs of the previous member of the family rather than those of the original pendulum orbit.

\section{Features of periodic trajectories born at bifurcations of pendula}
\label{s:features-newly-born-pend}

\begin{figure}[!h]
\centering
	\begin{subfigure}{0.45\textwidth}
	\includegraphics[width = \textwidth]{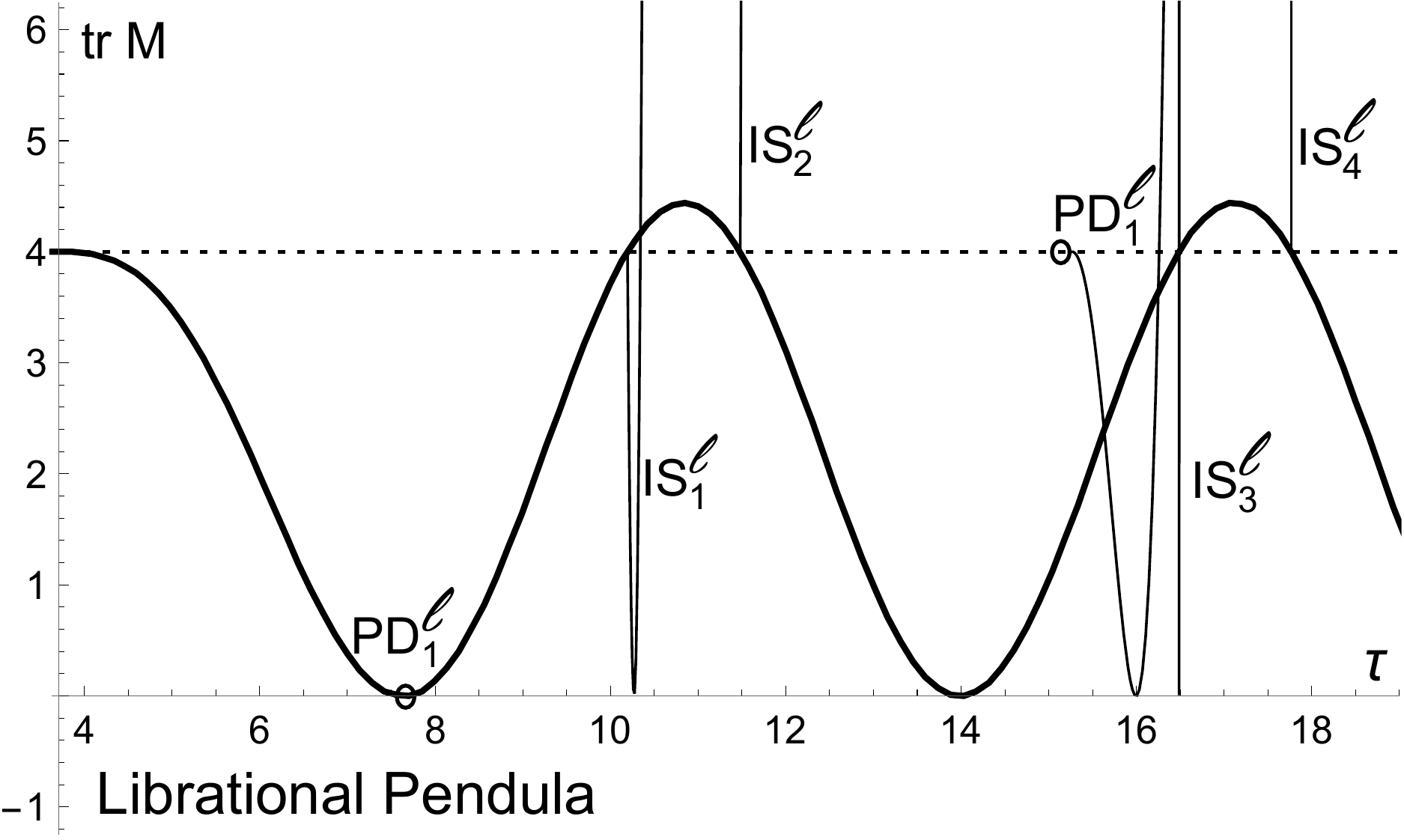}
	\caption{}
	\label{f:trM-v-T-lib-IS-PD}
	\end{subfigure}
\hfil
	\begin{subfigure}{0.45\textwidth}
	\includegraphics[width = \textwidth]{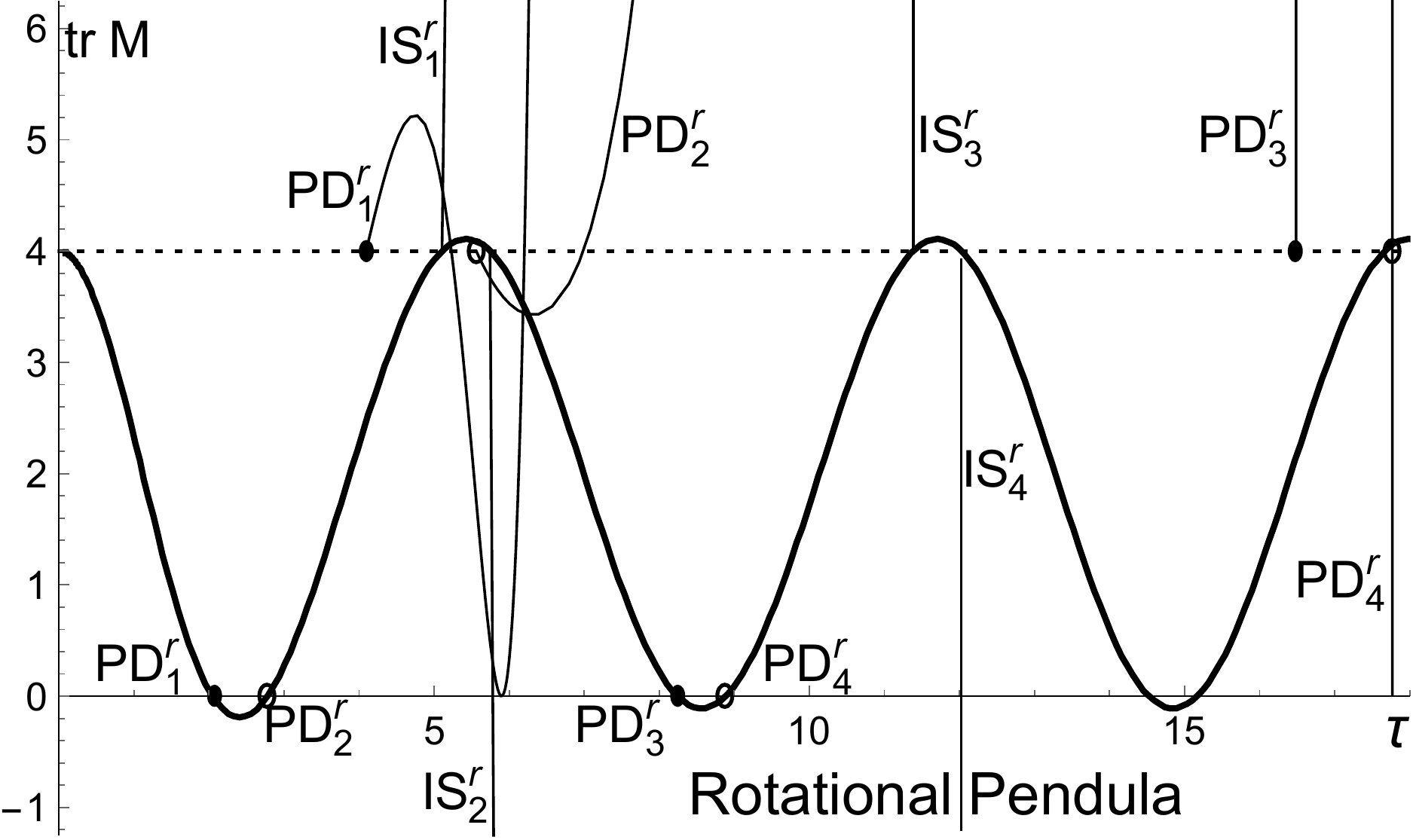}
	\caption{}
	\label{f:trM-v-T-rot-IS-PD}
	\end{subfigure}
	\caption{\small $\tr M$ vs time period $\tau$ for (a) librational and (b) rotational pendula and families of newly born periodic orbits at their isochronous and period-doubling bifurcations. The new orbits fall into seven classes as mentioned in \S \ref{s:Pendulum stability transition energies and time periods}. Upon including the newly born orbits, $\tr M(\tau)$ is asymptotically $4\pi$-periodic as $\tau \to \infty$. Unlike the corresponding bifurcations in the A orbits of He\'non-Heiles \cite{brck-omega}, $\tr M$ for $\IS^\ell_{1,3,5,\dots}$ is a nonlinear function of $\tau$: it reaches a minimum and then increases.}
	\label{f:trM-v-T-lib-rot-pend-IS}
\end{figure}

Using the algorithm of \S \ref{s:search-method-pend}, we find the periodic orbits born at the $\IS$ and $\PD$ bifurcations of pendula (see Fig~\ref{f:trM-v-T-lib-rot-pend-IS}). The bifurcations are shown to be forward fork-like \cite{brck-fork} as only one new family of periodic orbits is born and exists only for energies exceeding the bifurcation energy. In all cases, the newly born orbits have $\tr M = 4$ at the bifurcation point.

\subsection{Shapes of new periodic orbits in the $\al_1$-$\al_2$ plane}

At each bifurcation point, $\al_2$ for the newly born trajectory is essentially the same as for the corresponding pendulum trajectory $(\bar \al_2)$. On the other hand, $\al_1$ ($\bar \al_1 \equiv 0$ for pendula) displays progressively more oscillations as we proceed through the bifurcation cascade toward $E = 4g$. Near the bifurcation points, the time dependence of $\al_1 = \bar \al_1 + \del \al_1$ is given by the periodic Lam\'e functions \cite{Ince,erdelyi}  listed in Tables \ref{t:lib-al1-lame-fns-IS-PD} and \ref{t:rot-al1-lame-fns-IS-PD}, which are solutions of equations (\ref{e:lib-pend-lame-eqn}) and (\ref{e:rot-pend-lame-eqn}) obtained by linearizing the EOM around the pendulum orbits. For example, $\al_1$ for newly born periodic orbits near the $\IS^\ell_2$ bifurcation satisfies (\ref{e:lib-pend-lame-eqn}) with $k^2 \approx 1 - e^{-5.78}/4$ from Table \ref{t:lib-pend-trans-E-T}. In the notation of Refs.\cite{Ince,erdelyi}, the periodic solution of (\ref{e:lib-pend-lame-eqn}) is the second ($m=2$) Lam\'e sine function Es$^m_n(z)$ of order $n$ with Lam\'e eigenvalue $h = 1$. It has $m = 2$ zeros in the interval $0 \leq z \leq 2K(k)$ which is the period of this Lam\'e function.


\begin{table}
\begin{center}
\begin{tabular}{|c|c|c|c|}
\hline
new periodic & time period & Lam\'e function & stability \\
orbit &  of $\al_1 \approx \del \al_1$  & $\del \al_1 = $ Ex$^m_n$ & near bifur. \\
\hline
$\PD^\ell_1$& $2$ & $m = 3/2$ & stable \\
$\IS^\ell_1$ & $1/2$ &  Ec$_n^2$ & stable \\
$\IS^\ell_2$ & $1/2$ &  Es$_n^2$ & unstable \\
$\PD^\ell_2$& $2$ &  $m = 5/2$ & stable \\
$\IS^\ell_3$ & $1$ &  Ec$_n^3$ & stable \\
$\IS^\ell_4$ & $1$ &  Es$_n^3$ & unstable \\
$\PD^\ell_3$& $2$ &  $m = 7/2$ & stable \\
$\IS^\ell_5$ & $1/2$ &  Ec$_n^4$ & stable \\
$\IS^\ell_6$ & $1/2$ &  Es$_n^4$ & unstable \\
$\PD^\ell_4$& $2$ &  $m = 9/2$ & stable \\
$\IS^\ell_7$ & $1$ &  Ec$_n^5$ & stable \\
$\IS^\ell_8$ & $1$ &  Es$_n^5$ & unstable \\
\hline
\end{tabular}
\caption{\small Newly born periodic trajectories at isochronous and period-doubling bifurcations of librational pendula and their stability in increasing order of bifurcation energy $E = 4 g k^2$. The period of $\del \al_1$ is given in units of $\tau_\ell$ (\ref{e:time-per-lib-rot-pend}). The new orbits are $\tau_\ell$ and $2 \tau_\ell$ periodic for $\IS$ and $\PD$. Note that $n(n+1) = 2/3$ and the Lam\'e eigenvalue $h = 1$ in all cases while $m$ is the number of zeros of the solutions Ex$^m_n$ of (\ref{e:decoupled-lame-al1-al2}) in the periodicity interval $0 \leq \tl t < \tau_\ell/2$ of the coefficients. Ex$^m_n$ is $\tau_\ell/2$ or $\tau_\ell$ periodic according as $m$ is even or odd. The number of nodes of $\al_1$ in one period of the newly born trajectory is $2m$ for $\IS$ and $4m$ for $\PD$. We have not identified Lam\'e functions corresponding to $\PD^\ell_n$. We note that as $k \to 1$, it becomes harder to numerically evaluate periodic Lam\'e functions accurately using inbuilt routines in the Mathematica software package \cite{mathematica}.}
\label{t:lib-al1-lame-fns-IS-PD}
\end{center}
\end{table}


\begin{figure}[!h]
\centering
	\begin{subfigure}[h]{0.22\textwidth}
	\includegraphics[width=\textwidth]{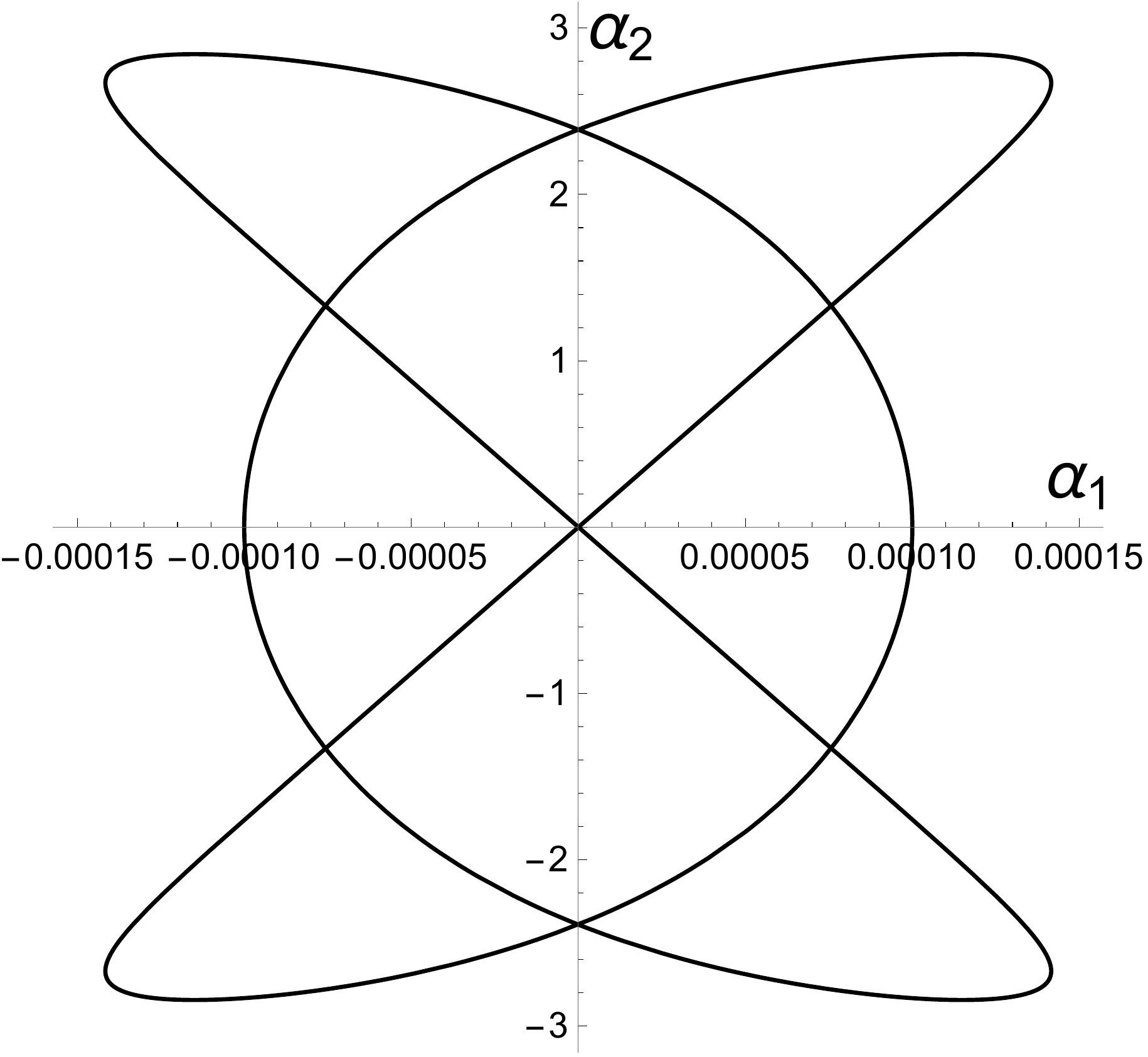}
	\caption{$\PD^\ell_1$}
	\end{subfigure}
\hfil
	\begin{subfigure}[h]{0.22\textwidth}
	\includegraphics[width=\textwidth]{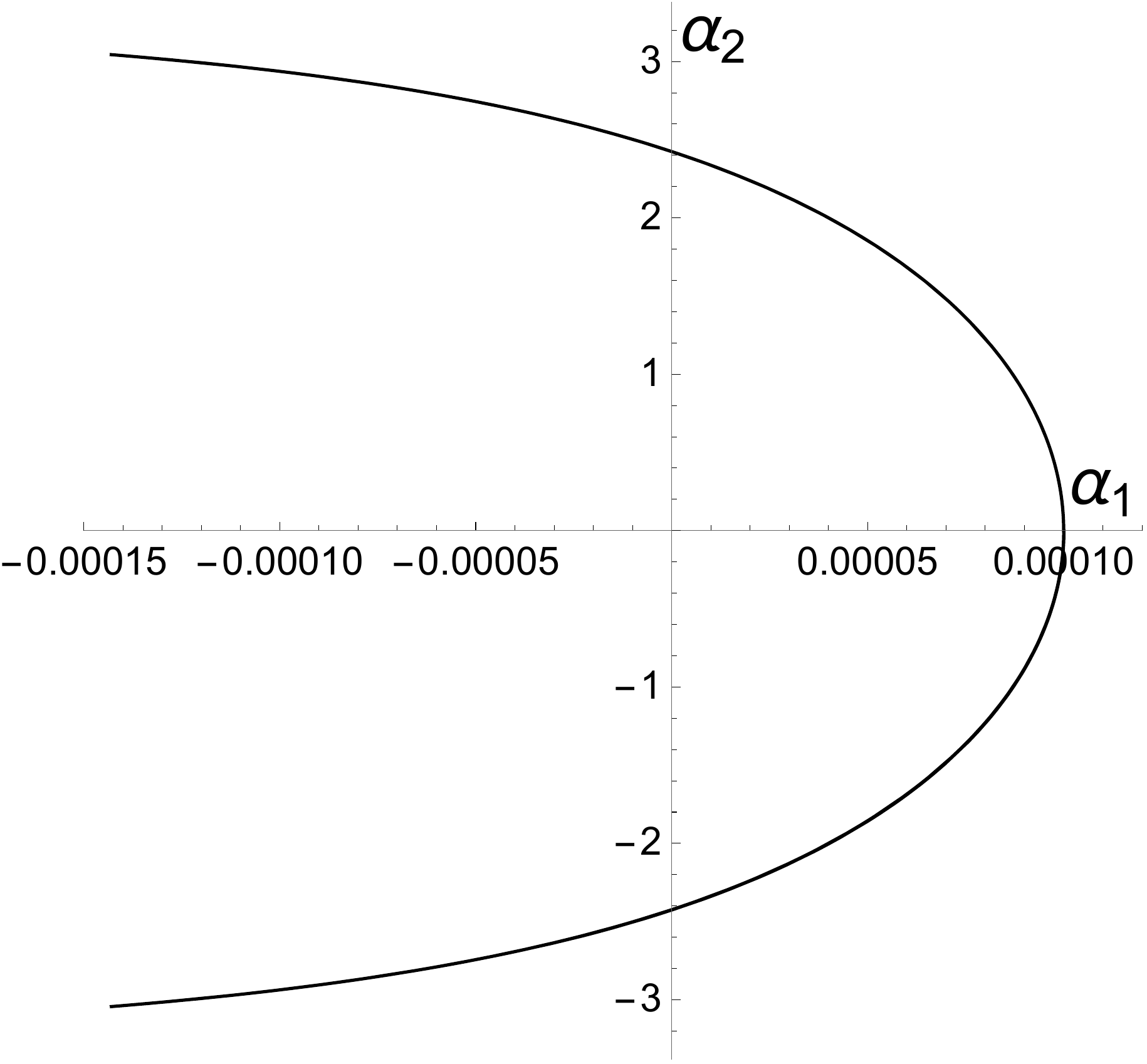}
	\caption{$\IS^\ell_1$}
	\end{subfigure}
\hfil
	\begin{subfigure}[h]{0.22\textwidth}
	\includegraphics[width=\textwidth]{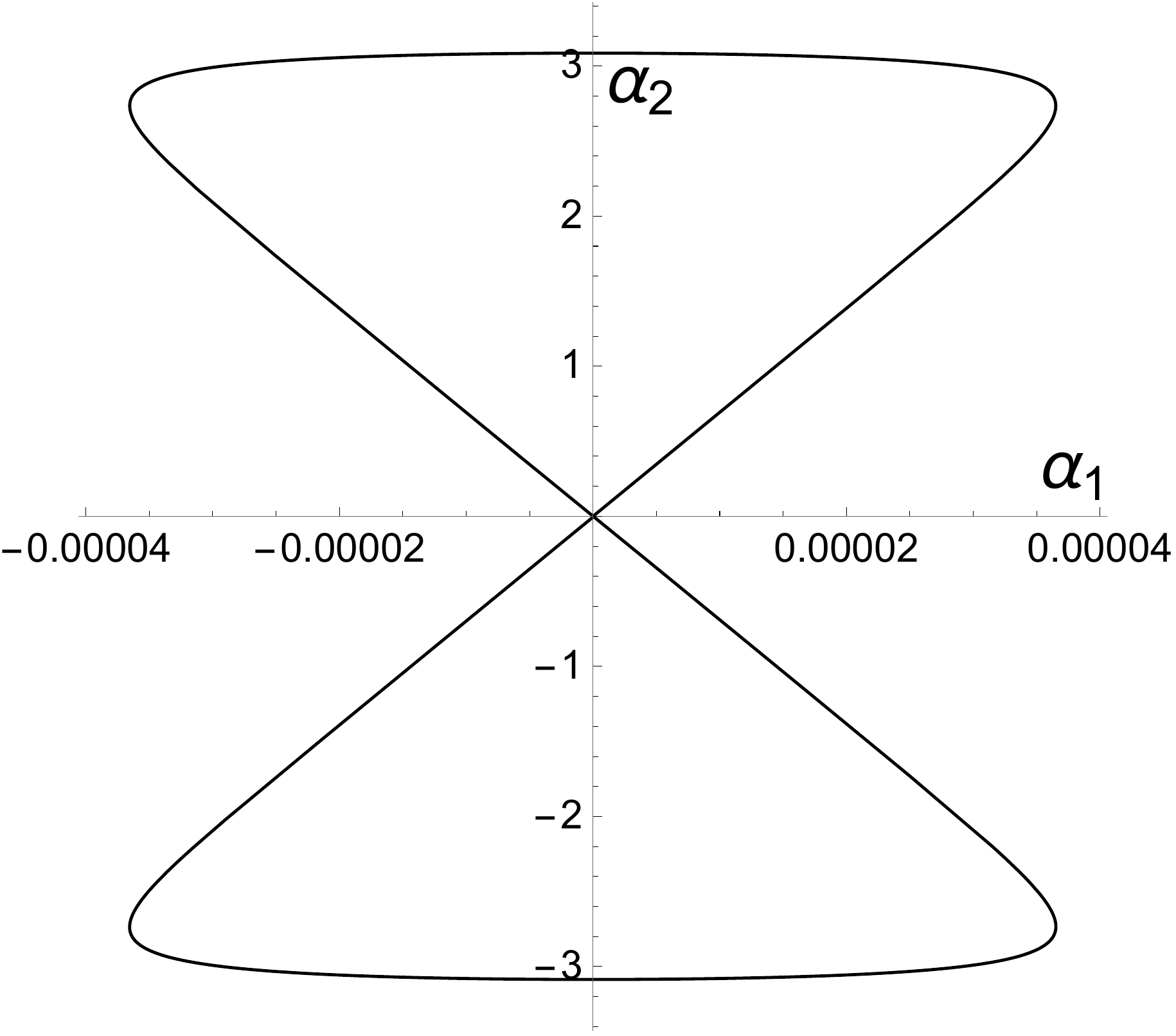}
	\caption{$\IS^\ell_2$}
	\end{subfigure}
\hfil
	\begin{subfigure}[h]{0.22\textwidth}
	\includegraphics[width=\textwidth]{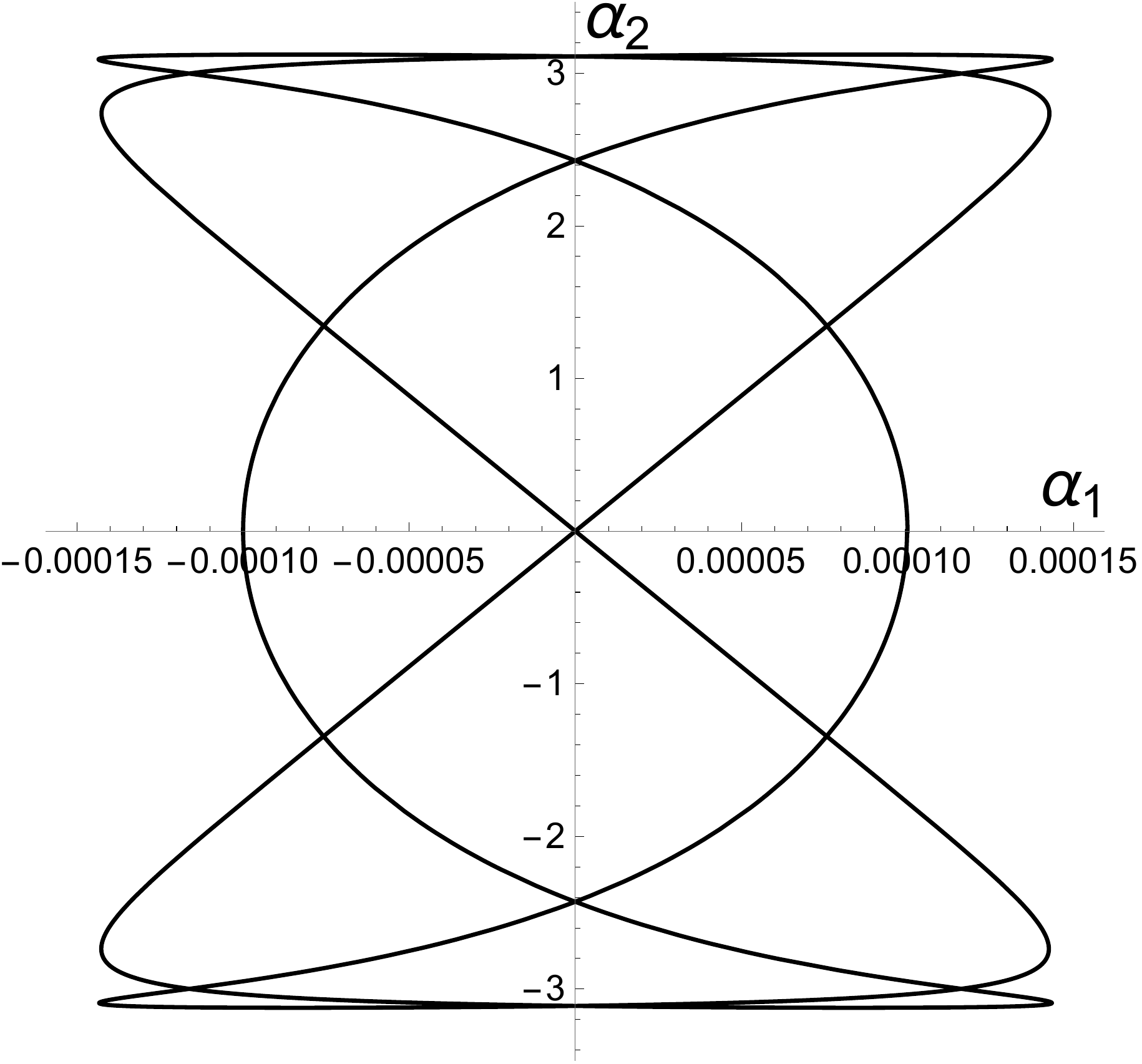}
	\caption{$\PD^\ell_2$}
	\end{subfigure}

	\begin{subfigure}[h]{0.22\textwidth}
	\includegraphics[width=\textwidth]{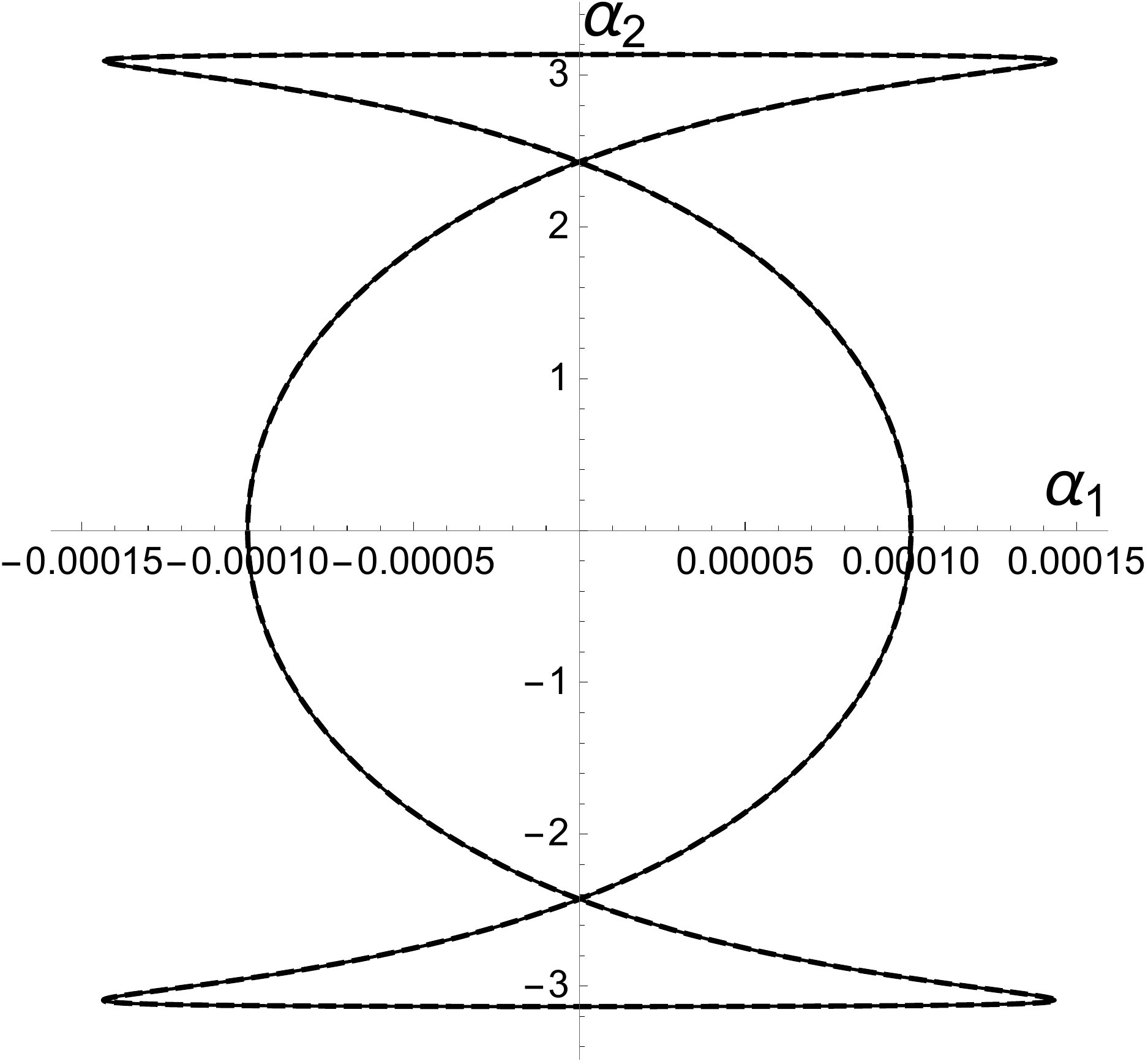}
	\caption{$\IS^\ell_3$}
	\end{subfigure}
\hfil
	\begin{subfigure}[h]{0.22\textwidth}
	\includegraphics[width=\textwidth]{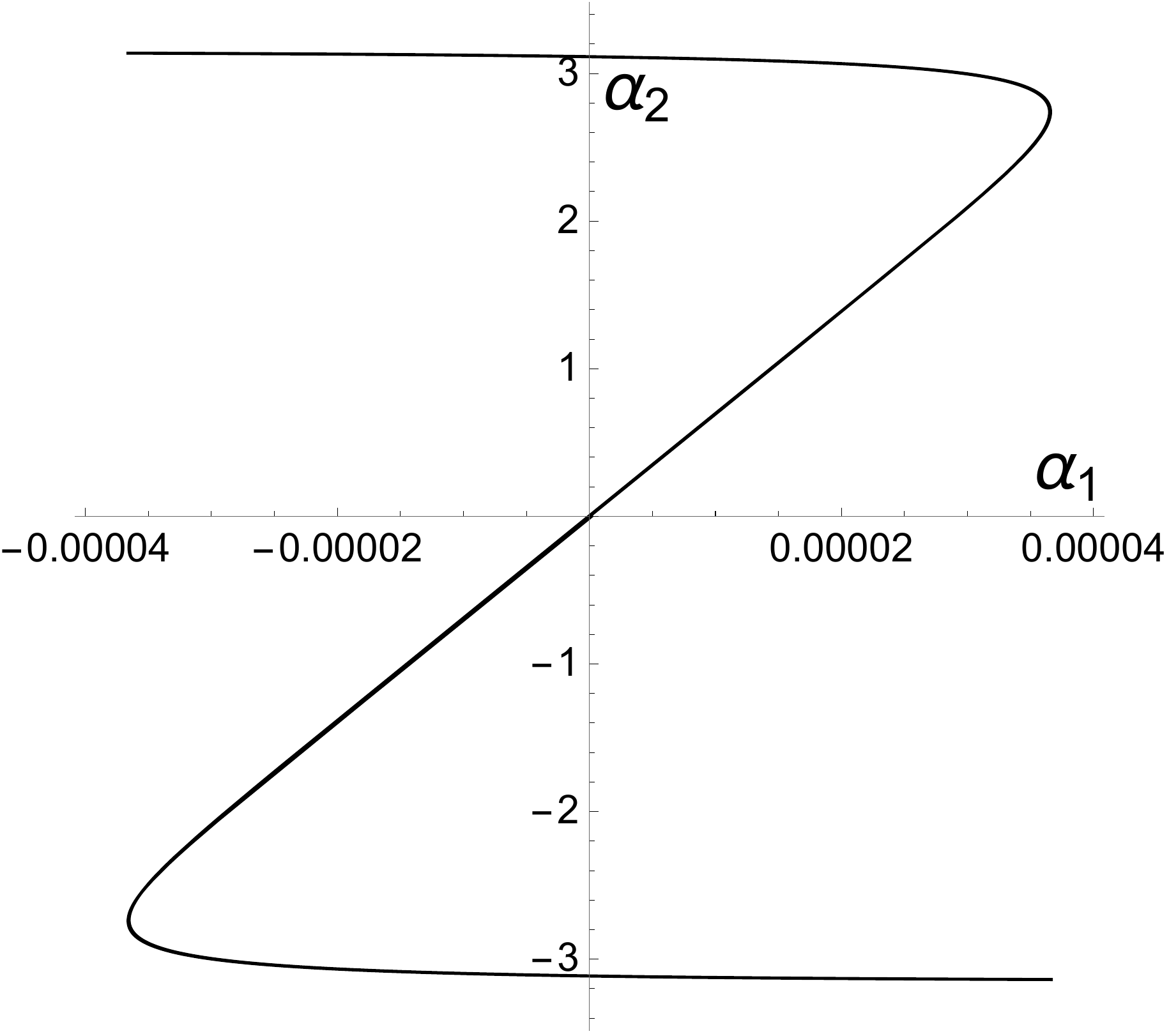}
	\caption{$\IS^\ell_4$}
	\end{subfigure}
\hfil
	\begin{subfigure}[h]{0.22\textwidth}
	\includegraphics[width=\textwidth]{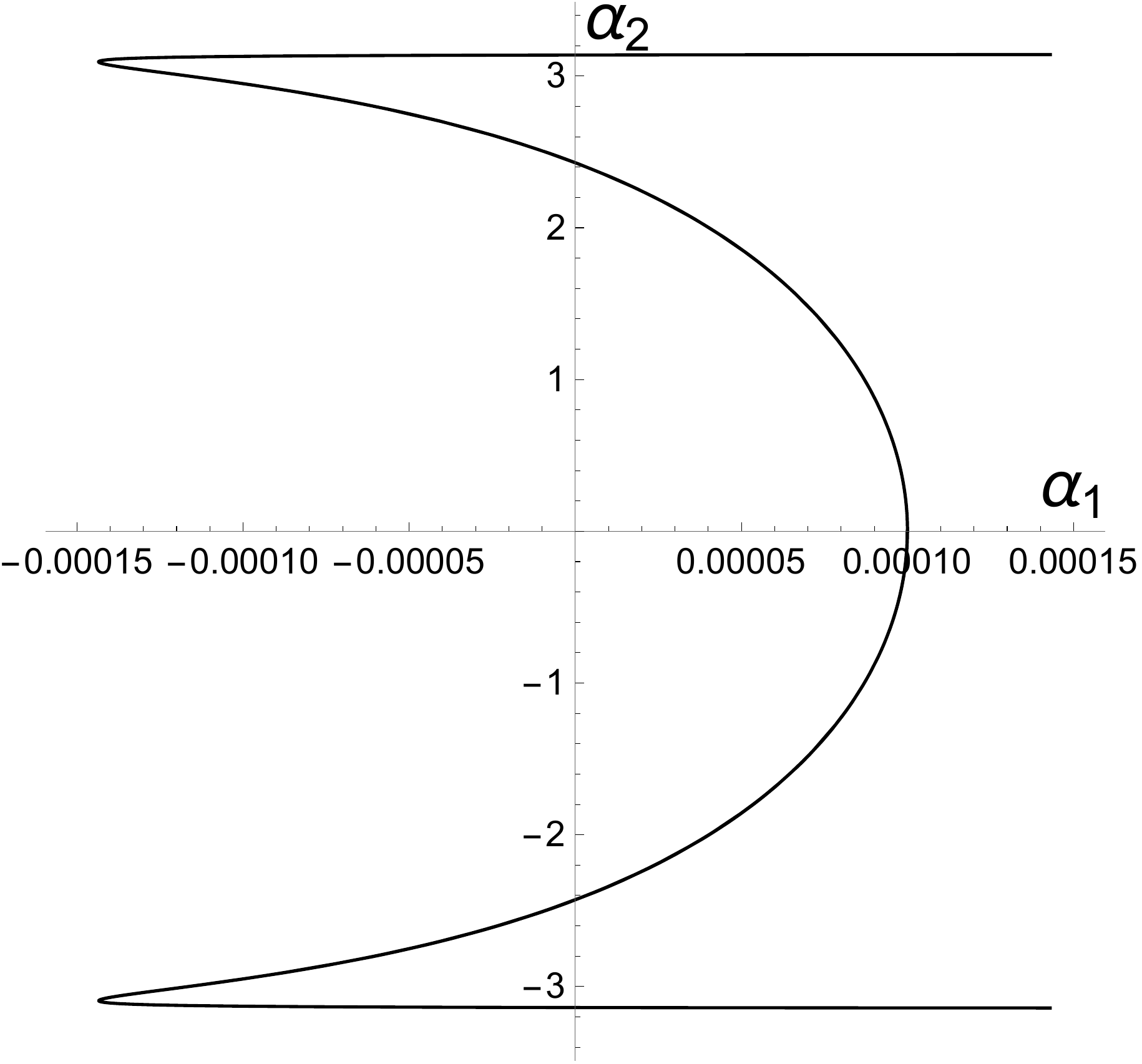}
	\caption{$\IS^\ell_5$}
	\end{subfigure}
\hfil
	\begin{subfigure}[h]{0.22\textwidth}
	\includegraphics[width=\textwidth]{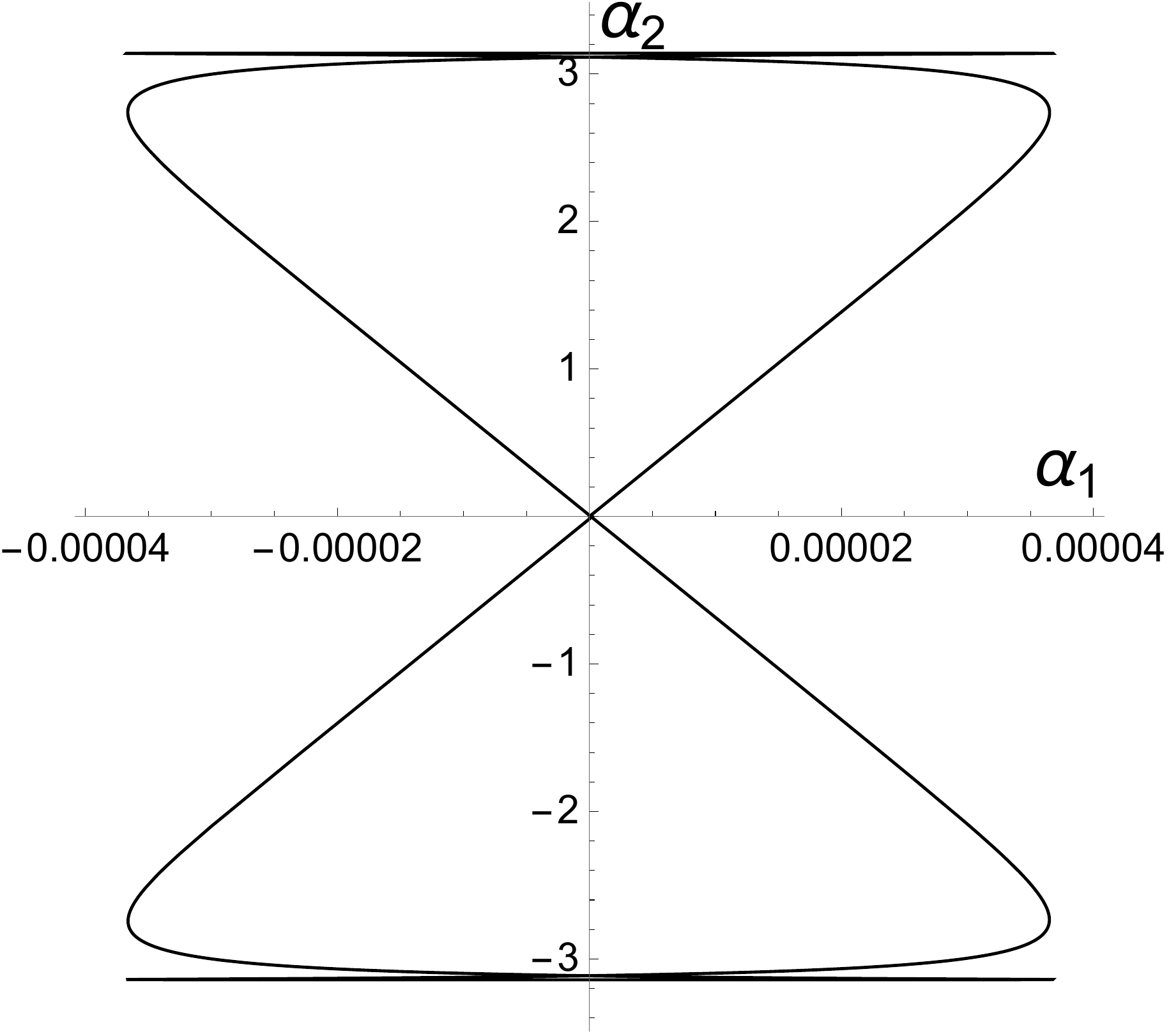}
	\caption{$\IS^\ell_6$}
	\end{subfigure}
	\caption{\small Shapes of newly born periodic orbits on the $\al_1$-$\al_2$ plane at bifurcations of librational pendula in increasing order of energy. They are obtained by solving the linearized equation for $\del \al_1$ while taking $\al_2 = \bar \al_2$. The resulting shapes agree with those obtained by solving the nonlinear EOM using the search method with the same ICs. This is demonstrated for $\IS^\ell_3$ where the dashed curve (nonlinear evolution) agrees with the solid curve (linearized evolution). The closed curves are traversed once per period while the `open' ones are traversed back and forth.}
	\label{f:pend-lib-new-traj}
\end{figure}

\begin{figure}[!h]
\centering
	\begin{subfigure}[h]{0.23\textwidth}
	\includegraphics[width=\textwidth]{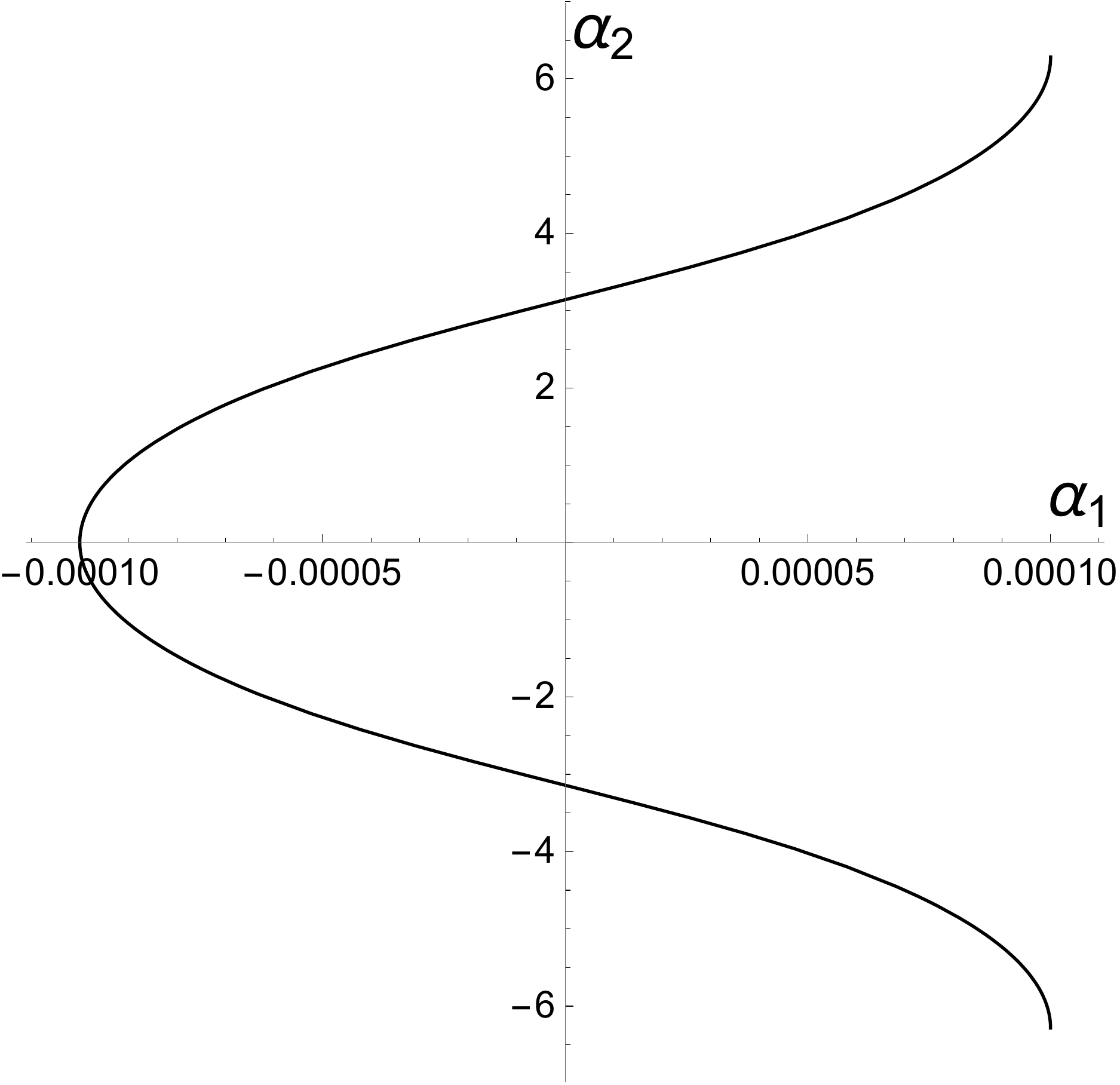}
	\caption{$\PD^r_1$}
	\end{subfigure}
\hfil
	\begin{subfigure}[h]{0.23\textwidth}
	\includegraphics[width=\textwidth]{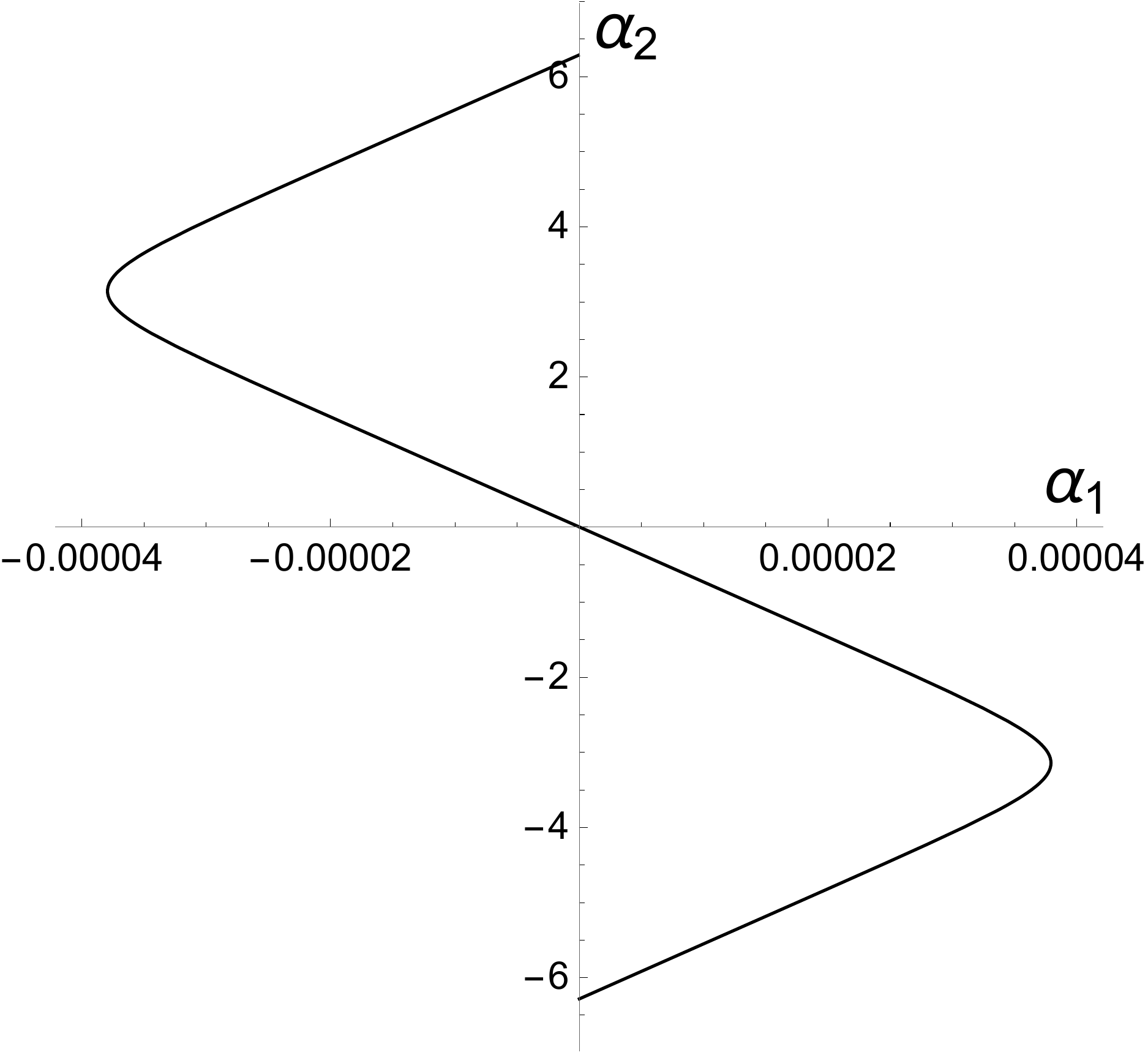}
	\caption{$\PD^r_2$}
	\end{subfigure}
\hfil
	\begin{subfigure}[h]{0.23\textwidth}
	\includegraphics[width=\textwidth]{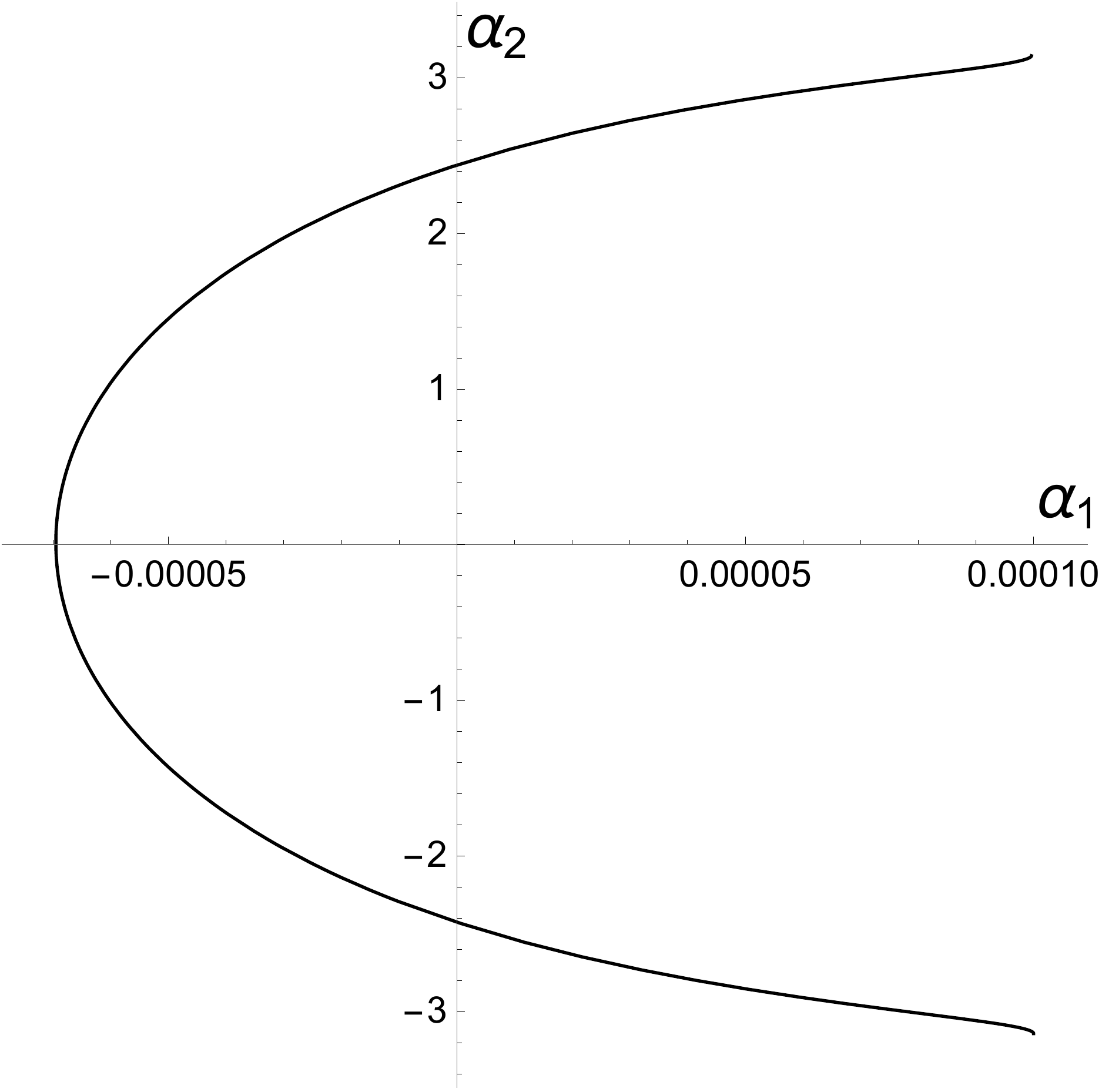}
	\caption{$\IS^r_1$}
	\end{subfigure}
\hfil	
	\begin{subfigure}[h]{0.23\textwidth}
	\includegraphics[width=\textwidth]{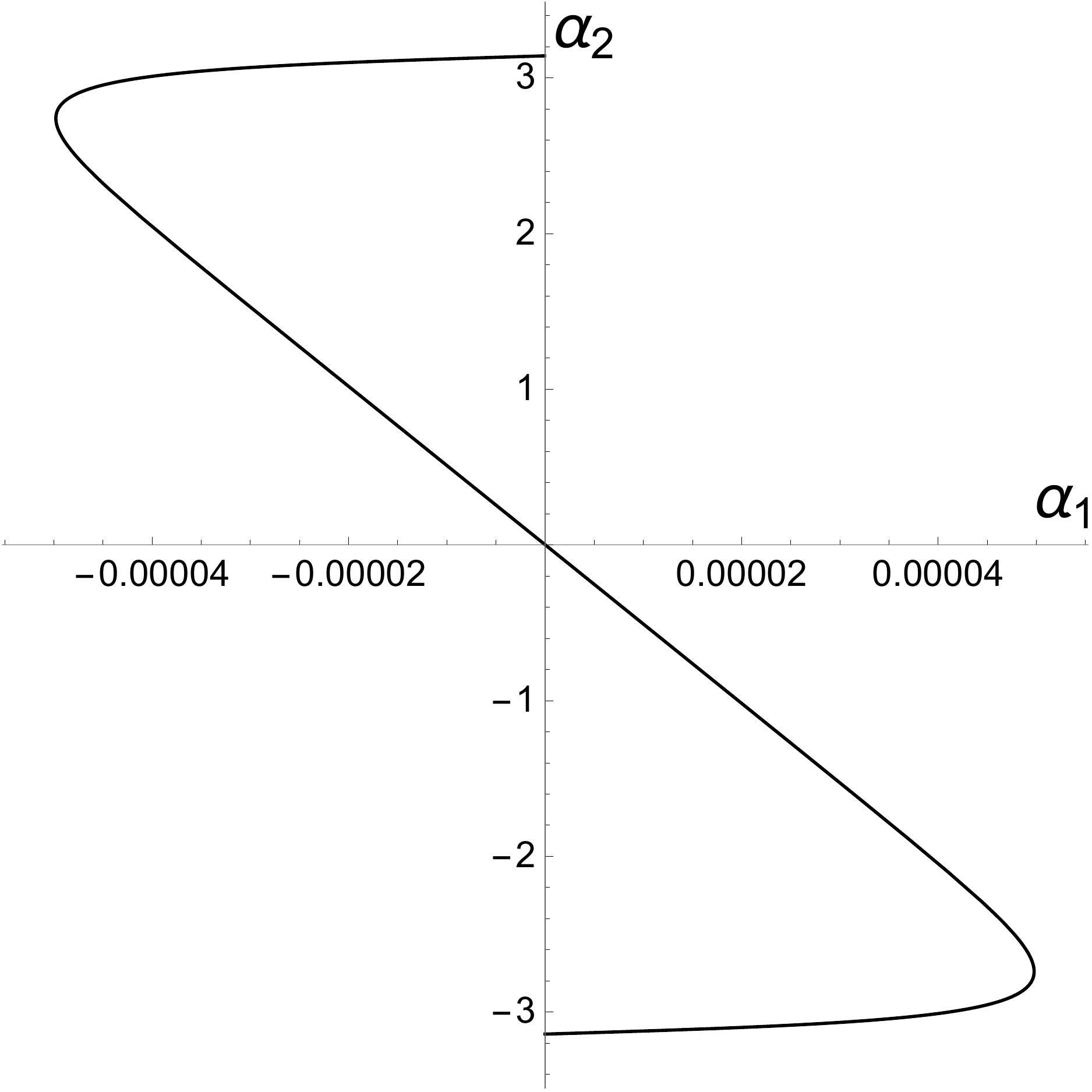}
	\caption{$\IS^r_2$}
	\end{subfigure}

	\begin{subfigure}[h]{0.23\textwidth}
	\includegraphics[width=\textwidth]{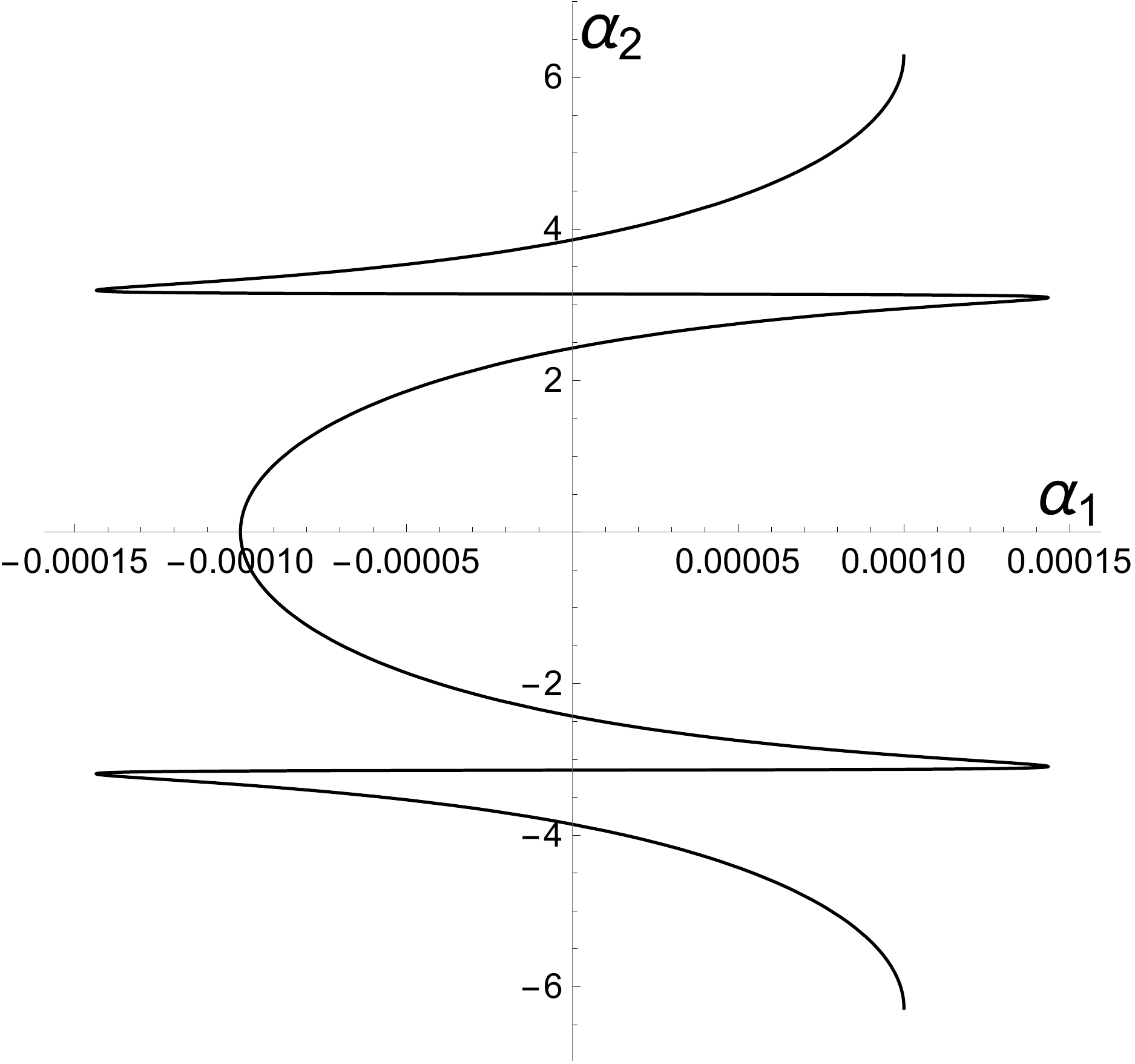}
	\caption{$\PD^r_3$}
	\end{subfigure}
\hfil
	\begin{subfigure}[h]{0.23\textwidth}
	\includegraphics[width=\textwidth]{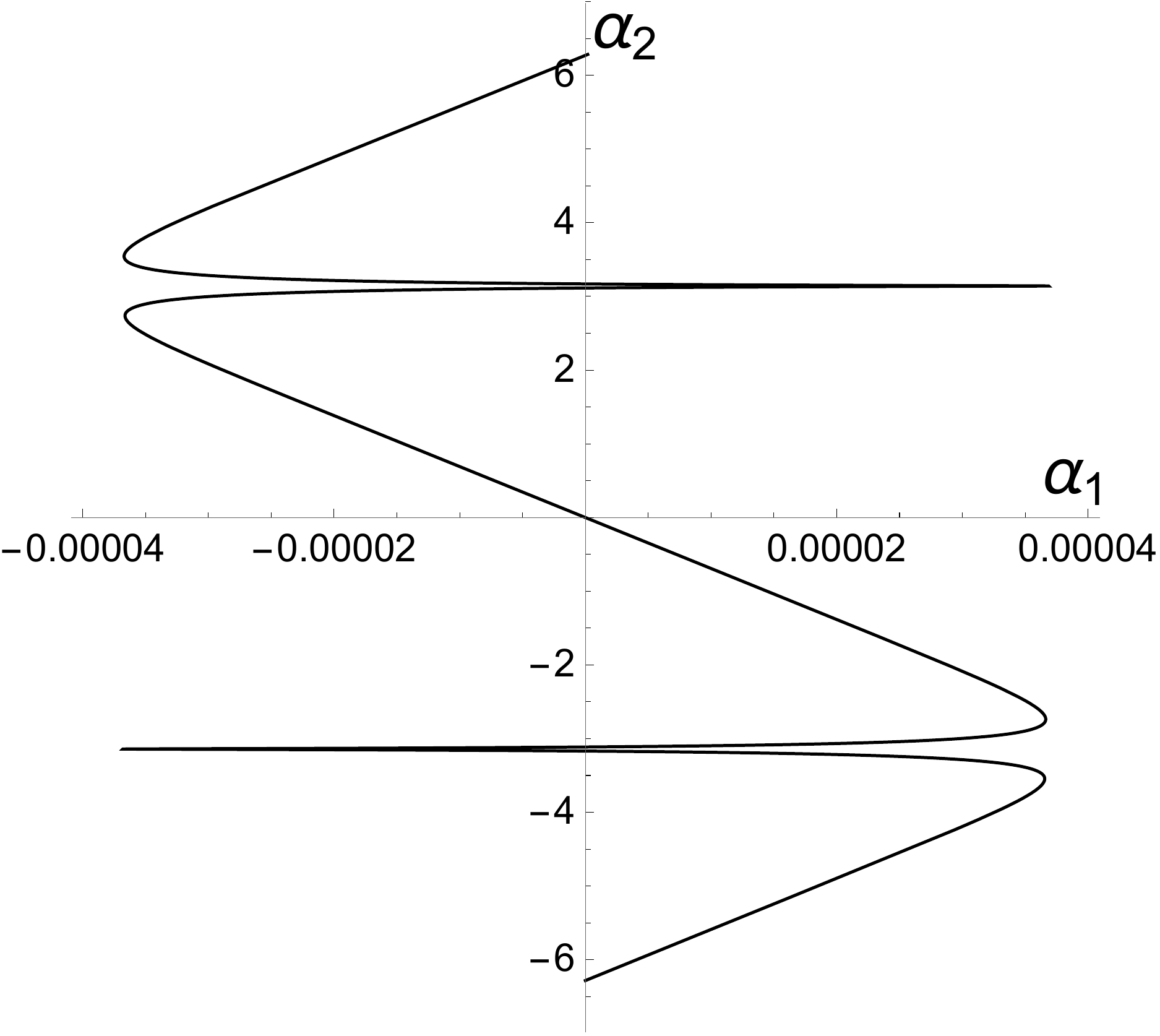}
	\caption{$\PD^r_4$}
	\end{subfigure}
\hfil
	\begin{subfigure}[h]{0.23\textwidth}
	\includegraphics[width=\textwidth]{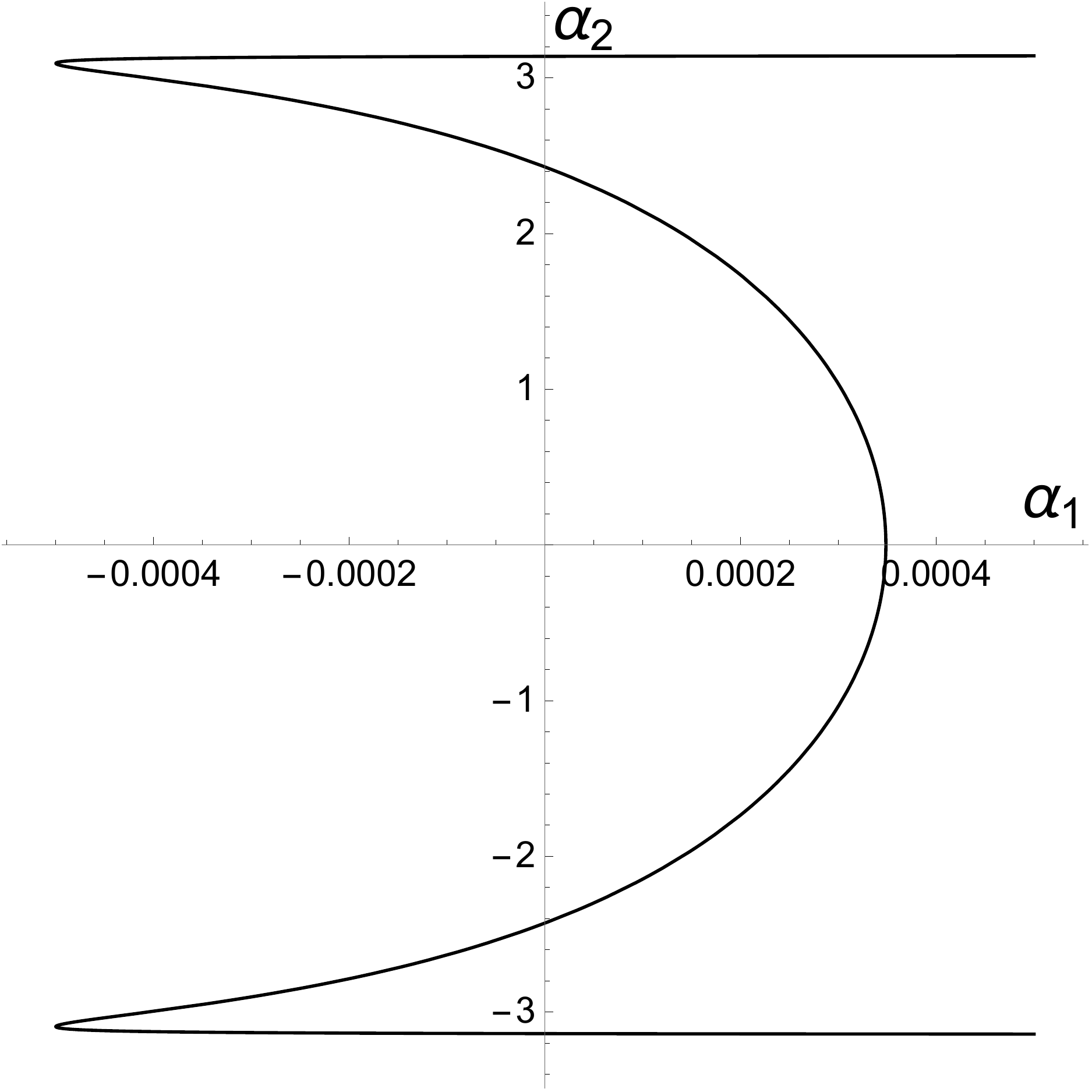}
	\caption{$\IS^r_3$}
	\end{subfigure}
\hfil
	\begin{subfigure}[h]{0.23\textwidth}
	\includegraphics[width=\textwidth]{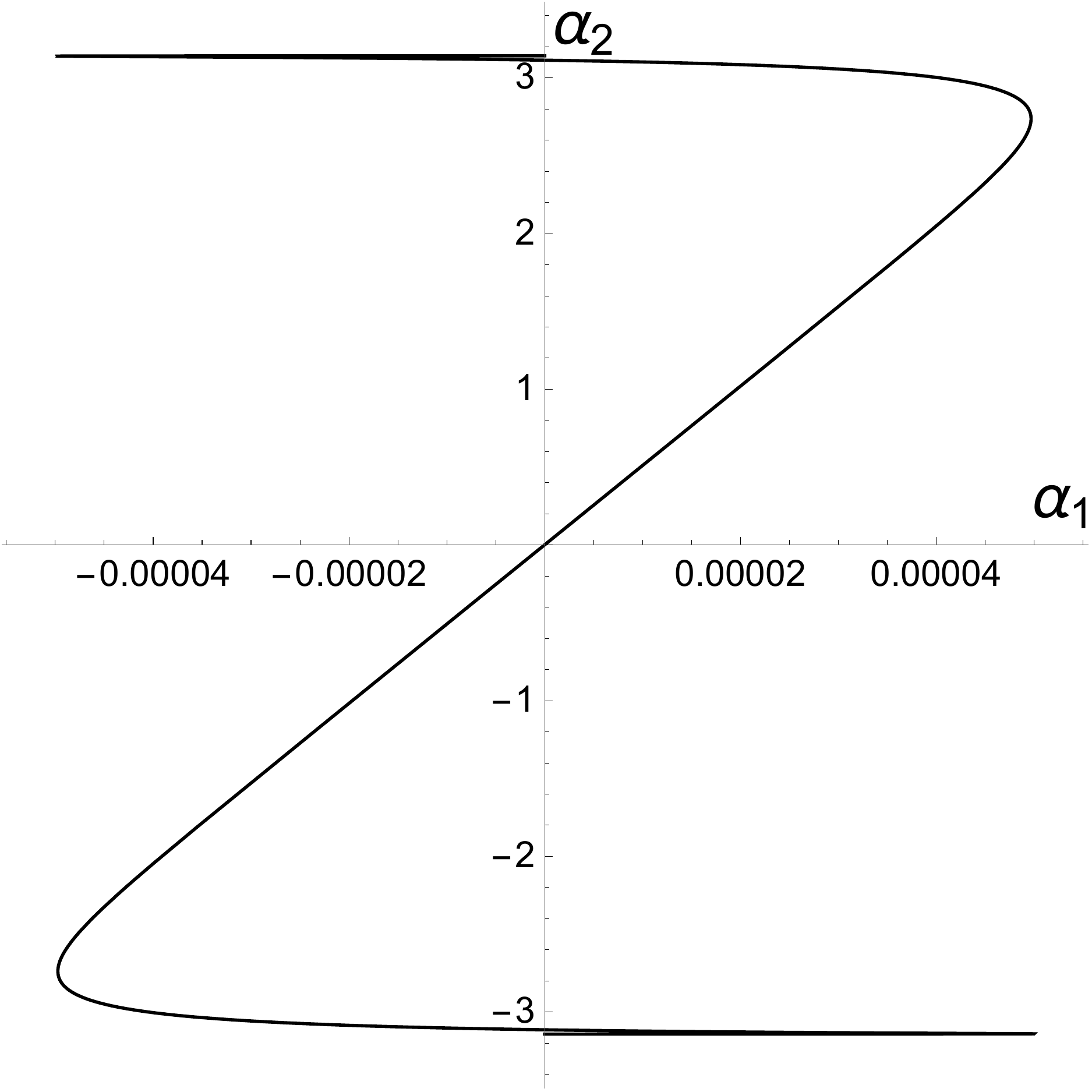}
	\caption{$\IS^r_4$}
	\end{subfigure}
\caption{\small Shapes of newly born periodic orbits on the $\al_1$-$\al_2$ plane at bifurcations of rotational pendula in decreasing order of energy. They are obtained by the same procedure as in Fig.~\ref{f:pend-lib-new-traj}. These newly born periodic trajectories are rotational, i.e., like winding modes around a cylinder (with axis parallel to the $\al_1$ axis) once we identify the points $\al_2 = \pm \pi$ (for $\IS$) and $\al_2 = \pm 2 \pi$ (for $\PD$). Qualitatively, the last four figures may be obtained from the first four by including an extra oscillation in the $\al_1$ direction at $\al_2 = \pm \pi$.}
\label{f:pend-rot-new-traj}
\end{figure}

Fig.~\ref{f:pend-lib-new-traj} and Fig.~\ref{f:pend-rot-new-traj} contain parametric plots of the newly born trajectories on the $\al_1-\al_2$ plane. The orbit shapes shown in these figures remind us of the successive excited state wavefunctions of a quantum system, with bifurcation energies playing the role of energy eigenvalues. Near each bifurcation point, the shape of a newly born orbit obtained via the search method of \S \ref{s:search-method-pend} matches that obtained by solving the linearized EOM for $\del \al_1$ (\ref{e:lame-eq-transv-al1-pi1}) with the same ICs, while taking $\al_2 = \bar \al_2$ at the bifurcation point. The linearized approximation of course cannot be trusted far from the bifurcation point.


\begin{table}
\begin{center}
\begin{tabular}{|c|c|c|c|}
\hline
new per. orb & time period of $\al_1$ & Lam\'e fn & stability \\
\hline
$\PD^r_1$& $2$ & Es$_n^1$ & unstable \\
$\PD^r_2$& $2$ & Ec$_n^1$ & stable \\
$\IS^r_1$ & $1$ & Ec$_n^2$ & unstable \\
$\IS^r_2$ & $1$ & Es$_n^2$ & stable \\
$\PD^r_3$& $2$ & Es$_n^3$ & unstable \\
$\PD^r_4$& $2$ & Ec$_n^3$ & stable \\
$\IS^r_3$ & $1$ & Ec$_n^4$ & unstable \\
$\IS^r_4$ & $1$ & Es$_n^4$ & stable \\
$\PD^r_5$& $2$ & Es$_n^5$ & unstable \\
$\PD^r_6$& $2$ & Ec$_n^5$ & stable \\

\hline
\end{tabular}
\caption{\small Newly born periodic trajectory at $\IS$ and $\PD$ bifurcations of rotational pendula and their stability in increasing order of $\ka = 1/k$ and decreasing order of bifurcation energies. The time period of $\al_1$ is given in units of $\tau_r$ (\ref{e:time-per-lib-rot-pend}). The integer $m$ is the number of nodes of $\al_1$ in the fundamental domain of duration $\tau_r$.}
	\label{t:rot-al1-lame-fns-IS-PD}
\end{center}
\end{table}


\subsection{Stability of newly born orbits}
\label{stblty-new-orb}

Aside from the period-doubling bifurcations of librational pendula, all the bifurcations can be classified into two types: stable to unstable transitions $(\IS^\ell_{1,3,5,\cdots}, \IS^r_{2,4,6,\cdots}$ and $\PD^r_{2,4,6,\cdots})$ and unstable to stable transitions $(\IS^\ell_{2,4,6, \cdots}, \IS^r_{1,3,5,\cdots}$ and $\PD^r_{1,3,5,\cdots})$ with increasing energy $E$. Soon after the bifurcation, the newly born family has a stability opposite to that of the surviving pendulum solution. Moreover, as shown in Fig.~\ref{f:trM-v-T-lib-rot-pend-IS}, near the bifurcations, if the newly born family is (un)stable, its $\tr M$ (increases)decreases linearly with its time period $\tau$. As we move farther away from the bifurcation along a stable family of new orbits ($\IS^\ell_{1,3,5, \ldots}$ and $\PD^r_{4,6,8, \ldots}$), we find that $\tr M(\tau)$ decreases from $4$, reaches a local minimum where $\tr M = 0$ and then increases beyond $\tr M = 4$ at which stage the family becomes unstable. By contrast, for newly born unstable orbits ($\IS^\ell_{2,4,6 , \ldots}$, $\IS^r_{1,3,5, \ldots}$ and $\PD^r_{3,5,7, \ldots}$), $\tr M$ appears to increase linearly with $E$ even far from the bifurcation. Finally, $\tr M(\tau)$ for the stable orbits $\IS^r_{2,4,6, \ldots}$ decreases linearly with $\tau$ eventually rendering them unstable when $\tr M$ becomes negative. Notably, the energies of all the newly born orbits always increase linearly with $\tau$ at least in the immediate vicinity of the bifurcations. 

We find that the newly born orbits at $\PD^r_1$ and $\PD^r_2$ are atypical in comparison with those at the other period-doubling bifurcations of rotational pendula. (1) The energies at these bifurcation points differ significantly from the asymptotic geometric progression: $\log((4-E(\PD^r_1))/(4-E(\PD^r_3))) = 10.59$, $\log((4-E(\PD^r_2))/(4-E(\PD^r_4))) = 10.49$ while $\log((4-E(\PD^r_{n}))/(4-E(\PD^r_{n+2}))) \approx 10.88$ for $n \geq 3$ (see Table \ref{t:rot-pend-trans-E-T}) (2) The graph of $\tr M(\tau)$ for pendula has a greater dip between $\PD^r_1$ and $\PD^r_2$ than elsewhere. (3) $\PD^r_{1,2}$ are not part of the rotation-libration duality of \S \ref{s:duality-lib-rot}. (4) The graphs of $\tr M(\tau)$ for the newly born orbits at these two bifurcations are qualitatively different from those at other period-doubling bifurcations.



Period-doubling bifurcations ($\PD^\ell_{1,2,3,\cdots}$) of librational pendula are all of the same sort. The newly born family at each of these bifurcations is neutrally stable as $\tr M - 4$ has a double zero at the bifurcation point as a function of $E$ or $\tau$. However, as one moves away from the bifurcation point, the newly born trajectory becomes stable (see Fig.~\ref{f:trM-v-T-lib-IS-PD}).

\subsection{Slopes of $\tr M$ at bifurcations}

At isochronous bifurcations (both librational and rotational), numerically, we find that
	\beq
	\tr M_{\rm pend}'(E) \approx -(1/2) \tr M_{\rm new}'(E),
	\label{FLB-slope-theorem}
	\eeq
where both derivatives are evaluated at the bifurcation energy $E(\IS^{\ell,r}_{1,2,3, \ldots})$. For example, both sides of (\ref{FLB-slope-theorem}) are $\approx 159$ and $\approx -43$ at $\IS^\ell_1$ and $\IS^r_1$. Such an equality is expected from the FLB slope theorem for fork-like bifurcations \cite{brck-fork}.

For period-doubling bifurcations, we find a similar relation. In the librational phase, $\tr M(E)$ has a double zero at the bifurcation point ($\PD^\ell_n$) for both the pendulum orbit as well as the newly born family of periodic orbits. For period-doubling bifurcations of rotational pendula $(\PD^r_{1,2,3 \cdots})$ we find
	\beq
	\tr M_{\rm pend}'(E) \approx (1/8) \tr M_{\rm new}'(E).
	\label{FLB-slope-theorem-PD}
	\eeq
For instance we find that $\tr M_{\rm new}'(E)/\tr M_{\rm pend}'(E)$ is $8.3, 7.9$ and $8.1$ for $\PD^r_{1,2,3}$. The corresponding ratio for the only period-doubling bifurcation of the breather family discussed in \S \ref{s:breather} is $\approx 7.99$. Based on these numerical results, we conjecture that this is a period-doubling analog of the FLB slope theorem for isochronous bifurcations.

\begin{figure*}
\centering
	\begin{subfigure}{0.32\textwidth}
	\includegraphics[width = \textwidth]{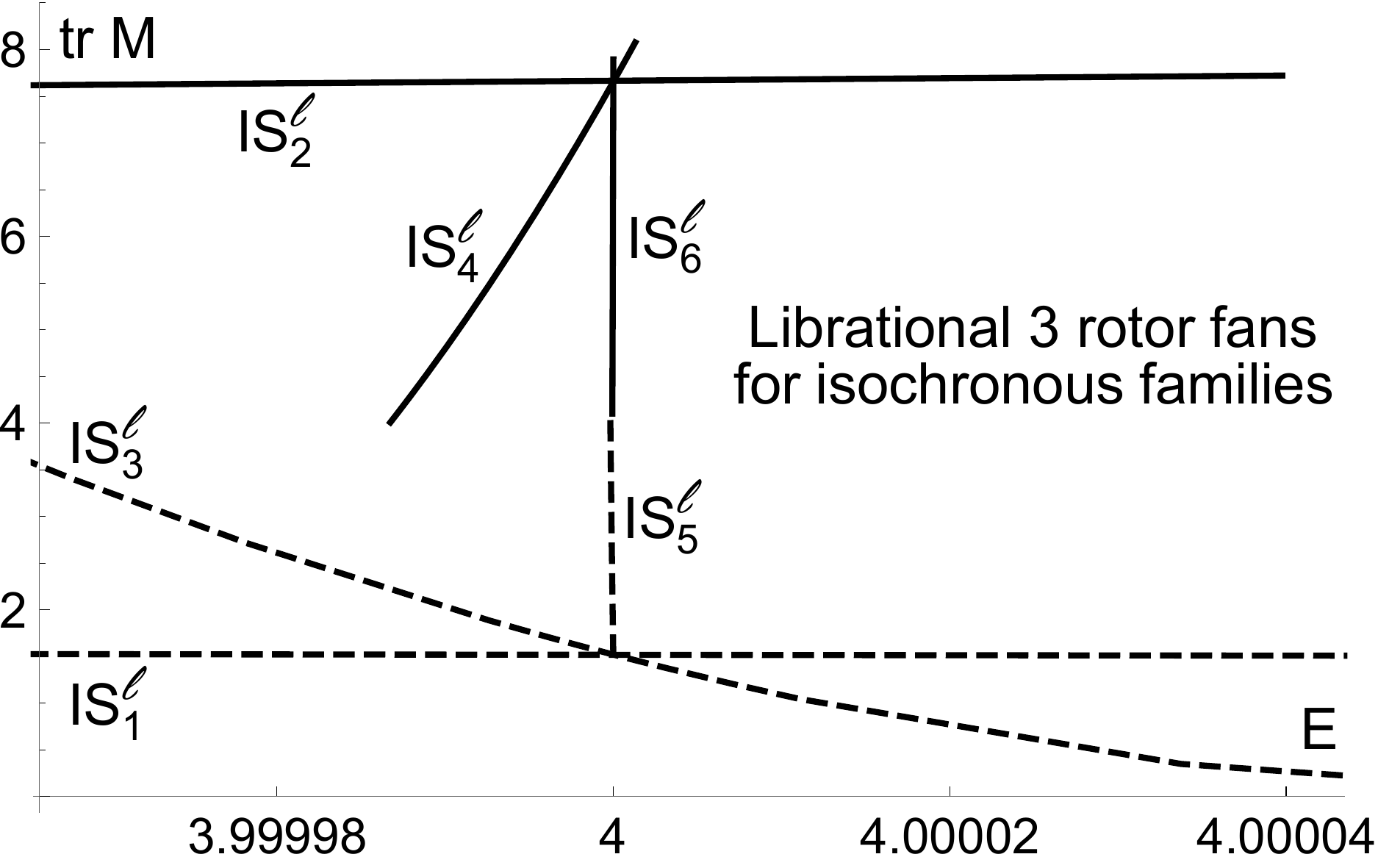}
	\caption{}
	\label{f:fan-IS-lib}
	\end{subfigure}
\hfil
	\begin{subfigure}{0.32\textwidth}
	\includegraphics[width = \textwidth]{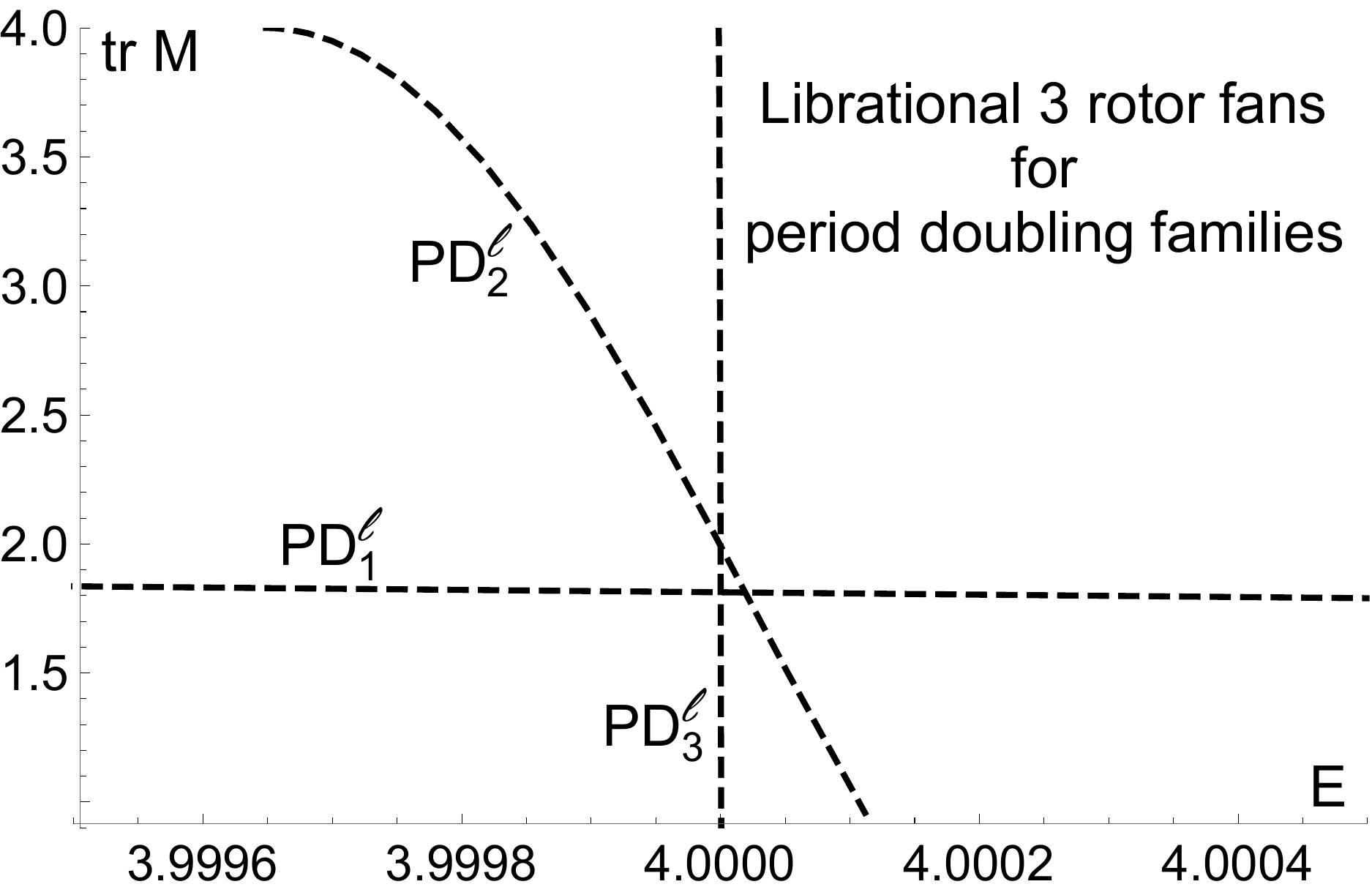}
	\caption{}
	\label{f:fan-PD-lib}
	\end{subfigure}
\hfil
	\begin{subfigure}{0.32\textwidth}
	\includegraphics[width = \textwidth]{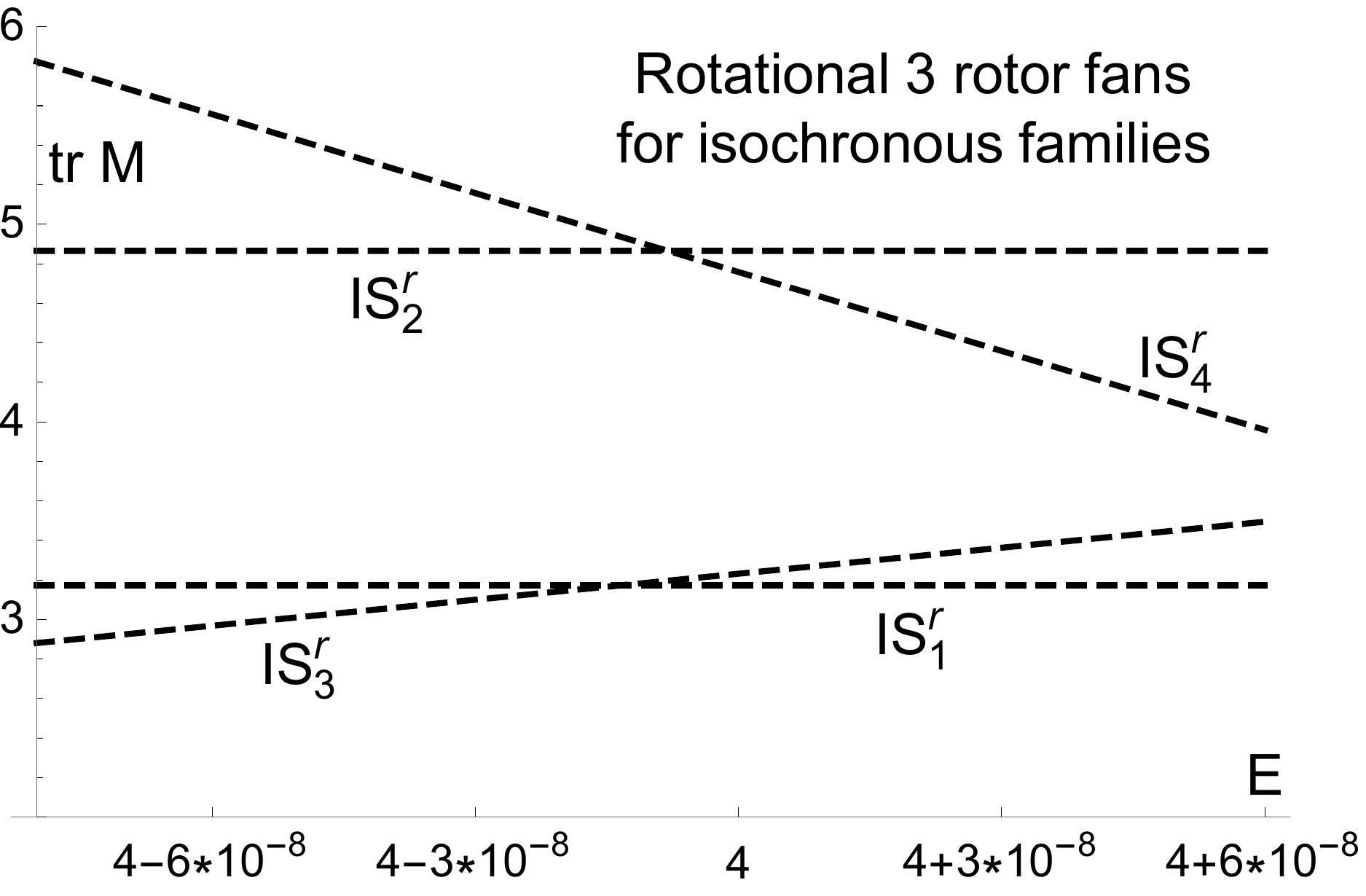}
	\caption{}
	\label{f:fan-IS-rot}
	\end{subfigure}
	\caption{\small Fans for families of newly born orbits at isochronous and period-doubling bifurcations of pendula showing intersections of $\tr M(E)$ graphs close to $E = 4g$. In (b) the slight departure from fan-like behavior may be because asymptotic behavior has not yet set in for $\PD^\ell_1$. In the isochronous rotational fans of (c), the graphs are linearly extrapolated backward in energy to find intersection points. Due to limited numerical data, only the first two members of each family are shown. The fact that they intersect close to $E = 4g$ is indicative of the formation of $\IS^r_{2n}$ and $\IS^r_{2n-1}$ fans. Note that these fan-like structures do not imply that the corresponding trajectories become coincident on the configuration/phase space at $E = 4g$. In fact, as shown in Figs.~\ref{f:pend-lib-new-traj} and \ref{f:pend-rot-new-traj}, their shapes are different.}
	\label{f:fan-IS-lib-rot}
\end{figure*}

\subsection{Three-rotor fans}

Interestingly, we find that the graphs of $\tr M(E)$ for the newly born families $\IS^\ell_{1,3,5, \ldots}$, $\IS^\ell_{2,4,6, \ldots}$, $\PD^\ell_{1,2,3,\ldots}$, $\IS^r_{1,3,5, \ldots}$ and $\IS^r_{2,4,6, \ldots}$ intersect at $E \approx 4g$ with $\tr M \approx 1.52, 7.55, 1.81, 3.17$ and $4.87$. These confluences are displayed in Fig.~\ref{f:fan-IS-lib-rot} and may be called three-rotor fans by analogy with similar phenomena in the H\'enon-Heiles system \cite{HH-fans-brack}. Note that in the rotational phase, the graphs are extrapolated `backward' in energy to look for fan-like intersections at $E \approx 4g$. This is because the energies of a family of orbits born at any rotational bifurcation always exceeds the corresponding bifurcation point energy $E_n$, which in turn is always greater than $4g$. By contrast, in the librational phase, there no need for such extrapolation as the newly born families include orbits with energies up to and beyond $4g$. In addition to the five fans mentioned above, we expect to find two more fans from the families of orbits born at the remaining two classes of bifurcations, viz. $\PD^r_{2n+2}$ and $\PD^r_{2n+1}$. As noted in \S \ref{stblty-new-orb}, the orbits born at $\PD^r_{1,2}$ are anomalous and are not expected to be part of fans.

\subsection{Scaling constants for self-similarity in newly born orbits} 

In this section, we define two more scaling constants $\al$ and $\beta$ associated with self-similarity in the $\al_1$ and $\al_2$ directions of the newly born orbits at the bifurcations of librational pendula (see Fig.~\ref{f:pend-lib-new-traj}). Subsequently, we will obtain analytical estimates for $\al$ and $\beta$ and compare them with numerical calculations from orbit shapes. The extension to the rotational regime will be touched upon at the end of this section. 

\paragraph*{Scaling constant $\al^\ell$ from the ratio of $\al_1$ amplitudes.} From Fig.~\ref{f:pend-lib-new-traj}, we begin to suspect that every successive period-doubling and every fourth isochronous bifurcation of librational pendula (e.g., $\PD^\ell_1, \PD^\ell_2, \ldots$ or $\IS^\ell_1, \IS^\ell_5, \ldots$ or $\IS^\ell_2, \IS^\ell_6, \ldots$) results in a similarly shaped new orbit with extra oscillations in $\al_1$ around $\al_2 \approx \pm \pi$. Upon appropriately zooming in near $\al_2 = \pm \pi$, the extra oscillation (say of $\PD_3^\ell$) has the same shape as the previous one ($\PD^\ell_2$). To make quantitative comparisons, we will evaluate each family of orbits in the sequence at the accumulation energy $4g$. Thus, we define the sequence of amplitude ratios
	\beq
	\al^\ell_{\PD}(n) = \frac{\al_1^{\rm max}(n;4g)}{\al_1^{\rm max}(n+1;4g)}, \quad
	\al^\ell_{\IS}(n) = \frac{\al_1^{\rm max}(n;4g)}{\al_1^{\rm max}(n+4;4g)}
	\label{al-seq}
	\eeq
for $n = 1,2,3,\ldots$. Here, $\al_1^{\rm max}(n;4g)$ refers to the amplitude in the $\al_1$ direction of the $n^{\rm th}$ family of new orbits evaluated at $E = 4g$. Taking a limit, we define the scaling constant
	\beq
	\al^\ell_{\PD, \IS} = \lim_{n \to \infty} \al^\ell_{\PD, \IS}(n).
	\label{al-def}
	\eeq

\paragraph*{Scaling constant $\beta^\ell$ from the ratio of $\al_2$ amplitudes.} The second scaling constant $\beta^\ell$ quantifies the scale invariance in the $\al_2$ direction when successive orbits are zoomed in around $\al_2 = \pi$. Thus, we focus on the small amount by which the amplitudes $\al_2^{\rm max}$ of the new orbits differ from $\pi$. Comparing these `deficits' at successive members of the sequence (e.g., $\IS^\ell_4, \IS^\ell_8, \IS^\ell_{12}, \ldots$) all evaluated at $E = 4g$ leads us to define
	\beqs
	\beta^\ell_{\IS,\PD} &=& \lim_{n \to \infty} \beta^\ell_{\IS,\PD}(n), \quad \text{where} \cr
	\beta^\ell_{\IS,\PD}(n) &=& \fr{\pi - \al_2^{\rm max}(n;4g)}{\pi - \al_2^{\rm max}(n+k;4g)}, \;\; k = 4,1 \text{ for $\IS, \PD$}. \;\quad
	\label{beta-al-2}
	\eeqs

\begin{figure}[!h]
	\begin{subfigure}{0.23\textwidth}
	\includegraphics[width = \textwidth]{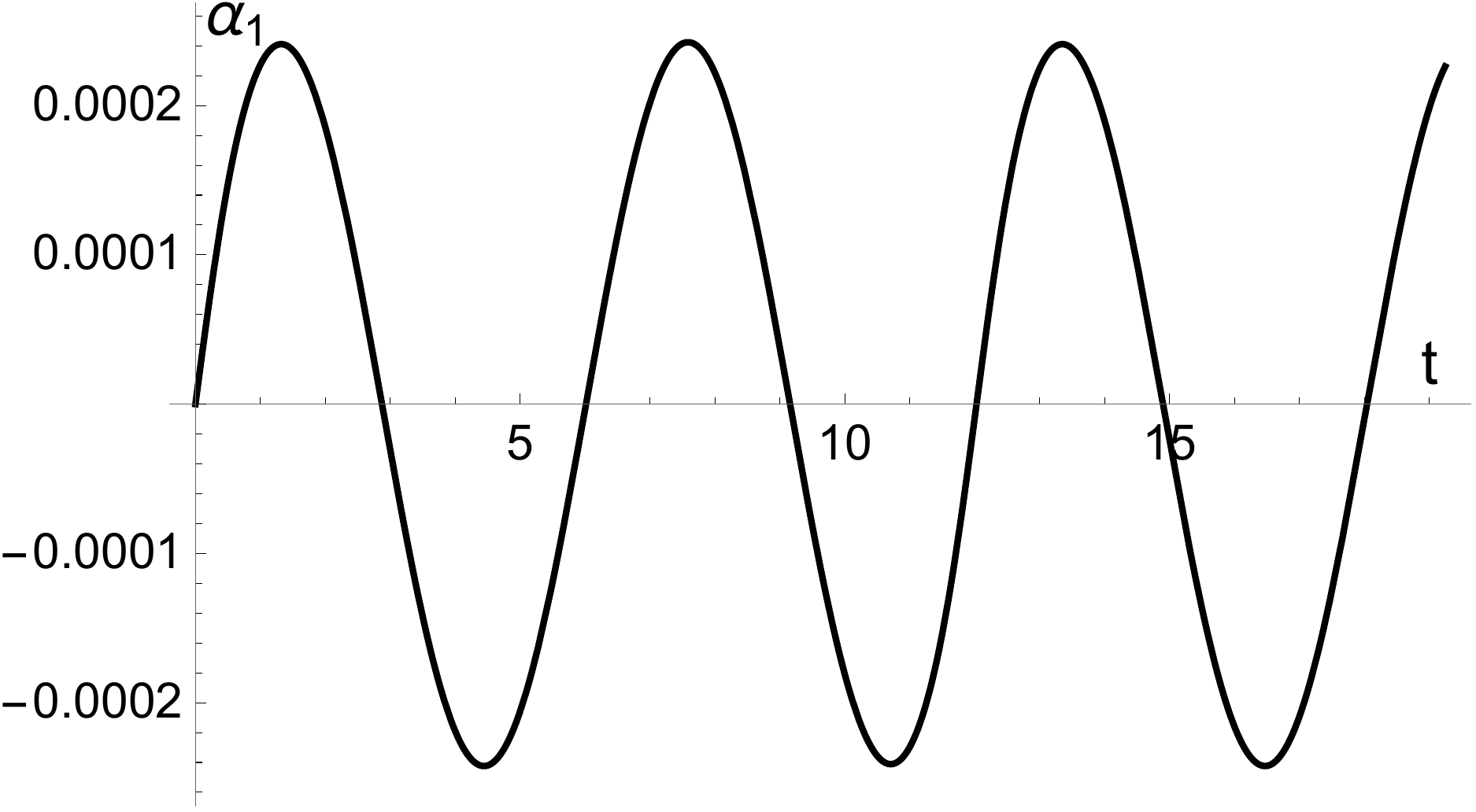}
	\caption{}
	\label{f:isl-6-al1}
	\end{subfigure}
	\begin{subfigure}{0.23\textwidth}
	\includegraphics[width = \textwidth]{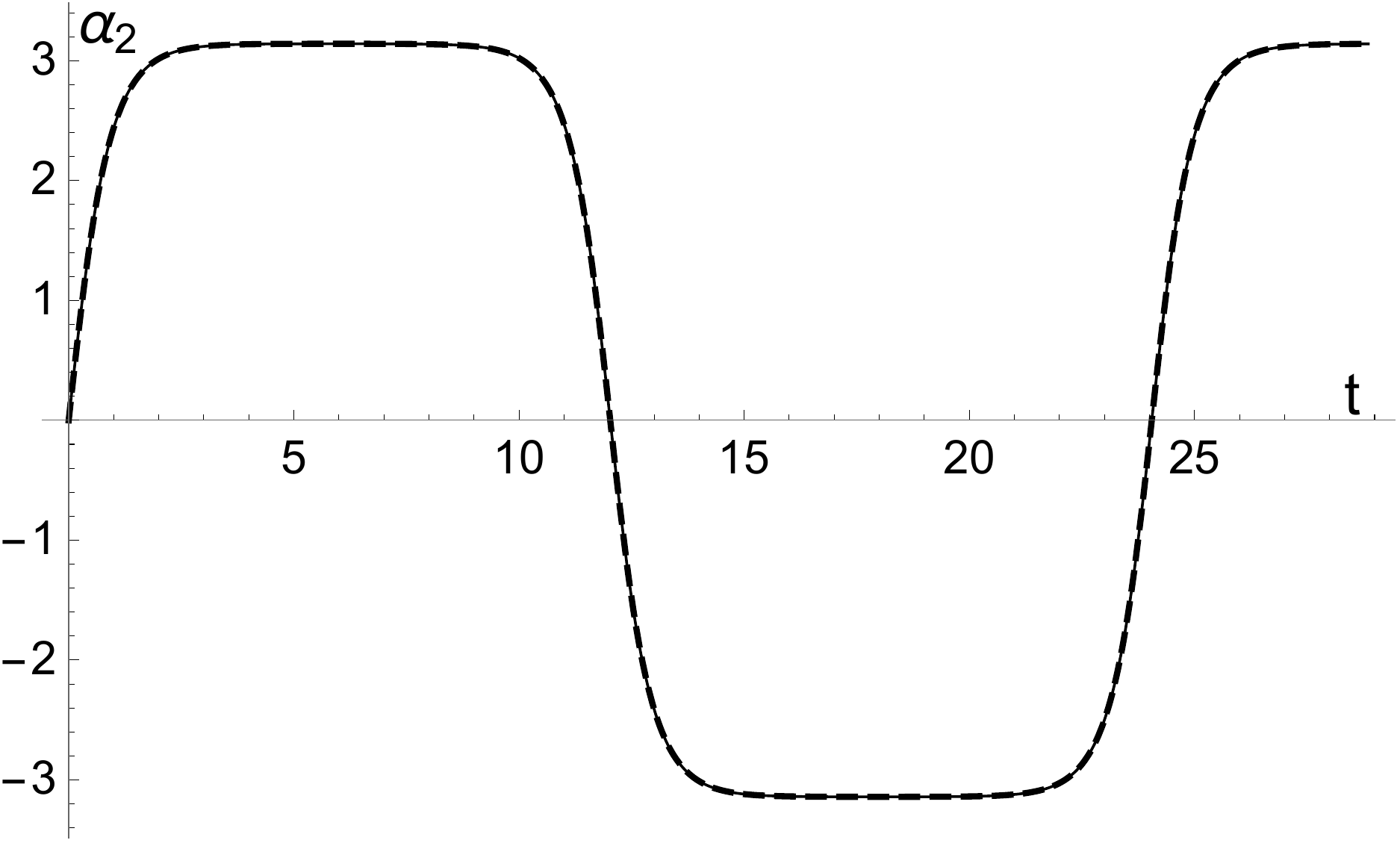}
	\caption{}
	\label{f:isl-6-al2}
	\end{subfigure}

	\begin{subfigure}{0.23\textwidth}
	\includegraphics[width = \textwidth]{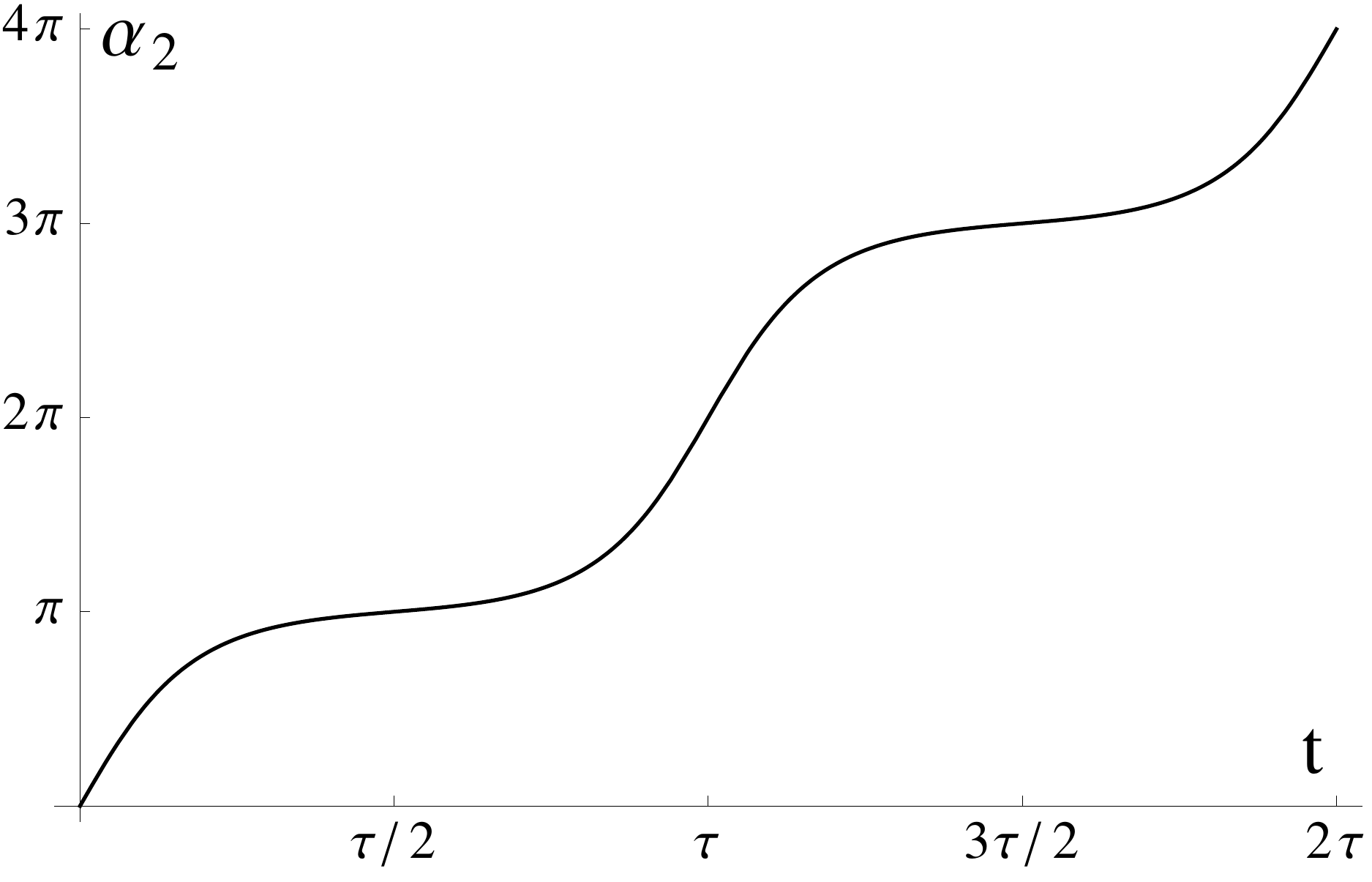}
	\caption{}
	\label{f:isr-1-al2}
	\end{subfigure}
\hfil	
	\begin{subfigure}{0.23\textwidth}
	\includegraphics[width = \textwidth]{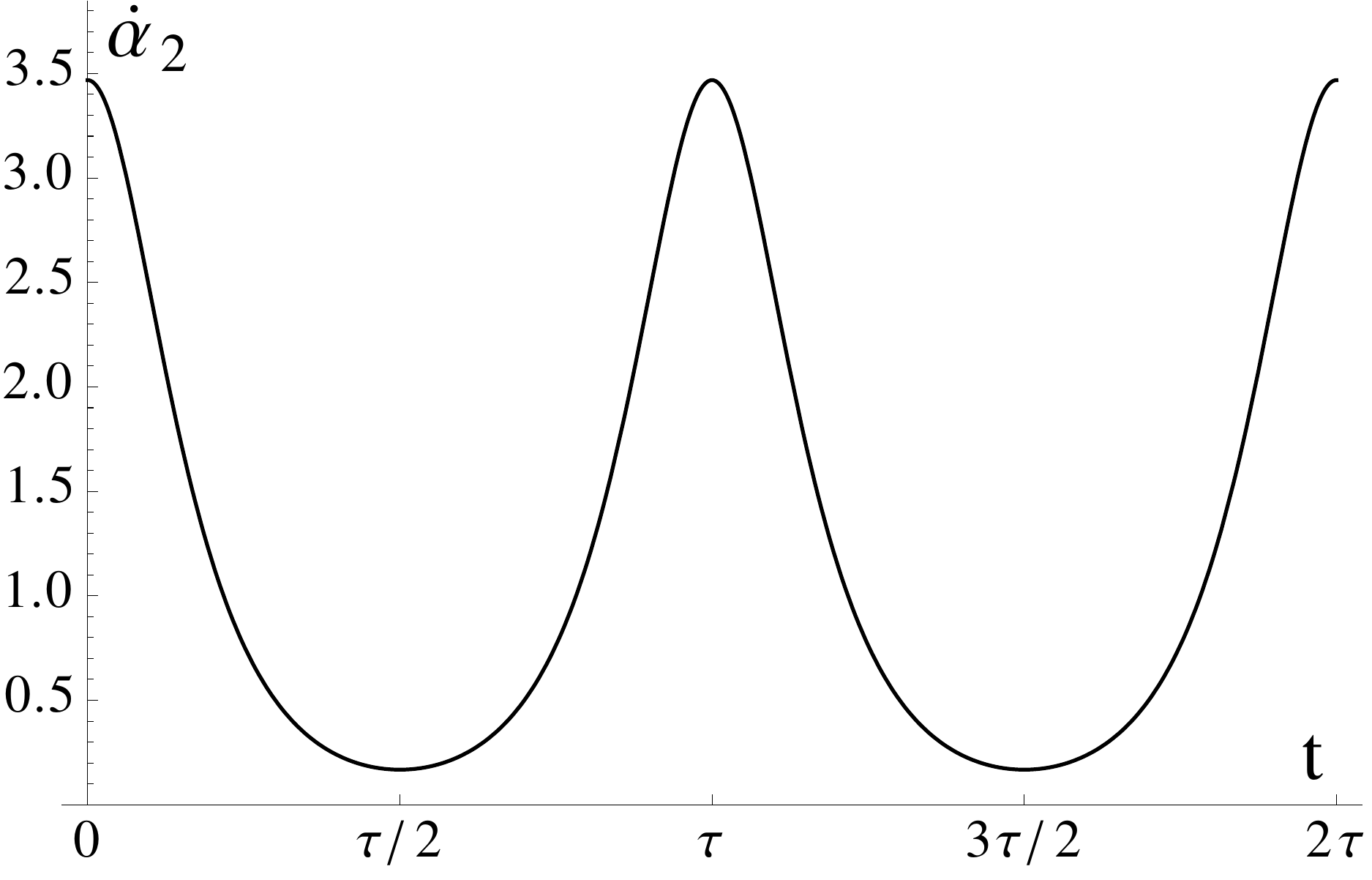}
	\caption{}
	\label{f:isr-1-al2-dot}
	\end{subfigure}
	\caption{\small (a) $\al_1(t)$ for $\IS^\ell_6$ (at $E = 4g$) is $\ll 1$ permitting it to be treated to leading order in (\ref{e:taylor-exp-newly-born-orbit-energy}). In (b), we see that the time series of $\al_2$ for $\IS^\ell_6$ at $E = 4g$ (solid line) and of $\bar \al_2$ for the pendulum at the bifurcation energy $E(\IS^\ell_6)$ (dashed line) are practically the same. This is exploited in Eqns. (\ref{e:decoupled-egy})--(\ref{al-2-max}). (c) $\al_2(t)$ for $\IS^r_1$ near the bifurcation point showing absence of turning points. (d) $\dot\al_2$ has the time period $\tau = (2\ka/\om_0) K(\ka)$ with minima occurring when $\al_2$ is an odd multiple of $\pi$.}
\end{figure}

\begin{figure*}
	\begin{subfigure}{0.4\textwidth}
	\includegraphics[width = \textwidth]{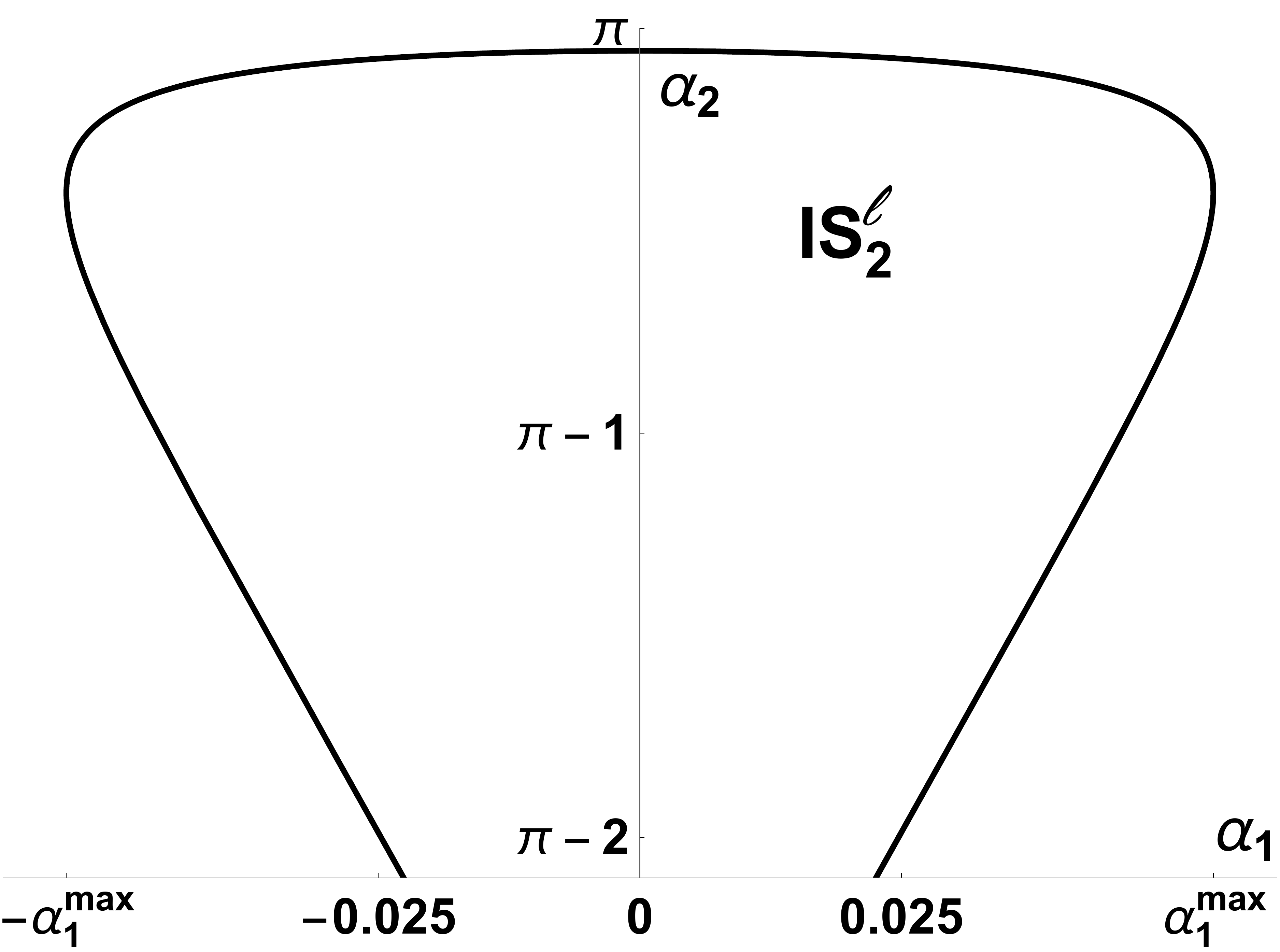}
	\caption{}
	\end{subfigure}
\hfil
	\begin{subfigure}{0.4\textwidth}
	\includegraphics[width = \textwidth]{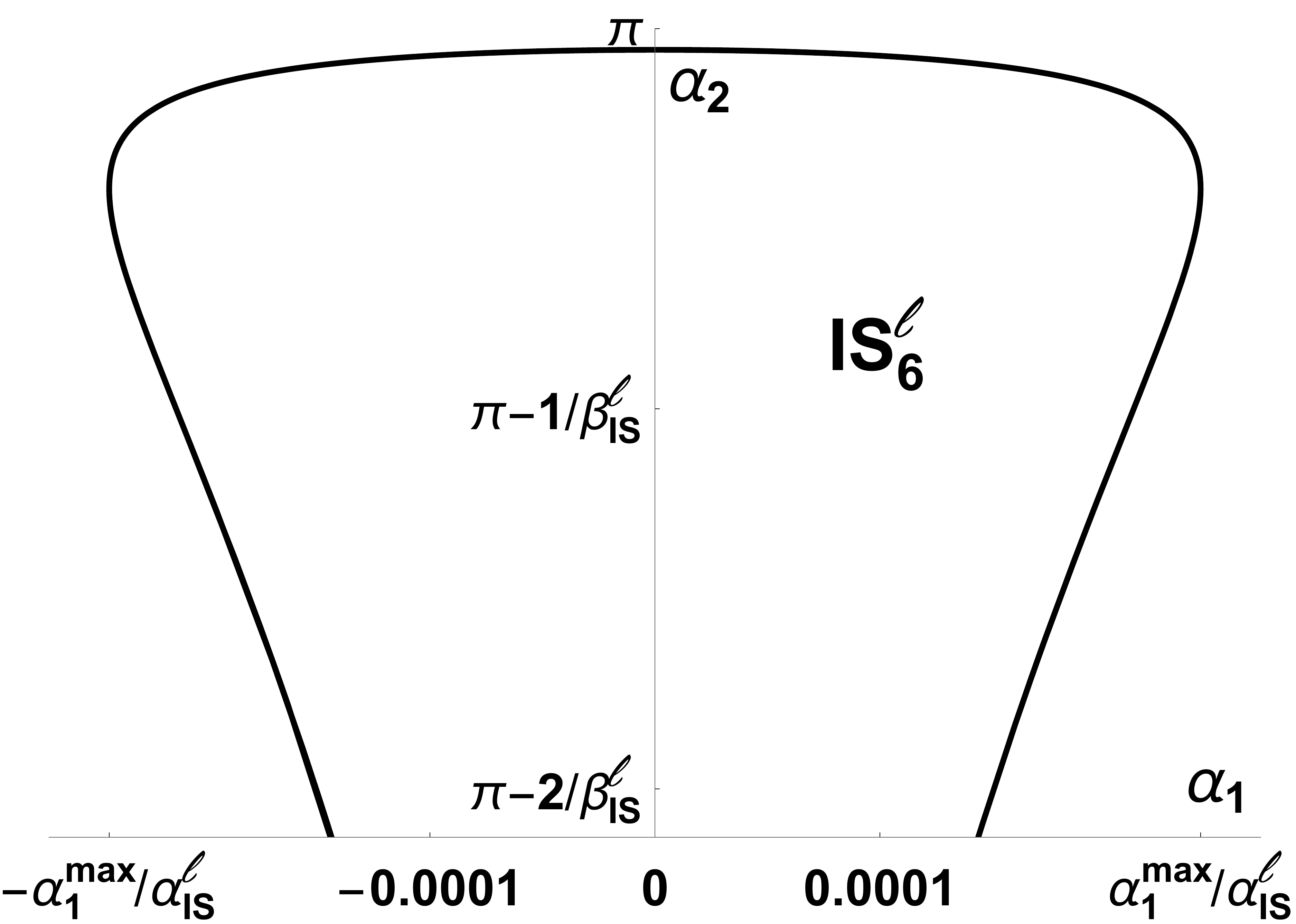}
	\caption{}
	\end{subfigure}
	
	\begin{subfigure}{0.4\textwidth}
	\includegraphics[width = \textwidth]{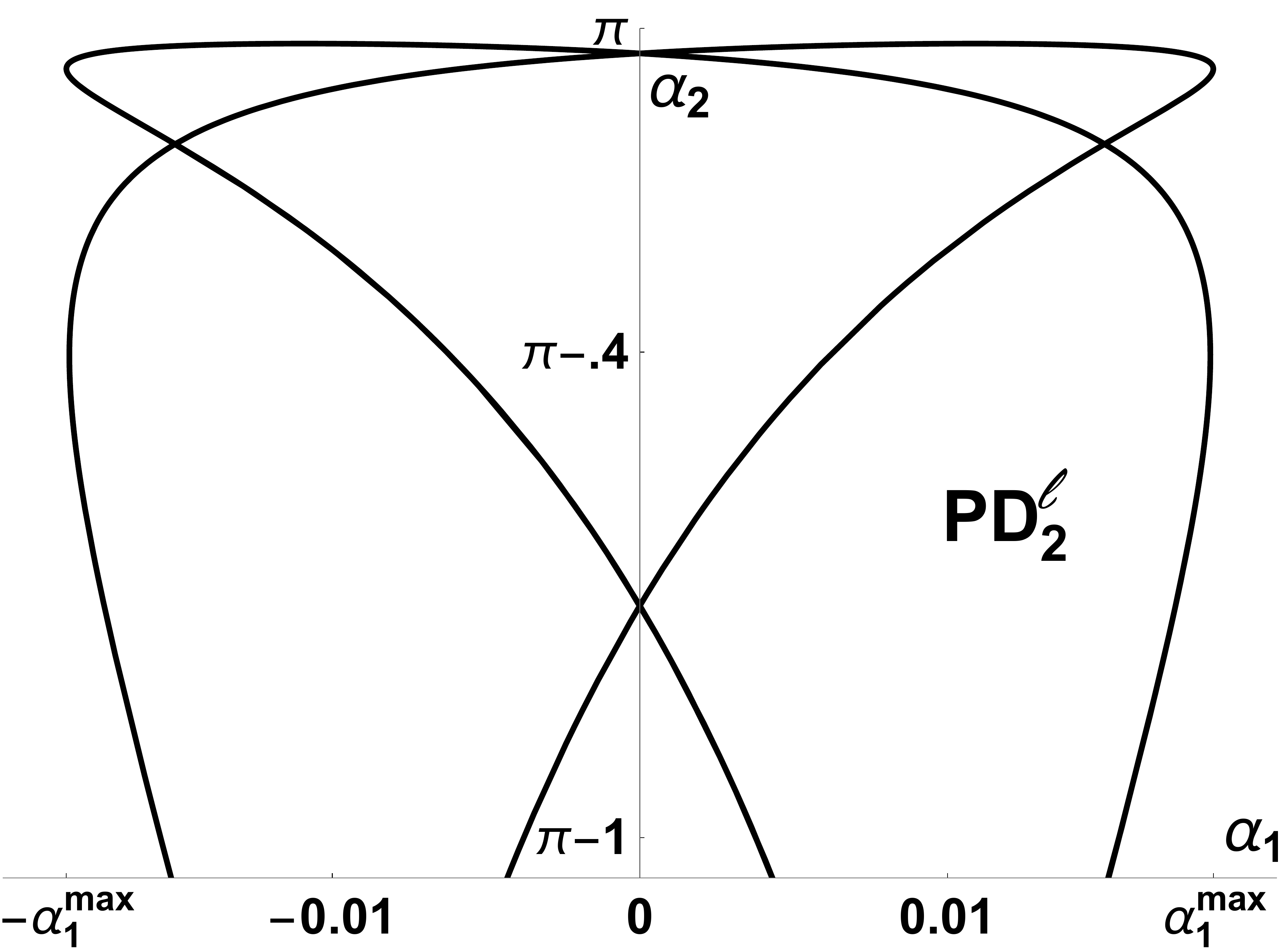}
	\caption{}
	\end{subfigure}
\hfil
	\begin{subfigure}{0.4\textwidth}
	\includegraphics[width = \textwidth]{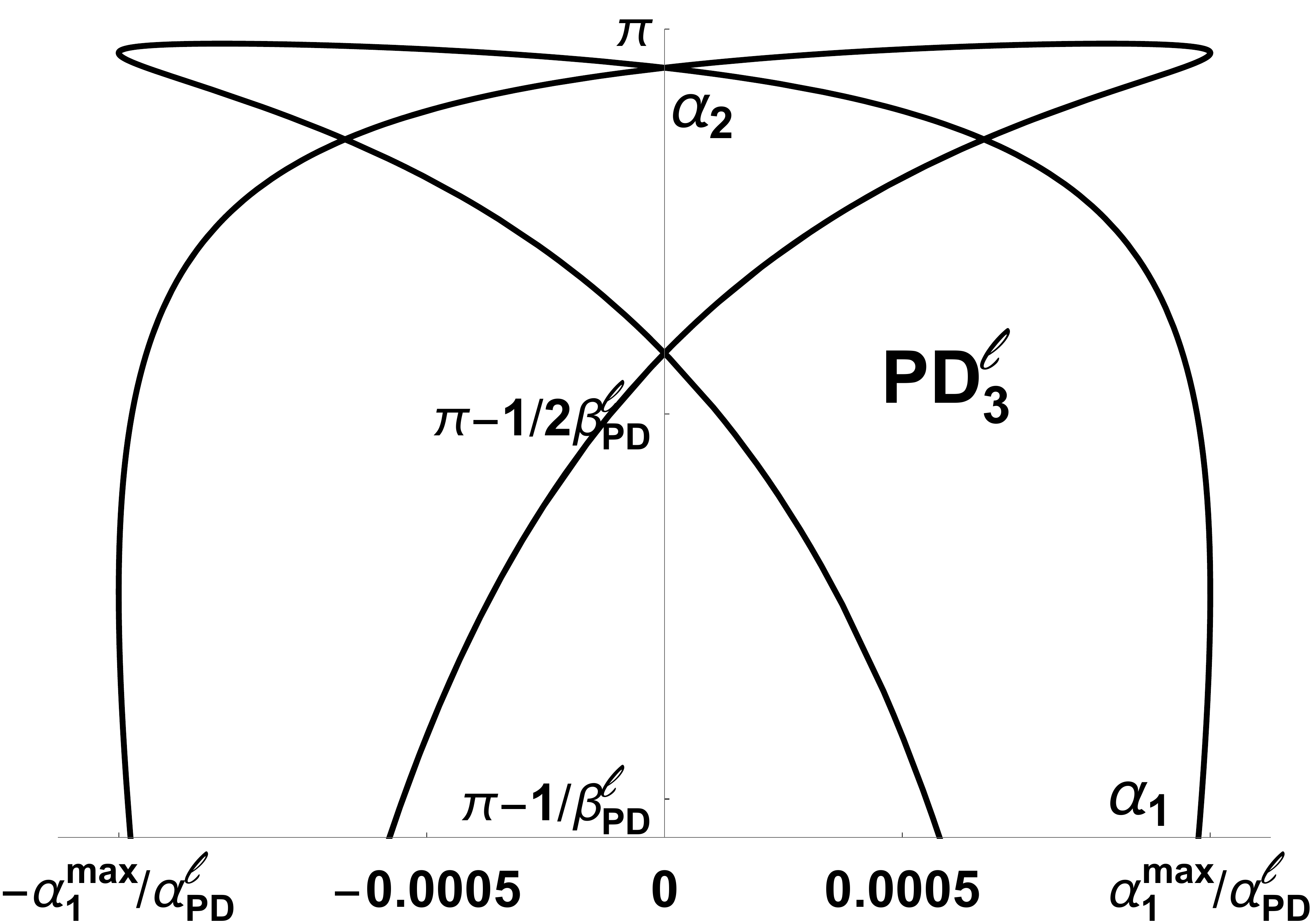}
	\caption{}
	\end{subfigure}
	\caption{\small Self-similarity between shapes (in the $\al_1 - \al_2$ plane at $E = 4g$) of new families of orbits born at bifurcations of librational pendula. The shape of $\IS^\ell_6$ shown in (b) is similar to that of $\IS^\ell_2$ shown in (a) provided the scales in the $\al_1$ and $\al_2$ directions in (b) are related to those in (a) by the scaling constants $\al^\ell_\IS$ and $\beta^\ell_\IS$. (c) and (d) show self-similarity between the period-doubling families $\PD^\ell_2$ and $\PD^\ell_3$ upon using the magnifications $\al^\ell_\PD$ and $\beta^\ell_\PD$ for the $\al_1$ and $\al_2$ axes.}
	\label{f:isl-2-6-e=4g-al1-al2-plane}
\end{figure*}

We have obtained numerical estimates for $\al$ and $\beta$ by examining the first few families of orbits in the self-similar bifurcation sequences. For instance, we find $\al^\ell_{\IS}(2) \approx 226$ and $\beta^\ell_\IS(2) \approx 230.42$ by comparing $\IS^\ell_2$ and $\IS^\ell_6$, while $\al^\ell_{\PD}(2) \approx 15.13$ and $\beta^\ell_\PD(2) \approx 15.19$ from $\PD^\ell_2$ and $\PD^\ell_3$. Pleasantly, Fig.~\ref{f:isl-2-6-e=4g-al1-al2-plane} shows the similarity in the shapes of $\IS^\ell_2$ and $\IS^\ell_6$ as well as in $\PD^\ell_2$ and $\PD^\ell_3$ upon zooming in by these factors of $\al$ and $\beta$. Next, we describe an approximation scheme that allows us to find the amplitudes in the $\al_1$ and $\al_2$ directions leading to analytical estimates for $\al$ and $\beta$.


\paragraph*{Analytical estimates for $\al_1$ and $\al_2$ amplitudes.}  Now, we will express the amplitudes of newly born periodic orbits in the $\al_1$ and $\al_2$ directions in terms of the bifurcation energies $E_n$. This will allow us to exploit the geometric nature of the bifurcation energy sequence $4g - E_n$ to estimate the scaling constants $\al$ and $\beta$.

To begin with, we observe that at the bifurcation point with energy $E_n$, the pendulum and newly born family coincide: they both have $\al_1 \equiv 0$ and $\al_2 = \bar \al_2(E_n)$ (\ref{al1=0-pend}). From there onward, as we move along the family of newly born orbits, $\al_1$ grows but its amplitude remains small even up to $E = 4g$ (e.g., as shown in Fig.~\ref{f:isl-6-al1}, for $\IS^\ell_6$, $|\al_1^{\rm max}| < 10^{-3}$ at $E = 4g$). Thus, the energy $E$ of the newly born family (for $E_n \leq E \leq 4g$) may be approximated by expanding (\ref{e:egy-V-3rot-al1-al2}) to leading order in $\del \al_1 \equiv \al_1$,
	\beq
	E \approx m r^2 (\dot \al_2^2/3 +  \del \dot \al_1^2) + g(2 (1 - \cos \al_2) + (2 + \cos \al_2) \del \al_1^2).
	\label{e:taylor-exp-newly-born-orbit-energy}
	\eeq
It is convenient to regard this energy as a sum of two contributions:
	\beqs
	E  &\approx& E_{\al_2} + E_{\del \al_1} \quad \text{where} \cr
	E_{\al_2} &=& (m r^2/3) \dot \al_2^2 + 2 g(1 - \cos\al_2) \quad \text{and}  \cr
	E_{\del \al_1} &=& m r^2 \del \dot \al_1^2 + g (2 + \cos \al_2) \del \al_1^2
	\label{egy-linear}
	\eeqs
are the energies in the $\al_2$ and $\del \al_1$ modes of the newly born orbits. We now argue that for this range of energies, $E_{\al_2}$ is nearly independent of $E$. To this end, notice that the equation for $\al_2$ (\ref{sec-ordr-eom}) is independent of $\al_1$ (treated to linear order) and reduces to that of pendula. Therefore, as long as $\al_1 = \del \al_1$ is small, $\al_2$ remains `frozen' at the bifurcation point solution $\bar \al_2(E_n)$. For example, as Fig.~\ref{f:isl-6-al2} shows, $\al_2$ for the newly born family $\IS^\ell_6$ at $E = 4g$ is nearly the same as the pendulum $\bar \al_2$ at the $\IS^\ell_6$ bifurcation point. Thus, we may approximate $E_{\al_2}$ by $E_n$. Furthermore, all the bifurcation energies $E_n$ are quite close to $4g$ (see Table \ref{t:lib-pend-trans-E-T}), where pendula spend most of their time at the `bottlenecks' near $(\al_1 = 0, \al_2 = \pm \pi)$ (see \S \ref{s:asymp-beha-time-per-egy-pend-bifur}). Therefore, we may take $\bar \al_2 \approx \pm \pi$ as $E_n \to 4g$. As a consequence, we may rewrite (\ref{egy-linear}) as
	\beqs
	 E &\approx& E_n + E_{\del \al_1} \quad \text{which implies that} \cr
	 E - E_n &\approx& m r^2 \del \dot \al_1^2 + g (2 + \cos \left(\bar \al_2 = \pm \pi )\right) \del\al_1^2 \; \;\text{or} \cr
	 E - E_n &\approx& m r^2 \del \dot \al_1^2 + g \del \al_1^2 \quad \text{as} \quad E_n \to 4g.
	 \label{e:decoupled-egy}
	\eeqs
Here, by $\del \al_1 \equiv \del \al_1(n;E)$ we mean the $\al_1$ coordinate of the trajectory of the $n^{\rm th}$ family evolved to energy $E$ (with $E_n \leq E \leq 4g$). The amplitude in $\del \al_1$ is obtained by specializing to a turning point:
	\beq
	\del \al_1^{\rm max}(n; E) \approx \sqrt{(E - E_n)/g} \quad \text{as} \quad E_n \to 4g.
	\label{al1-asymptotic}
	\eeq
This asymptotic formula is good to about a percent even for the smallest values of $n$. For instance, (\ref{al1-asymptotic}) predicts that $\del \al_1^{\rm max}(\IS^\ell_2; 4g) =  0.0555$ and $\del \al_1^{\rm max}(\IS^\ell_6; 4g) =$ 0.000 239 9, while the numerically obtained values are 0.0548 and 0.000 242 4.

Similarly, we may obtain an approximate expression (in terms of $4g - E_n$) for $\pi$ minus the amplitude in the $\al_2$ direction. To this end, we recall from (\ref{egy-linear}) that the energy of the newly born orbit was split as $E = E_{\al_2} + E_{\del \al_1}$, where $E_{\del \al_1}$ was relevant to the scaling constant $\al$. For reasons mentioned above, we equate the remaining energy $E_{\al_2}$ to the bifurcation point energy $E_n$:
	\beq
	E_{\al_2} = (m r^2/3) \dot \al_2^2 + 2 g(1 - \cos\al_2) \approx E_n.
	\label{al2-mode-egy}
	\eeq
To focus on the amplitude $\al_2^{\rm max}$, we put $\dot \al_2 = 0$ to get
	\beq
	E_n \approx E_{\al_2} = 2g(1 - \cos \al_2^{\rm max}).
	\eeq
Taylor expanding $E_{\al_2}$ around $\pi$ (an expansion around $-\pi$ leads to the same result),
	\beq
	E_n \approx 2 g(1 - \cos(\pi- \pi + \al_2^{\rm max})) \approx g(4 - (\pi - \al_2^{\rm max})^2)
	\eeq
leading to
	\beq
	\pi - \al_2^{\rm max}(n;4g) \approx \sqrt{4 - E_n/g}.
	\label{al-2-max}
	\eeq
The numerical values of $\pi - \al_2^{\rm max}$ for $\PD^\ell_2$ and $\PD^\ell_3$ are 0.018 75 and 0.001 234 45, which are very close to the values obtained by the above approximation: 0.018 76 and 0.001 234 43.


\paragraph*{Estimates for scaling constants $\al$ and $\beta$.} Finally, we use (\ref{al1-asymptotic}) and (\ref{al-2-max}) along with the definitions (\ref{al-def}) and (\ref{beta-al-2}) to estimate $\al$ and $\beta$. As for $\al$, we may take $\al_1 = \del \al_1$ in (\ref{al-def}) and use (\ref{al1-asymptotic}) and (\ref{egy-asymp}) to get
	\beqs
	\al^\ell_{\PD} &=& \lim_{n \to \infty} \sqrt{\fr{4 - E_n/g}{4 - E_{n+1}/g}} = e^{\om_0 \D \tau/4} = e^{\sqrt 3 \pi/2} = \sqrt{\del^\ell} \approx 15.19 \cr
	\al^\ell_{\IS} &=& \lim_{n \to \infty} \sqrt{\fr{4 - E_n/g}{4 - E_{n+4}/g}} = e^{\om_0 \D \tau/2} = e^{\sqrt 3 \pi} = \del^\ell \approx 230.76. \;\;\;\,\quad
	\label{e:al-IS-PD-analytical}
	\eeqs
In fact, the numerically obtained ratio of amplitudes $\al^\ell_\IS(2)$ is $0.0548/0.0002424 \approx 226$ and $\al^\ell_\PD(2)$ is $0.0186/0.00123 \approx 15.13$, which are reasonably close to our analytical estimates for $\al^\ell_\IS$ and $\al^\ell_\PD$. 

On the other hand, to estimate $\beta$ using the definition (\ref{beta-al-2}), we substitute for $\pi - \al_2^{\rm max}$ from (\ref{al-2-max}) and for $E_n$ from (\ref{egy-asymp}) to obtain
	\beqs
	\beta^\ell_{\PD} &=& \lim_{n \to \infty} \sqrt{\fr{4 - E_n/g}{4 - E_{n+1}/g}} = e^{\om_0 \D \tau/4} = e^{\sqrt 3 \pi/2} = \sqrt{\del^\ell} \approx 15.19, \cr
	\beta^\ell_{\IS} &=& \lim_{n \to \infty} \sqrt{\fr{4 - E_n/g}{4 - E_{n+4}/g}} = e^{\om_0 \D \tau/2} = e^{\sqrt 3 \pi} = \del^\ell \approx 230.76. \;\;\;\,\quad
	\eeqs
This is quite close to the numerical values for the ratios in (\ref{beta-al-2}), which for $\beta^\ell_\PD(2)$ is $\approx 15.19$ and for $\beta^\ell_\IS(2)$ is $\approx 230.42$.

These scaling constants are used to demonstrate the self-similarity in shapes of newly born families of orbits in Fig.~\ref{f:isl-2-6-e=4g-al1-al2-plane}. It is noteworthy that the scaling exponents $\al, \beta$ and $\del$ satisfy certain relations: $\al^\ell_\PD \times \beta^\ell_\PD = \del^\ell$ and $\al^\ell_\IS \times \beta^\ell_\IS = (\del^\ell)^2$. Such relations remind us of Widom scaling in critical phenomena and have been observed in H\'enon-Heiles as well \cite{brck-omega}.


\paragraph*{Limitations in scaling constant definitions.} Notice that in defining $\al$ (\ref{al-def}) and $\beta$ (\ref{beta-al-2}), we evolved the newly born families up to $E = 4g$ where the amplitudes in $\al_{1,2}$ were evaluated. Unfortunately, this cannot be done in the rotational regime as all the orbits born at rotational bifurcations only exist at energies greater than $4g$. Furthermore, in defining $\beta$ we used a turning point of the new periodic orbit in the $\al_2$ direction. Such a turning point does not exist for new orbits in the rotational regime. Thus our definitions of scaling constants do not directly extend to the rotational regime. To circumvent this, we now offer an alternate definition of $\al$ that may be generalized to the rotational phase and also propose a new definition for $\beta$ in the rotational phase.

As for $\al$, instead of extending the new families up to the accumulation energy $E = 4g$ we will extend them up to the next bifurcation energy at which a similar family is born (e.g., $\IS^\ell_1$ to $\IS^\ell_5$). Thus we are led to define new sequences of ratios that replace the ones appearing in (\ref{al-seq}) leading to new scaling constants
	\beq
	\tl \al_\IS^\ell \equiv \lim_{n \to \infty} \frac{\al_1^{\rm max}(n;E_{n+4})}{\al_1^{\rm max}(n+4;E_{n+8})}, \quad \tl \al_\PD^\ell \equiv \lim_{n \to \infty} \frac{\al_1^{\rm max}(n;E_{n+1})}{\al_1^{\rm max}(n+1;E_{n+2})}.
	\label{e:alpha-tilde-IS-PD}
	\eeq
Upon analytically estimating their values as in (\ref{e:al-IS-PD-analytical}), we find that they reduce to our estimates for $\al_{\IS,\PD}$. The advantage of $\tl \al$ is that they may be extended to the rotational regime.

\paragraph*{Scaling constants in the rotational phase.} In the rotational regime, there are four sequences of newly born families: $\IS^r_{2n-1}, \IS^r_{2n}, \PD^r_{2n-1}$ and $\PD^r_{2n}$ whose shapes (see Fig.~\ref{f:pend-rot-new-traj}) display a self-similar pattern analogous to the librational ones discussed above Eqn. (\ref{al-seq}). For example, $\IS^r_3$ is similar to $\IS^r_1$ and $\PD^r_4$ to $\PD^r_2$ with the former in each pair having extra oscillations near $\al_2 = \pm \pi$. Here, we define scaling constants associated with this scale-invariance in the $\al_1$ and $\al_2$ directions:
	\beq
	\al^r \equiv \lim_{n \to \infty} \fr{\al_1^{\rm max}(n;E_{n-2})}{\al_1^{\rm max}(n+2;E_{n})}
	\;\; \& \;\;
	\beta^r \equiv \lim_{n \to \infty} \fr{\dot \al_2^{\rm min}(n;E_{n-2})}{\dot \al_2^{\rm min}(n+2;E_{n})}.
	\label{e:alpha-beta-r}
	\eeq
The definition of $\al^r$ is motivated by that of $\tl \al$ (\ref{e:alpha-tilde-IS-PD}). The constant $\beta^r$ is the limiting magnification factor to relate the extra oscillations in $\IS^r_{n+2}$ to $\IS^r_{n}$. In the librational case, $\al_2$ had turning points near $\al_2 = \pm \pi$. In the rotational case, $\al_2$ does not have any turning points but $\dot \al_2$ does have turning points. We focus on the turning points at $\al_2 = \pm \pi$ where $\dot \al_2$ is minimal in magnitude (see Fig.~\ref{f:isr-1-al2-dot}).

Although we have extended the definitions of $\al$ and $\beta$ to the rotational phase, we have not been able to numerically evaluate these scaling constants. This is due to the difficulty in accurately extending, say, the newly born family $\IS^r_3$ to the energy of $\IS^r_1$ as these periodic orbits are very unstable [$\tr M$ is a very rapidly increasing/decreasing function of energy (see Figs.~\ref{f:fgb-rot-self-simi-trM-vs-E-1}, \ref{f:fgb-rot-self-simi-trM-vs-E-2})]. Despite this numerical difficulty, we may estimate $\al^r$ and $\beta^r$ by the methods adopted around Eqn. (\ref{e:taylor-exp-newly-born-orbit-energy}). For the former, we take $\al_1 \approx \del \al_1$ in (\ref{e:alpha-beta-r}) and use (\ref{al1-asymptotic}) and (\ref{egy-asymp-rot}) to obtain
	\beq
	\al^r \approx \lim_{n \to \infty} \sqrt{\frac{E_{n-2}-E_n}{E_n - E_{n+2}}} = e^{\om_0 \D \tau/2} = e^{\sqrt 3 \pi} \approx \sqrt{\del^r}.
	\eeq
To estimate $\beta^r$, we put $\al_2 = \pm \pi$ in (\ref{al2-mode-egy}) to get
	\beq
	 \dot \al_2^{\rm min}(n) \approx \sqrt{3(E_n - 4g)/m r^2}.
	\eeq
Using (\ref{egy-asymp-rot}) for the bifurcation energies, we find
	\beq
	\beta^r \approx \lim_{n \to \infty} \sqrt{\frac{E_n - 4g}{E_{n+2} - 4g}} = e^{\om_0 \D \tau/2} = e^{\sqrt 3 \pi} \approx \sqrt{\del^r}.
	\eeq
Unlike in the librational phase, all four rotational  bifurcation sequences share the same scaling constants, which satisfy the common relation $\al^r \times \beta^r = \del^r$. Moreover, we notice that our estimates for $\al^r$ and $\al^\ell_\IS$ are the same as are those for $\beta^r$ and $\beta^\ell_\IS$ while the scaling constants for the librational period-doubling families apparently have no rotational counterpart. This observation is perhaps not surprising in the light of the upcoming duality between librational isochronous and both isochronous and period-doubling rotational bifurcations.


\subsection{Duality between librational and rotational bifurcations}
\label{s:duality-lib-rot}

We discover an asymptotic (as $E \to 4g$) relation between the energies of librational isochronous bifurcations and rotational bifurcations (both isochronous and period-doubling) of pendula:
	\beqs
	4g - E(\IS^\ell_1) &\approx& E(\IS^r_1) - 4g \approx g e^{-4.7}, \cr
	4g - E(\IS^\ell_2) &\approx& E(\IS^r_2) - 4g \approx g e^{-5.78}, \cr
	4g - E(\IS^\ell_3) &\approx& E(\PD^r_3) - 4g \approx g e^{-10.118}, \cr
	4g - E(\IS^\ell_4) &\approx& E(\PD^r_4) - 4g \approx g e^{-11.229}, \cr
	4g - E(\IS^\ell_5) &\approx& E(\IS^r_3) - 4g \approx g e^{-15.5598}, \cr
	4g - E(\IS^\ell_6) &\approx& E(\IS^r_4) - 4g \approx g e^{-16.6701} \cr
	4g - E(\IS^\ell_7) &\approx& E(\PD^r_5) - 4g \approx g e^{-21.0012}, \cr
	4g - E(\IS^\ell_8) &\approx& E(\PD^r_6) - 4g \approx g e^{-22.1116} \cr
	4g - E(\IS^\ell_9) &\approx& E(\IS^r_5) - 4g \approx g e^{-26.4426}, \cr
	4g - E(\IS^\ell_{10}) &\approx& E(\IS^r_6) - 4g \approx g e^{-27.5529} \ldots. \qquad
	\eeqs
Evidently, the relation is increasingly accurately satisfied as $E \to 4g$. Thus, we propose the following dualities between bifurcations
	\beq
	\IS^\ell_1 \leftrightarrow \IS^r_1, \quad
	\IS^\ell_2 \leftrightarrow \IS^r_2, \quad
	\IS^\ell_3 \leftrightarrow \PD^r_3, \quad
	\IS^\ell_4 \leftrightarrow \PD^r_4, \ldots.
	\eeq
More generally for $j = 0,1,2,\ldots$ we have the dualities
	\beqs
	\IS^\ell_{4j+1} \leftrightarrow \IS^r_{2j+1}, \quad
	\IS^\ell_{4j+2} \leftrightarrow \IS^r_{2j+2}, \cr
	\IS^\ell_{4j+3} \leftrightarrow \PD^r_{2j+3} \quad
	\text{and} \quad
	\IS^\ell_{4j+4} \leftrightarrow \PD^r_{2j+4}.
	\eeqs
Consequently, the elliptic moduli $k_\ell$ at a librational $\IS$ bifurcation is related to $k_r$ at the dual rotational bifurcation via the formula
	\beq
	k_{\rm \ell}^2 + k_{r}^2 \approx 2.
	\eeq
Thus the elliptic moduli at dual bifurcations lie on a circle centered at the origin of the $k_{\ell}-k_{r}$ plane with radius $\sqrt{2}$. They accumulate at $(k_\ell =1, k_r = 1)$.

We also find that the periodic transverse Lam\'e functions which govern the shapes of the newly born periodic orbits near dual bifurcation points are related (see Tables \ref{t:lib-al1-lame-fns-IS-PD} and \ref{t:rot-al1-lame-fns-IS-PD}). For example, for the duality $\IS^\ell_1 \leftrightarrow \IS^r_1$,
	\beq
	 \al_1^\ell(t) = \text{Ec}^2_n( \om_0 t; k_\ell) \quad
	\text{while} \quad \al_1^r(t) = \text{Ec}^2_n( \om_0 k_r t; 1/k_r)
	\eeq
where $n(n+1) = 2/3$.

Notably, the first two bifurcations of rotational pendula $\PD^r_{1,2}$, which display atypical features as discussed in \S \ref{s:features-newly-born-pend}, are not part of this duality. Moreover, the above duality does not extend to the function $\tr M(E)$ (for pendula) as a whole. For instance, $\tr M$ has a double zero at each PD bifurcation in the librational regime but a simple zero at the PD bifurcations in the rotational regime. Furthermore, we have not found rotational duals to the librational period-doubling bifurcations.

\section{Period-doubling bifurcation in the rotational isosceles breather family}
\label{s:breather}

\begin{figure}
\centering
	\begin{subfigure}{0.22\textwidth}
	\includegraphics[width = \textwidth,height=75pt]{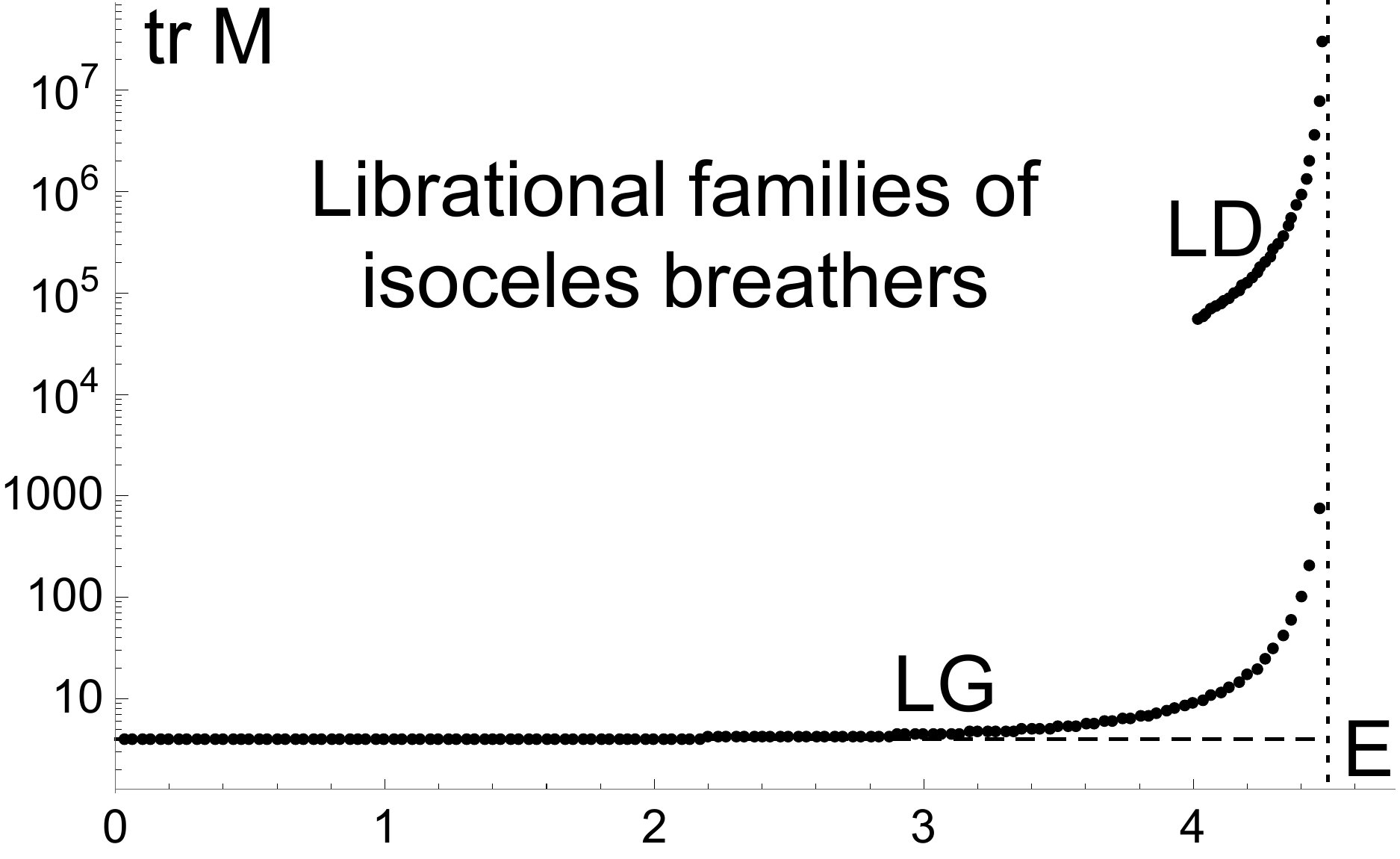}
	\caption{}
	\label{f:LG-LD-breather-trMvsE}
	\end{subfigure}
\hfil
	\begin{subfigure}{0.22\textwidth}
	\includegraphics[width = \textwidth,height=75pt]{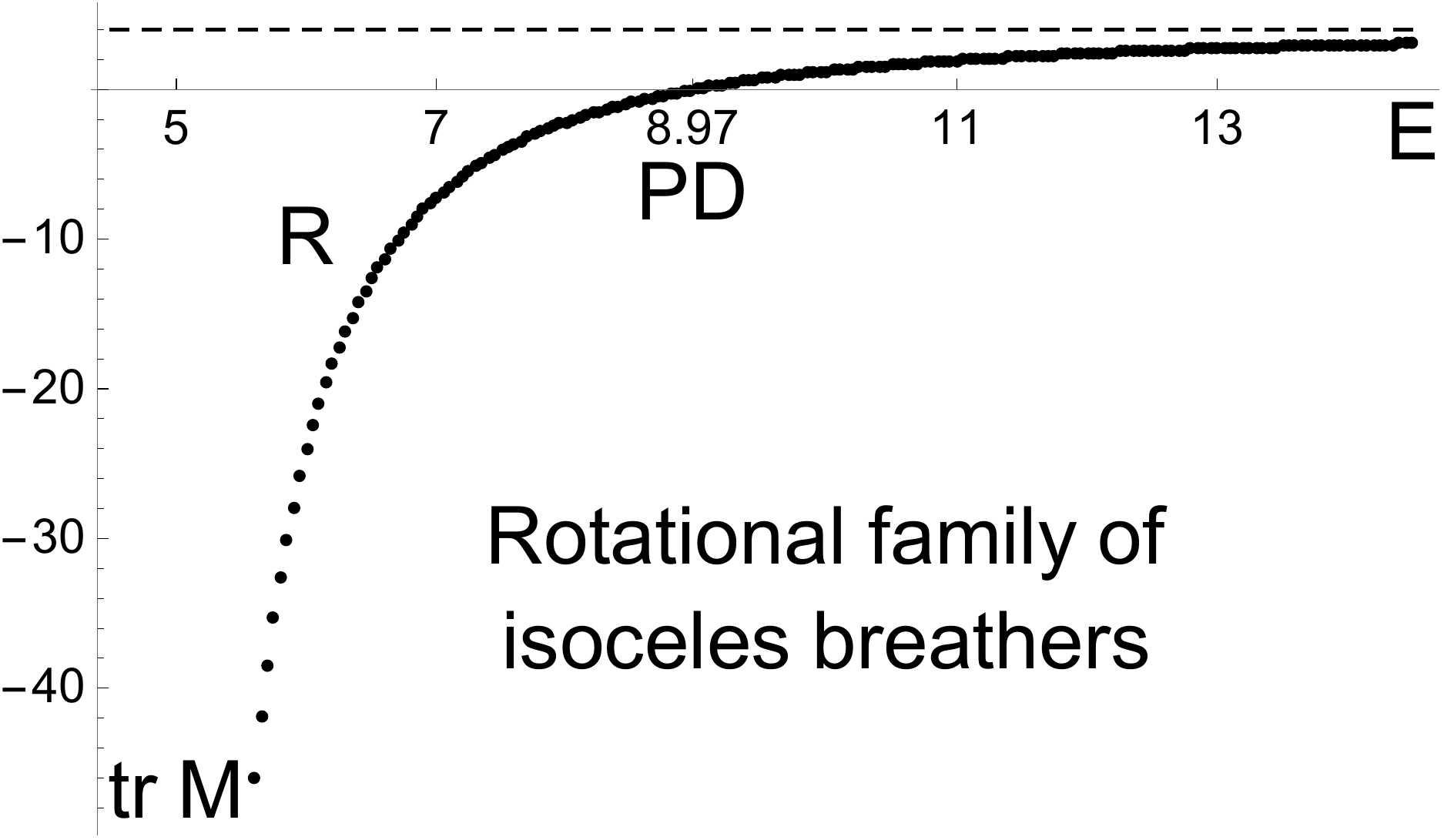}
	\caption{}
	\label{f:R-breather-trMvsE}
	\end{subfigure}
\hfil
	\begin{subfigure}{0.22\textwidth}
	\includegraphics[width = \textwidth,height=75pt]{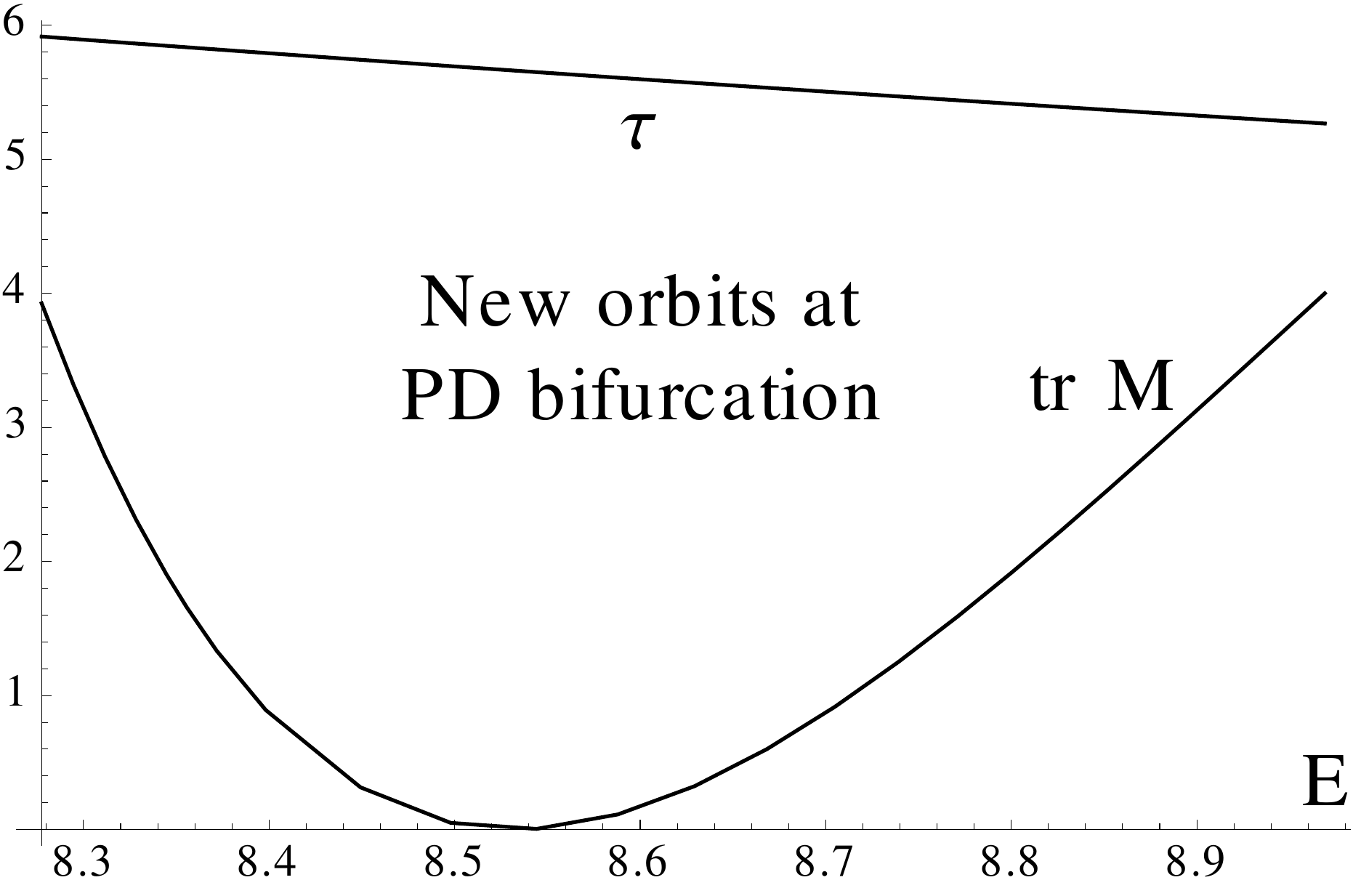}
	\caption{}
	\label{f:breather-pd-trMvE}
	\end{subfigure}
\hfil
	\begin{subfigure}{0.22\textwidth}
	\includegraphics[width = \textwidth,height=75pt]{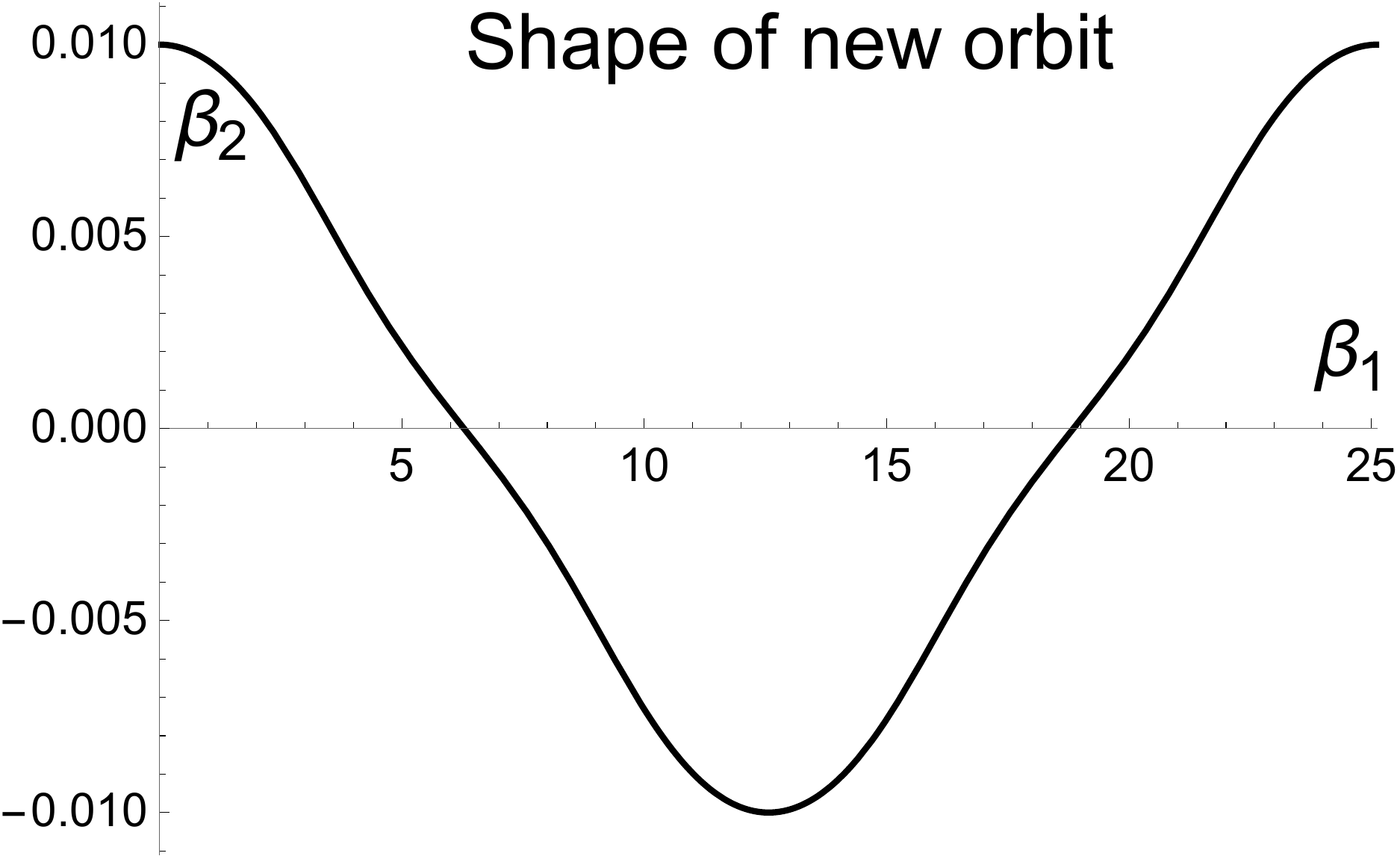}
	\caption{}
	\label{f:breather-new-born-orbit}
	\end{subfigure}
\caption{\small (a) Librational breathers ($\rm LG$ and $\rm LD$) are always unstable: $\tr M(E)$ increases from $4$ to $\infty$ as $E$ goes from $0$ to $4.5g$. Rotational breathers (b) are unstable for $4.5g < E < 8.97g$ and become stable thereafter with $\tr M \to 4^-$ as $E \to \infty$. (c) Time period $\tau(E)$ and $\tr M(E)$ for periodic orbits born at the period-doubling bifurcation at $E \approx 8.97g$. Unlike with $\PD$ bifurcations of pendula, the energy of the newly born orbits decreases as we move away from the bifurcation point, signaling a backward fork-like bifurcation. (d) Shape of newly born trajectory on the $\beta_1$-$\beta_2$ plane. By contrast, breathers are represented by horizontal segments ($\beta_2 = 0$).}
\label{f:breather-pd-monod-vs-E}
\end{figure}

There are three isosceles breather families of periodic solutions. Here we restrict attention to the one with $\vphi_1 = \vphi_2 \equiv \vphi$. The EOM in terms of $\tl t = t \sqrt{g/m r^2} $ (\ref{e:dim-less-t-mom}) reduces to Eqn. (46) of Ref.\cite{gskhs-3rotor}:
	\beq
	\fr{d^2\vphi}{d \tl t^2} = -( \sin \vphi + \sin 2 \vphi).
	\label{e:breather-eom-phi-2nd-order}
	\eeq
The dimensionless conjugate momenta
	\beq
	\tl p_{1} = \ov{3} \left( 2 \DD{\vphi_1}{\tl t} + \DD{\vphi_2}{\tl t} \right), \quad \tl p_{2} = \ov{3} \left( \DD{\vphi_1}{\tl t} + 2 \DD{\vphi_2}{\tl t} 		\right)
	\eeq
for this breather family are equal and we denote them by $\tl p_1 = \tl p_2 \equiv \tl p$. For small perturbations
	\beq
	\vphi_{1,2} = \vphi + \del \vphi_{1,2}, \quad \tl p_{1,2} = \tl p + \del \tl p_{1,2},
	\eeq
the second order perturbation equations are
	\beq
	\fr{d^2}{d\tl t^2} \begin{smmat} \del \vphi_1 \\ \del \vphi_2 \end{smmat} = - \begin{smmat} 2 \cos \vphi + \cos 2 \vphi & -\cos \vphi + \cos 2 \vphi  \\ -\cos \vphi + \cos 2 \vphi  & 2 \cos \vphi + \cos 2 \vphi  \end{smmat} \begin{smmat} \del \vphi_1 \\ \del \vphi_2 \end{smmat},
	\label{e:breather-pert-2nd-order}
	\eeq
while the first order perturbation equations are
	\beq
	\DD{}{\tl t} \begin{smmat} \del \vphi_1 \\ \del \tl p_1\\ \del \vphi_2 \\ \del \tl p_2  \end{smmat} 
	= - \begin{smmat} 0 &  -2 & 0 & 1 \\ 2 \cos \vphi + \cos 2 \vphi  & 0 & - \cos \vphi + \cos 2 \vphi  & 0 \\ 	0 & 1 & 0 & -2 \\ - \cos \vphi + \cos 2 \vphi  & 0 & 2 \cos \vphi + \cos 2 \vphi  & 0 \end{smmat} 
	\begin{smmat} \del \vphi_1 \\ \del \tl p_1\\ \del \vphi_2 \\ \del \tl p_2  \end{smmat}.
	\label{e:breather-pert}
	\eeq
Equations (\ref{e:breather-pert-2nd-order}) may be decoupled
	\beq
	\fr{d^2}{d\tl t^2} \begin{pmatrix} \del \beta_1 \\ \del \beta_2 \end{pmatrix} = - \begin{pmatrix}
	 \cos \vphi + 2 \cos 2 \vphi & 0 \\ 0 & 3 \cos \vphi \end{pmatrix} \begin{pmatrix} 
	 \del \beta_1 \\ \del \beta_2 \end{pmatrix}
	\eeq
by defining the canonically conjugate variables $\beta_{1,2} = \vphi_1 \pm \vphi_2$ and $\pi_{1,2} = (1/2) ( \tl p_1 \pm \tl p_2)$ with $\{ \beta_i, \pi_j \} = \del_{ij}$. The Hamiltonian in these variables is
	\beq
	H = \pi_1^2 + 3 \pi_2^2 + 3 - \cos \left( \fr{\beta_1 + \beta_2}{2}\right) - \cos\left( \fr{\beta_1 - \beta_2}{2}\right) - \cos \beta_1.
	\eeq
The equations of motion are
	\beqs
	\dot \beta_1 = && 2 \pi_1, \quad 
	\dot \beta_2 = 6 \pi_2, \cr
	\dot \pi_1 =  && - \half ( \sin ( (\beta_1 + \beta_2)/2) + \sin ( (\beta_1 - \beta_2)/2) + 2 \sin \beta_1 )
	\cr
	\dot \pi_2 =  && - \half ( \sin ( (\beta_1 + \beta_2)/2) - \sin ( (\beta_1 - \beta_2)/2)).
	\eeqs
The breather solution (\ref{e:breather-eom-phi-2nd-order}), corresponds to $\bar \beta_2 = \bar \pi_2 = 0$ and $\bar \beta_1 = 2 \vphi$. Equations for perturbations to the breathers are
	\beq
	\DD{}{\tl t} \begin{smmat} \del \beta_1 \\ \del \pi_1\\ \del \beta_2 \\ \del \pi_2  \end{smmat} 
	= - \begin{smmat} 0 &  -2 & 0 & 0 \\ \half \cos (\bar \beta_1 /2) + \cos \bar \beta_1  & 0 & 0 & 0 \\ 	0 & 0 & 0 & -6 \\ 0 & 0 & \half \cos (\bar \beta_1 /2) & 0 \end{smmat} 
	\begin{smmat} \del \beta_1 \\ \del \pi_1\\ \del \beta_2 \\ \del \pi_2  \end{smmat}.
	\eeq
The monodromy matrix is block diagonal in this new basis and $\tr M_1= \tr M_\parallel = 2$ so that $\tr M = 2 + \tr M_2 = 2 + \tr M_\perp$. We plot $\tr M$ in Fig.~\ref{f:LG-LD-breather-trMvsE} and \ref{f:R-breather-trMvsE}. Evidently, for $E < 8.97g$ breathers are unstable while they are stable for $E > 8.97g$. Since $\tr M$ vanishes at $E = 8.97g$, we expect the rotational breather with period $\tau \approx 2.63$ (in units where $m = r = g = 1$) to undergo a period-doubling bifurcation.

At $E \approx 8.97 g$, the monodromy matrix $M$ has two eigenvalues $\pm 1$, each with multiplicity two. However, there is only one linearly independent eigenvector corresponding to each eigenvalue: for $1$ it is the sliding eigenvector $(1,0,0,0)^t$ and for $-1$ it is the transverse eigenvector $(0,0,1,0)^t$. As before, these eigenvectors pertain to $M$ evaluated at the basepoint $G$ on the breather orbits.

To find the newly born trajectory at the bifurcation point, we perturb the IC of the breather ($\bar \beta_1(0) = 0, \bar \beta_2(0) = \bar \pi_2(0) = 0$ and $\bar \pi_1(0) = \sqrt{E}$, in units where $m = g =r = 1$) along the transverse eigenvector with amplitude $\del \beta_2$. To search for a newly born periodic trajectory we consider the initial conditions $\beta_1(0) = 0, \beta_2(0) = \del \beta_2, \pi_1(0) = \bar \pi_1(0) + \del \pi_1, \pi_2(0) = 0$. The resulting trajectory is evolved untill a time $t_* \approx 2 \tau$ when $\beta_1$ reaches the value $8 \pi$ as $\beta_1$ has a periodicity of $4\pi$ for the breather. In order that the resulting trajectory be periodic, we use our search algorithm to adjust the value of $\pi_1(0)$ to minimize the `error'
	\beqs
	{\rm Er}(\pi_1(0)) &=& [(\beta_2(t_*) - \beta_2(0))^2 + (\pi_1(t_*) - \pi_1(0))^2
	\cr && + (\pi_2(t_*) - \pi_2(0))^2]^{1/2}.
	\eeqs
Proceeding in this manner we discover a family of stable ($\tr M < 4$) newly born rotational periodic trajectories which exist for energies $E < 8.97g$. Thus, by contrast with the bifurcations of the pendulum family, this is a `backward' fork-like bifurcation. In fact, we find that the slopes of $\tr M(E)$ of the newly born and parent breather family at the bifurcation point satisfy the period-doubling FLB slope theorem (\ref{FLB-slope-theorem-PD}). As we move away from the bifurcation point, the time periods of this new family increase from the period-doubling value $2 \tau$ while $\tr M$ decreases from $4$. These properties are illustrated in Fig.~\ref{f:breather-pd-monod-vs-E}, where the shape of the newly born trajectory on the $\beta_1-\beta_2$ plane is also shown.

\section{Are there stable periodic orbits in the band of global chaos?} 
\label{s:global-chaos-band}

Numerical evidence presented in Ref. \cite{gskhs-3rotor,gskhs-3rotor-ergodicity} indicated a band of `global chaos' in the 3-rotor system for $5.33g\lesssim E \lesssim 5.6g$. To be truly globally chaotic, there must not be any stable periodic orbit in this energy range. On more detailed inspection, a numerically suggested band of global chaos may fail to satisfy this condition. For instance, in Chirikov's standard map, although numerically there appears to be global chaos for sufficiently large values of coupling $k$, arbitrarily close to any such $k$, the dynamics is known\cite{elliptic-isles} to admit plenty of elliptic islands. In this paper, we have discovered families of periodic orbits that are born at fork-like bifurcations of periodic pendulum and breather orbits. We may, thus, ask whether any of these parent or daughter periodic orbits is stable in the above energy band. Remarkably, our results on the stability index ($\tr M$) for these orbits indicate that none of them is likely to be stable in this band of chaos. To begin with, pendula and breathers are unstable in this band. In fact, Figs.~\ref{f:TrMvT-pendl} and \ref{f:breather-pd-monod-vs-E} show that $|\tr M-2|>2$ for them. What is more, we will now argue that the families of periodic orbits born at the $\IS$ and $\PD$ bifurcations of pendula become unstable at energies much smaller than $5.33g$ and tend to become increasingly unstable as the above band is approached. For instance, Fig.~\ref{f:trM-v-T-lib-rot-pend-IS} shows that the sequences $\IS^\ell_{2n}$, $\IS^r_{2n-1}$ and $\PD^r_{2n+1}$ for $n = 1,2,3,\ldots$, are born unstable and remain unstable. As for the remaining four sequences ($\PD^\ell_n$, $\IS^\ell_{2n-1}$, $\IS^r_{2n}$ and $\PD^r_{2n+2}$), it is possible to see that all families in a sequence are unstable if the first one is unstable in the band of global chaos. This is because, as we go down a sequence, the graphs of $\tr M(E)$ are similar except that their slopes increase in magnitude. What is more, the first members $\PD^\ell_1, \IS^\ell_1, \IS^r_2$ and $\PD^r_4$ become unstable at $E \approx 4.09g, 4.04g, 4.02g$ and  $4.0001g$. Above these energies they seem to remain unstable. Hence, we expect all orbits in these four sequences to be unstable in the chaotic band. It remains to examine the two exceptional families $\PD^r_1$ and $\PD^r_2$. $\PD^r_1$ exists only for energies above those of the chaotic band while $\PD^r_2$ becomes unstable at $E \approx 4.55g$ and appears to remain unstable thereafter. Thus, pendula and their daughter periodic trajectories born at stability transitions appear to be unstable in the chaotic band. Finally, we consider the family of periodic orbits born at the reverse fork-like $\PD$ bifurcation of the breather family at $E \approx 8.97g$. Figure \ref{f:breather-pd-trMvE} shows that this family is stable down to $E \approx 8.27g$ but becomes unstable at lower energies. In conclusion, none of the periodic orbits we have examined appears to be linearly stable in the chaotic band. This provides additional evidence for the globally chaotic nature of this band. Of course, there are many other periodic orbits that we have not examined, including the new families born at bifurcations of the above-mentioned daughter trajectories.

\section{Discussion}
\label{Discussion}

A summary of our results may be found in \S \ref{s:summary-results}. Here, we discuss some open questions arising from our work on the three-rotor problem.

1. Pendula undergo a doubly infinite geometric cascade of bifurcations as $E \to 4g$ from librational as well as rotational phases. The accumulation energy is also the energy at which widespread chaos sets in \cite{gskhs-3rotor}. Moreover, the band of global chaos $(5.33g \lesssim E \lesssim 5.6g)$ lies in the energy interval between $\PD^r_1$ and $\PD^r_2$ where pendula are unstable. Intriguingly, the band of global chaos appears to terminate at the energy of the last pendulum stability transition $E(\PD^r_1) \approx 5.6g$, beyond which pendula are stable. Thus, it would be interesting to further explore a possible link between chaos and the cascade of pendulum bifurcations.


2. This pendulum bifurcation cascade may remind the reader of the cascade of period-doubling bifurcations in the logistic map \cite{FGB} and area preserving maps of the plane \cite{FGB-area-pres}. However, there is a distinction: while we follow successive (isochronous and period-doubling) bifurcations of a fixed parent pendulum family, in these maps one follows period-doubling bifurcations of the new stable orbits born at each bifurcation. By analogy with the universality of the Feigenbaum constants, one wonders whether there is a class of systems that share scaling constants with the 3-rotor system and whether one can develop a renormalization group method to address this behavior. In this context, it is noteworthy that the scaling constant $\del$ for the pendulum bifurcation cascade depends on (i) the curvature $\om_\perp^2$ (in the stable direction) of the potential $V(\al_1, \al_2)$ at the saddle point $D_3$ (\ref{shm-apprx}) and on (ii) the prefactor $N$ [-$2/\om_0$ for libration (\ref{asy-TvE-lib}) and -$1/\om_0$ for rotation (\ref{asy-TvE-rot})] of the logarithm in the asymptotic pendulum time period  $\tau(E)$. In fact, for pendula, $\del = e^{-2\pi/(\om_\perp N)}$. Consequently, other systems with a similar cascade can have the same Feigenbaum constant if they share the value of $\om_\perp N$. Incidentally, $\del$ for the A orbits of H\'enon-Heiles \cite{brck-omega} are not the same as for the 3-rotor pendula.

3. For the newly born families of orbits $\PD^r_n$ and $\IS^{\ell, r}_n$, the variable in the transverse direction $\al_1$ is either $2 K(k)$ or $4 K(k)$ periodic as a function of $z = \om_0 t$ (\ref{e:lib-pend-lame-eqn}) and $z = \om_0 t / \ka$ (\ref{e:rot-pend-lame-eqn}). In these cases, we have expressed the solution $\al_1$ of the transverse perturbation equation in terms of the periodic Lam\'e functions $\Ec$ and $\Es$ \cite{Ince, erdelyi}. However, for the $\PD^\ell_n$ families, $\al_1$ is $8K(k)$ periodic. It would be interesting to find suitable expressions for the latter in terms of periodic Lam\'e functions. This would generalize the results of Ref.~\cite{brck-lame} for period-doubling bifurcations of the analogous orbits in a quartic anharmonic oscillator.

4. We have proposed an asymptotic duality between isochronous bifurcations of librational pendula and bifurcations at all stability transitions of rotational pendula (except $\PD^r_{1,2}$). It would be nice to explain this duality via a symmetry and also identify the `missing' rotational bifurcations dual to period-doubling bifurcations of librational pendula.

5. We have presented numerical evidence for 3-rotor fans: confluences of graphs of $\tr M(E)$ for families of newly born orbits of a given class such as $\IS^\ell_{1,3,5}$. Can we explain these fan-like confluences analytically, perhaps by deriving (asymptotic) formulas for $\tr M(E)$ for the newly born orbits?

6. At low energies $E \gtrsim 0$, we are aware of three families of periodic trajectories: pendula, breathers and choreographies \cite{gskhs-3rotor}. We would like to know if there is a sense in which the static solution $G$ (at $E = 0$) bifurcates into these periodic orbits as $E$ is increased. We also hope to extend our results on the stability and bifurcations of pendula and breathers to choreographies. In particular, one would like to understand the nature of a possible bifurcation that nonrotating choreographies undergo at the edge of the band of global chaos ($E \approx 5.33g$).



7. We observed scale-invariance in the stability indices and shapes of new periodic orbits in the pendulum bifurcation cascade. It would also be interesting to look for `local scale-invariance' in Poincar\'e sections at these bifurcation energies as reported\cite{santhanam} for Hamiltonians with homogeneous potentials.


8. Finally, we would like to investigate quantum manifestations of chaos in the three-rotor problem. The classes of periodic orbits we have found should help in addressing this question in the semiclassical approximation.

\begin{acknowledgments}

We thank S R Jain, K Kumari, A Lakshminarayan, J D Meiss, H Senapati and an anonymous referee for helpful discussions and comments. This work was supported in part by the Infosys Foundation and grants (MTR/2018/000734, CRG/2018/002040) from the Science and Engineering Research Board, Govt. of India.

\end{acknowledgments}

\appendix
\section{Connection to a superconducting persistent current qubit}
\label{s:JJ-qubit}

In this Appendix, we relate the relative dynamics of the 3-rotor system to that of a three Josephson junction superconducting persistent current `flux' qubit. This relation could lead to an experimental realization of our model. In fact, from Eqn.~(1) of Ref.~\cite{mooij}, we observe that the Josephson coupling energy (with the sign of $\vphi_2$ reversed) 
	\beq
	U = E_j [2 + \al - \cos \vphi_1 - \cos \vphi_2 -\al \cos( 2\pi f + \vphi_1 + \vphi_2)]
	\eeq
reduces to our potential energy $V$ (\ref{e:lagr-poten-phi1-phi2}) if we identify $g$ with the Josephson coupling $E_j$, take all junction capacitances to be equal ($\al = 1$) and assume that there is no external magnetic flux in the loop ($f = 0$). On the other hand, up to an additive constant, the capacitive charging energy of the three junctions and two gates [(3) and (4) of Ref.~\cite{mooij} with $\vphi_2 \to - \vphi_2$] is
	\beq
	T = \fr{C}{2} \left(\fr{\hbar}{2 e} \right)^2 \colvec{2}{\dot \vphi_1}{\dot \vphi_2}^t \colvec{2}{1+\al +\g & \al}{\al & 1+ \al + \g} \colvec{2}{\dot \vphi_1}{\dot \vphi_2}
	\eeq
where $e$ is the electron charge, $C$ the capacitance of junctions 1 and 2 and $\al C$ that of the third junction (see Fig.~1 of Ref.~\cite{mooij}). This too reduces to our kinetic energy $(mr^2/3)(\dot \vphi_1^2 + \dot \vphi_2^2 + \dot \vphi_1 \dot \vphi_2)$ (\ref{e:lagr-poten-phi1-phi2}) if we identify $mr^2/3 = C (\hbar/2 e)^2$, assume that there are no gate capacitances ($\g =0$) and take $\al=1$. Thus, capacitances play the role of masses. Despite this equality of rotor and qubit energies, there is an important difference. While our rotors ($\tht_{1,2,3}$) have definite masses, it is the junctions $(\vphi_{1,2,3})$ (rather than superconducting segments) that have definite capacitances. In other words, while the mass matrix is diagonal in the $\dot \tht_{1,2,3}$ basis, the capacitance matrix is diagonal in the $\dot \vphi_{1,2,3}$ basis.

\section{Stability indices of pendula in terms of Lam\'e functions}
\label{s:new-lame-fns}

Here we obtain a formula for the stability index $\tr M_\perp$ for transverse perturbations to the pendulum family of orbits with period $\tau$. Recall from (\ref{e:lib-pend-lame-eqn}) and (\ref{e:rot-pend-lame-eqn}) that these perturbations are governed by the Lam\'e equation
	\beq
	\del \ddot \al_1 (z) + (h - n (n + 1)\, k^2 \sn^2 (z,k)) \del \al_1(z) = 0.
	\label{e:std-lame-eqn-again}
	\eeq
Here dots denote $z$-derivatives where $z = \om_0 t = \sqrt{3} \tl t$, $n(n+1) = 2/3$ and $k^2 = E/4g$ or $4g/E$ for libration and rotation. The Lam\'e eigenvalue $h = 1$ and $k^2$ for libration and rotation.

For our current purposes, it is convenient to denote two linearly independent solutions of (\ref{e:std-lame-eqn-again}) by $\Ls_n(z,k;h)$ and $\Lc_n(z,k;h)$. Although not necessary for the results in (\ref{e:delal1-intermsof-Ls-Lc}), for definiteness and to facilitate comparison with Mathieu functions in the limit $k \to 0$, we may suppose that they satisfy the ICs $\Ls_n(0,k;h) = 0$ and $\dot \Lc_n(0,k;h) = 0$. The functions $\Lc$ and $\Ls$ which are defined for arbitrary `eigenvalues' $h$ are to be distinguished from the (periodic) Lam\'e functions that appear in the works of Ince and Erd\'elyi \cite{Ince,erdelyi}. These authors define the $m^{\rm th}$ Lam\'e functions of order $n$ denoted $\Ec_n^m(z,k)$ and $\Es_n^m(z,k)$. They correspond to a discrete set of Lam\'e eigenvalues denoted $h = a_n^m(k), b_n^m(k)$. In these Lam\'e functions, $m$ plays the role that $h$ did in $\Lc$ and $\Ls$. $\Ec$ and $\Es$ are $2K(k)$ periodic for even $m = 2, 4, 6, \ldots$ and $4K(k)$ periodic if $m$ is odd. In either case, $m$ is the number of zeros of $\Ec$ and $\Es$ in the interval $0 \leq z < 2K(k)$ \cite{brck-lame}.

For small $k^2$, (\ref{e:std-lame-eqn-again}) reduces to the Mathieu equation
	\beq
	\del \ddot \al_1 (z) + \left[h - \fr{n(n+1)}{2}k^2 + \fr{n(n+1)}{2}k^2 \cos 2 z\right] \del \al_1(z) = 0.
	\label{e:mathieu-limit-of-lame}
	\eeq
Comparing with the standard form\cite{abramowitz-stegun} $\xi''(z) + (a - 2 q \cos 2z) \xi(z) = 0$, we read off the Mathieu parameters $a = h - n(n+1)k^2/2$ and $q = - n(n+1)k^2/4$. Two linearly independent solutions of this equation are the Mathieu sine and cosine functions $\Ms_{a,q}(z)$ and $\Mc_{a,q}(z)$ which satisfy $\Ms(0) = 0$ and $\Mc'(0) = 0$. Thus, for small $k^2$, our Lam\'e functions $\Ls_n(z,k;h)$ and $\Lc_n(z,k;h)$ reduce to these Mathieu functions.

Now, $\tr M_\perp$ is equal to the trace of the fundamental matrix solution $U(z,0)$ (with $U(0,0) = I$) of (\ref{e:std-lame-eqn-again}) evaluated at $\om_0 \tau$:
	\beq
	U(\om_0 \tau, 0) = \begin{smmat} \del \al_1^1(\om_0 \tau) & \del \al_1^2(\om_0 \tau) \\ \del \dot \al_1^1(\om_0 \tau) & \del \dot \al_1^2(\om_0 \tau) \end{smmat}.
	\eeq
[Caution: $M_\perp \ne U(\om_0 \tau,0)$ since the former is in the $(\del \al_1, \del \tl \pi_1)$ basis while $U$ is in the $(\del \al_1, \del \dot \al_1 = (1/2\sqrt3) \del \tl \pi_1)$ basis (\ref{e:dim-less-t-mom}).] Interestingly, solutions of (\ref{e:std-lame-eqn-again}) with ICs ($\del \al_1^1(0) = 1, \del \dot \al_1^1(0) = 0, \del \al_1^2(0) = 0, \del \dot \al_1^2(0) = 1$) can be expressed as \footnotesize
	\beqs
	\del \al_1^1(z) &=& \fr{\Ls_n(z,k;h) \dot \Lc_n(0,k;h) - \Lc_n(z,k;h) \dot \Ls_n(0,k;h)}{W(\Ls_n, \Lc_n)(z,k;h)} \quad \text{and} 
	\cr
	\quad 
	\del \al_1^2 (z) &=& \fr{\Ls_n(0,k;h) \Lc_n(z,k;h) - \Ls_n(z,k;h) \Lc_n(0,k;h)}{W(\Ls_n, \Lc_n)(z,k;h)}.
	\label{e:delal1-intermsof-Ls-Lc}
	\eeqs \normalsize
Here we used the fact that the Wronskian $W(\Ls_n, \Lc_n)(z,k;h) = \Ls_n(z,k;h) \dot \Lc_n(z,k;h) - \Lc_n(z,k;h)$ $ \dot \Ls_n(z,k;h)$ is independent of $z$. Now, taking a trace,
	\beqs
	\tr M_\perp &=& \tr U(\om_0 \tau,0) = \del \al_1^1(\om_0 \tau) + \del \dot \al_1^2(\om_0 \tau) \cr
	&=& (1/W(\Ls_n,\Lc_n)) [ \Ls_n(0,k;h) \dot \Lc_n(\om_0 \tau,k;h) \cr && - \dot \Ls_n(0,k;h) \Lc_n(\om_0 \tau,k;h) - \dot \Ls_n(\om_0 \tau,k;h)
	\cr && \Lc_n(0,k;h) + \Ls_n(\om_0 \tau,k;h) \dot \Lc_n(0,k;h) ].
	\eeqs
The computational utility of this formula would be enhanced once approximate values of the Lam\'e functions $\Lc_n$ and $\Ls_n$ are available as with $\Ec$ and $\Es$. 


\end{document}